\documentclass[twocolumn,twocolappendix,tighten,resetfootnote]{aastex701}
\usepackage{amsmath}
\usepackage{comment}
\usepackage{enumitem}
\usepackage{graphicx}
\usepackage{lmodern}
\usepackage{booktabs}
\usepackage{multirow}
\usepackage{rotating}
\usepackage{placeins}
\usepackage{hyperref}

\providecommand{\apjl}{ApJL}

\providecommand{\aap}{A\&A}

\newcommand{\UMich}{Department of Astronomy, University of Michigan, Ann Arbor, MI 48109, USA}
\newcommand{\CHARA}{The CHARA Array of Georgia State University, Mount Wilson Observatory, Mount Wilson, CA 91023, USA}

\shorttitle{CHARA Near-IR Limb Darkening Benchmark}
\shortauthors{Anugu et al.}
\graphicspath{{.}{Figures_old/}}

\begin{document}

\title{Empirical H- and K-band Limb Darkening for 31 CHARA Stars: A Near-Infrared Benchmark for Stellar-Atmosphere Models}
\correspondingauthor{Narsireddy Anugu}

\author[0000-0002-2208-6541]{Narsireddy Anugu}
\affiliation{\CHARA}
\email[show]{nanugu@gsu.edu}  

\author[0000-0002-3380-3307]{John D. Monnier}
\affiliation{\UMich}
\email{monnier@umich.edu}

\author[0000-0003-2125-0183]{Antoine M\'erand}
\affiliation{European Southern Observatory, Karl-Schwarzschild-Str. 2, 85748 Garching, Germany}
\email{amerand@eso.org}

\author[0009-0005-8004-2351]{Becky Flores}
\affiliation{\CHARA}
\email{bflores5@gsu.edu}

\author[0000-0001-7853-4094]{Alexandre Gallenne}
\affiliation{Instituto de Alta Investigaci\'on, Universidad de Tarapac\'a, Casilla 7D, Arica, Chile}
\email{agallenne@academicos.uta.cl}

\author[0000-0001-8537-3583]{Douglas R. Gies}
\affiliation{\CHARA}
\email{dgies@gsu.edu}

\author[0009-0006-0225-4444]{Mayra Gutierrez}
\affiliation{\UMich}
\email{mgutie60@ucsc.edu}

\author[0000-0002-4313-0169]{Robert Klement}
\affiliation{European Organisation for Astronomical Research in the Southern Hemisphere (ESO) Casilla 19001, Santiago 19, Chile}
\affiliation{Université Côte d’Azur, Observatoire de la Côte d’Azur, CNRS, Boulevard de l’Observatoire, CS 34229, 06304 Nice Cedex 4, France}
\affiliation{\CHARA}
\email{robertklement@gmail.com}

\author[0000-0001-6017-8773]{Stefan Kraus}
\affiliation{Astrophysics Group, Department of Physics \& Astronomy, University of Exeter, Stocker Road, Exeter, EX4 4QL, UK}
\email{S.Kraus@exeter.ac.uk}

\author[0000-0002-2488-7123]{Jayadev Rajagopal}
\affiliation{NSF's National Optical-Infrared Astronomy Research Laboratory, 950 N.\ Cherry Ave., Tucson, AZ 85719, USA}
\email{jayadev.rajagopal@noirlab.edu}

\author[0000-0002-9288-3482]{Rachael M. Roettenbacher}
\affiliation{\UMich}
\email{rmroett@umich.edu}

\author[0000-0001-5415-9189]{Gail H. Schaefer}
\affiliation{\CHARA}
\email{gschaefer@gsu.edu}

\begin{abstract}
Limb darkening, the decrease in stellar intensity from the disk center to the
limb, encodes the temperature structure and opacity of stellar atmospheres.
Direct spatially resolved measurements of this center-to-limb variation remain
scarce, especially in the near-infrared.
We present interferometric limb-darkening measurements for 31 stars observed
simultaneously in the $H$ and $K$ bands with the CHARA Array.
The sample spans spectral types F--M and luminosity classes IV--I.
The targets are well resolved in $H$ and, for most targets, also in $K$.
This coverage constrains the visibility curvature associated with limb
darkening in joint $H+K$ fits.
We fit the combined $H{+}K$ squared visibilities with four analytic limb-darkening laws and compare the resulting coefficients with bandpass-matched predictions from five stellar-atmosphere grids (Kurucz, MPS1, MPS2, Stagger, and spherical SATLAS, with reported coefficients placed on the Rosseland-radius convention) spanning one-dimensional plane-parallel, three-dimensional radiation--hydrodynamic, and spherical low-gravity models. The associated limb-darkened angular diameters are measured with median
formal precisions of $\simeq 0.2$--0.3\%.
The CHARA results show expected weaker limb
darkening at longer wavelengths and also with increasing
$T_{\rm eff}$. The clearest discrepancy with the atmosphere grids is in the
wavelength dependence: the median fractional decrease in the power-law
coefficient from $H$ to $K$ is $\simeq 39\%$ in the CHARA sample, compared with
only $\simeq 17$--$22\%$ with model predictions. Relative to the MPS2 comparison, the
empirical coefficients are higher by $\simeq 21\%$ in $H$ and lower by
$\simeq 5\%$ in $K$. 
Synthetic recovery tests indicate that the excess scatter is unlikely to be explained solely by incomplete spatial-frequency coverage or numerical instability in the fitting pipeline.
These results provide multi-band interferometric limb darkening as a
near-infrared benchmark for stellar-atmosphere models. 
\end{abstract}


\section{Introduction}\label{sec:introduction}

Limb darkening (LD), the center-to-limb intensity variation (CLV), is a
direct manifestation of the temperature gradient, opacity structure, and
geometric extension of a stellar atmosphere. Because it shapes the emergent
intensity profile across the stellar disk, limb darkening enters a wide range of
astrophysical inferences, including interferometric angular diameters, eclipsing
binary analyses, exoplanet transit light curves, and high-precision stellar
characterization. In practice, most of these applications rely on
model-atmosphere predictions \citep[e.g.][]{claret_new_2000,Howarth2011MNRAS.418.1165H} rather than direct empirical measurements,
especially in the near-infrared where spatially resolved constraints remain
limited.

The need for accurate limb-darkening prescriptions has become more pressing in
the era of precision space-based photometry and spectroscopy. Observations from
facilities such as JWST \citep[e.g.,][]{Rustamkulov2023} can reach a level of
precision at which systematic uncertainties in the adopted limb-darkening law
are no longer negligible \citep{Coulombe2024AJ,Verma2024}. 
At the same time, widely used stellar-atmosphere
grids do not always agree with one another, and in several cases empirical
measurements have shown mismatch with model predictions
\citep[e.g.,][]{claret_gravity_2011,Magic2015,Kervella2017}.

Most empirical constraints on limb darkening have come from transit, eclipsing
binaries, and microlensing event observations. These methods probe the stellar disk
indirectly, either by occulting part of the star or by differentially magnifying
the stellar surface \citep[e.g.,][]{Southworth2008,Afonso2000}.
Long-baseline optical interferometry provides a complementary and more direct
test for sufficiently resolved stars. The complex visibility is the Fourier
transform of the sky brightness distribution, so the measured squared visibility $V^2$ samples the
radial brightness profile of the stellar disk
\citep[e.g.,][]{Aufdenberg2005,Lacour2008A&A...485..561L,
Merand2010A&A...517A..64M,Kervella2017,2026MNRAS.548ag719C}. Once a star is
resolved beyond the first visibility null, the curvature of the $V^2$ profile
becomes especially sensitive to the underlying CLV, with different degeneracies
from transit or eclipse light-curve analyses.

Direct near-infrared interferometric limb-darkening measurements remain
relatively scarce, especially for evolved stars, with only a small number of dedicated case studies
\citep[e.g.,][]{Wittkowski2001,Lacour2008A&A...485..561L,
Merand2010A&A...517A..64M, Kervella2017}. Many previous interferometric studies measured
angular diameters while adopting fixed, model-based limb-darkening corrections
rather than fitting the coefficients directly in the visibility domain
\citep[e.g.,][]{HanburyBrown1974,van_belle_directly_2009,boyajian_stellar_2012,Baines2025}.
A systematic comparison between visibility-domain limb darkening and modern
stellar-atmosphere predictions, using matched passbands and stellar parameters,
has not yet been made for a larger target sample.

In this paper, we measure limb darkening for a sample of 31
resolved bright stars (\mbox{Table~\ref{tab:target_properties}}) spanning spectral types F--M and 
luminosity classes IV--I with the Center for High Angular Resolution Astronomy (CHARA) Array \citep{ten_brummelaar_first_2005}. 
We use simultaneous $H$- and $K$-band observations with MIRC-X \citep{Anugu2020} and
MYSTIC \citep{Setterholm2023}.
We fit analytic limb-darkening laws in the visibility domain, derive empirical
coefficients and angular diameters, and compare them to bandpass-matched
predictions from five atmosphere-model grids (see \mbox{Tables~\ref{tab:methods}} and \ref{tab:model_grids}): Kurucz ATLAS \citep{Kurucz1993}, Merged Parallelised Simplified-ATLAS set 1 (MPS1) and set 2 (MPS2) \citep{Kostogryz2022}, Stagger \citep{Magic2013,Magic2015}, and spherical ATLAS/SATLAS \citep{Neilson2011,Neilson2013}. These grids span one-dimensional plane-parallel, three-dimensional radiation--hydrodynamic, and spherical low-gravity atmosphere models. Our goal is to ask a direct question: do the limb-darkening profiles measured by CHARA agree with current atmosphere models, especially in how they change from $H$ to $K$ band?

The paper is organized as follows. In Section~\ref{sec:target_sample}, we
describe the target selection. Section~\ref{sec:obs_data_redu} describes the
observations and data reduction. In Section~\ref{sec:methods}, we present the
visibility-domain fitting of CHARA data, the adopted limb-darkening laws, the
atmosphere-grid coefficients, and the fixed-CLV diameter fits. In
Section~\ref{sec:results}, we compare the empirical CHARA coefficients with the
model predictions, quantify the wavelength-dependent residuals, assess possible
systematics, and compare the analytic-law and fixed-CLV fit angular diameters. In
Section~\ref{sec:discussion}, we summarize the implications for
stellar-atmosphere models and for applications that rely on model-based
limb-darkening coefficients. Conclusions are given in
Section~\ref{sec:conclusion}.

\begin{figure}[htbp]
\centering
\includegraphics[width=0.499\textwidth]{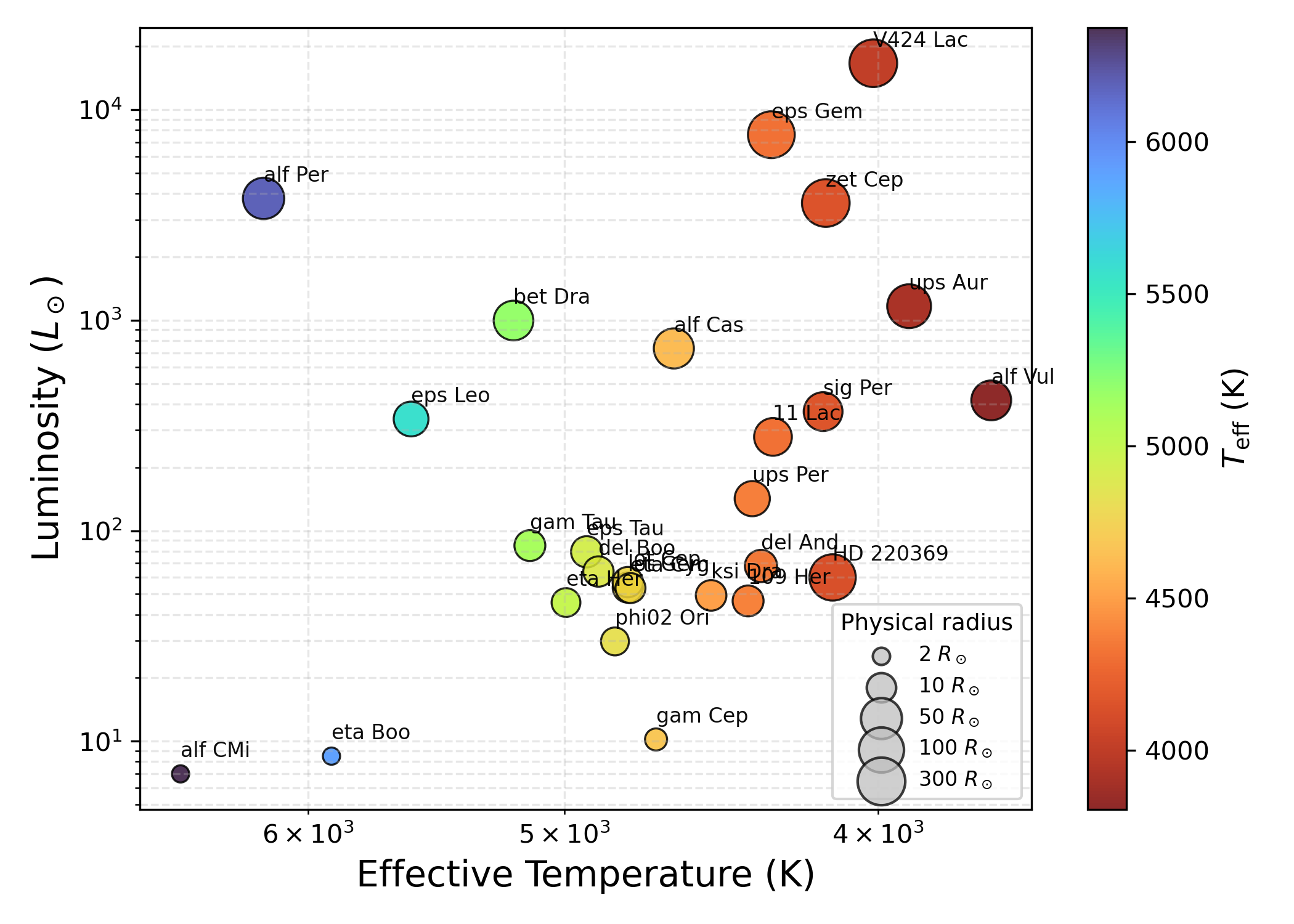}
\caption{Hertzsprung--Russell diagram of the CHARA limb-darkening target sample.
Points are colored by effective temperature and scaled by the physical star radius 
(see \mbox{Table~\ref{tab:target_properties}}). Targets span
$T_{\rm eff}\sim3690-6570$~K and include subgiants, giants and supergiants.}
\label{fig:Targets}
\end{figure}

\section{Target Sample}\label{sec:target_sample}

Our observed target sample consists of 31 bright and nearby stars.
\mbox{Table~\ref{tab:target_properties}} summarizes the target stellar properties,
including spectral type, effective temperature $T_{\rm eff}$, surface gravity $\log g$, and
metallicity $[{\rm M/H}]$. It also lists the combined H+K power-law limb-darkened angular diameter,
$\theta_{\rm PL}$, from our primary CHARA fit, and the stellar radius
derived from $\theta_{\rm PL}$ and Gaia EDR3 distance estimates (see  Section~\ref{sec:methods}). The final 31-target sample spans
$T_{\rm eff}\approx3690$--$6570$~K, $\log g=0.32$--$4.04$, and
$[{\rm M/H}]\approx-0.54$ to $+0.25$~dex. The full radius range is
$2.05$--$309.12\,R_\odot$, but the sample is dominated by stars with
$R<50\,R_\odot$; 25 of the 31 targets lie between $2.05$ and
$44.79\,R_\odot$. \mbox{Figure~\ref{fig:Targets}} places the targets in the
Hertzsprung--Russell diagram.
The literature values of $T_{\rm eff}$, $\log g$, and $[{\rm M/H}]$
listed in \mbox{Table~\ref{tab:target_properties}} are taken from \citet{Baines2025} and
\citet{Soubiran2016}, while the listed stellar radii are recomputed here from
our fitted $\theta_{\rm PL}$ values and Gaia EDR3 distances.
The sample is a benchmark set, selected for reliable visibility-domain
sensitivity to the CLV rather than for completeness.

The targets span spectral types F through M and are primarily subgiants and
giants. The sample is therefore intended as a benchmark set covering a range
of atmospheric regimes. The evolved subset is especially favorable for
interferometry because its larger apparent diameters and more extended
atmospheres provide sensitive tests of stellar-atmosphere grids
\citep[e.g.,][]{Neilson2013,Magic2015,Kostogryz2024,Hauschildt2025A&A...698A..47H}.

\begin{figure*}[htbp]
\centering
\includegraphics[width=\textwidth]{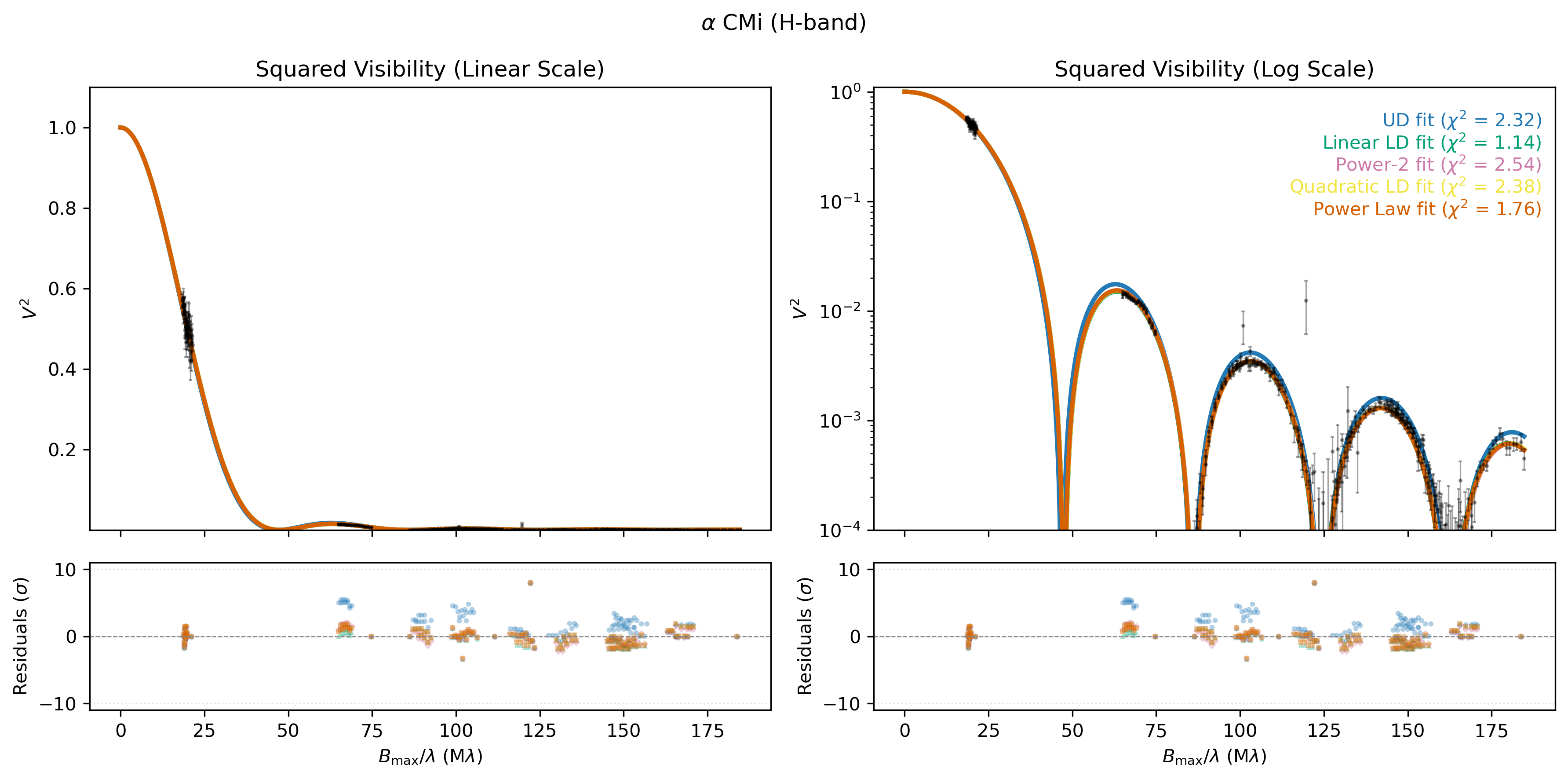}
\includegraphics[width=\textwidth]{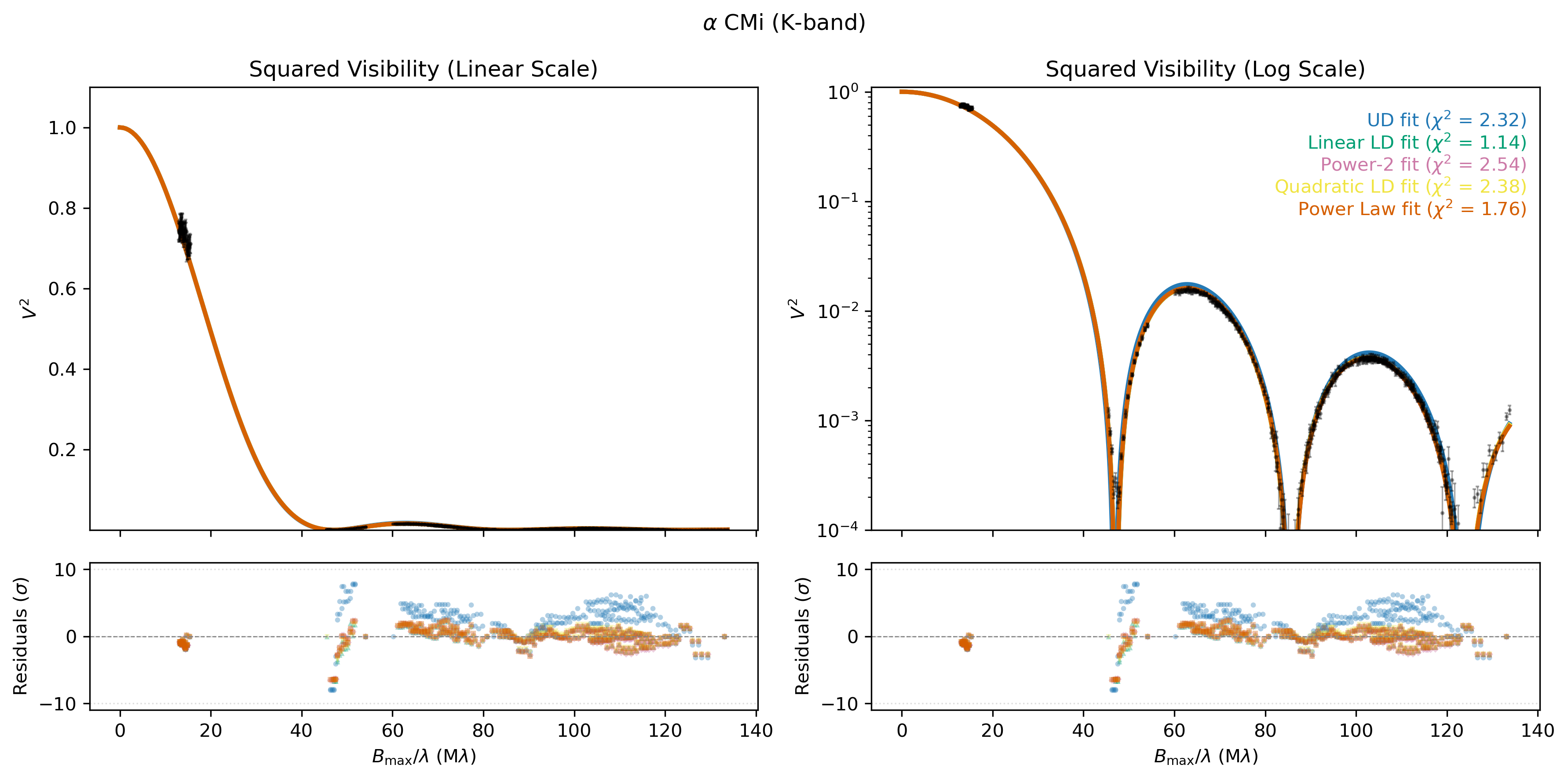}
\caption{Squared visibility $V^2$ and residuals for the limb-darkening fits of $\alpha$ CMi (HD 61421, Procyon, angular diameter $5.417 \pm 0.005$ mas) in the $H$ band (top row; MIRC-X) and $K$ band (bottom row; MYSTIC).
For each band, $V^2$ is shown as a function of spatial frequency
($B_{\rm max}/\lambda$) on linear (left) and logarithmic (right) scales. 
The linear-scale panels highlight the short-baseline $V^2$ quality. The E1
telescope was unavailable for this observation because of delay-line
constraints.
Observed data are plotted in black with $1\sigma$ uncertainties, and best-fit models are
overplotted for the uniform-disk baseline and the analytic limb-darkening laws
considered in this work (see Eq.~\ref{eq:laws}). Residuals are shown in the bottom panel. Multi-lobe coverage makes the visibility curvature sensitive to
limb darkening, allowing different prescriptions to be distinguished directly in
the measurement domain, although the laws can still look very similar for some
targets because they share the same null structure and overall envelope. The complete figure set (42 target/epoch pairs; 84 band-specific plots) is available in the online version of the article and in the Zenodo figure-set archive \citep{AnuguZenodo2026}. }
\label{fig:V2_fit}
\end{figure*}

We selected targets that are both bright enough ($H<3$ mag) for stable
fringe tracking and for calibrated CHARA $V^2$ measurements at the
$\sim10^{-3}$ precision level, and sufficiently resolved in both $H$ and $K$
to constrain the curvature of the squared-visibility function.
We therefore prioritized nearby large stars with angular diameters
large enough for the data to extend beyond the first visibility null and into
the second visibility lobe (example, \mbox{Figure~\ref{fig:V2_fit}}). In the final sample, all 31 targets have two-null
coverage in the $H$ band, and 24 have two-null coverage in the $K$ band.
Here the null/lobe coverage is referenced to the fitted angular diameter $\theta_{\rm PL}$ reported later in \mbox{Table~\ref{tab:ld_summary}}.
Our synthetic data analysis shows that a diameter threshold of about
1.7~mas is required to recover the limb-darkening coefficients reliably
(Appendix~\ref{app:alpha_diameter_recovery_test}); all targets in the final
sample are larger than 2.1~mas (see \mbox{Table~\ref{tab:ld_summary}}). This spatial-frequency
coverage allows us to distinguish between limb-darkening profiles, rather than
merely measuring a uniform-disk angular diameter.

The initial candidate list was assembled using the JMMC Stellar Diameters Catalogue version 2 \citep[JSDC;][]{Chelli2016,Bourges2017}.
We chose stars with diameters between 2 and 6~mas, with $H$ or $K<4$~mag, $|H-R|<4$~mag, and declination
$>-15^\circ$. We kept stars with clean JSDC calibrator flags, allowing only
unflagged entries with a high diameter-fit $\chi^2$, while rejecting
known close doubles and object types that indicate possible binarity or
pulsation. These cuts yield 744 potential CHARA-observable targets. 
We observed 50 candidate targets with CHARA. 
From the observed sample, 19 targets were excluded for the following reasons: 5 showed evidence for
surface-brightness structure, 2 showed signatures consistent with binarity,
and 12 lacked sufficient data quality and/or spatial-frequency
leverage to provide robust limb-darkening constraints. For the retained sample,
the closure-phase companion search shows no evidence for companions within an interferometric field of view of roughly $100$--$270$~mas, nor for large-scale
surface-brightness asymmetries at the level of $\gtrsim1\%$ of the total flux (see Section~\ref{sec:results}).
A separate manuscript will present the subsample of targets showing evidence for binarity or
resolved surface structure.

\startlongtable
\begin{deluxetable}{r l l l r r r r r r}
\tablecaption{Target properties with a shared index ($ID$). Targets are sorted by effective temperature (hot to cool). Stellar parameters are taken from \citet{Baines2025} and \citet{Soubiran2016}. The Mass column lists the adopted stellar masses used for the SATLAS model selection; bracketed numbers are references for masses. Radii ($R/R_\odot$) are computed from the power-law angular diameters (Table~\ref{tab:ld_summary}) and Gaia EDR3 distance estimates \citep{BailerJones2021}.}\label{tab:target_properties}
\tablewidth{0pt}
\tablehead{
\colhead{$ID$} & \colhead{Target} & \colhead{HD} & \colhead{Spectral type} & \colhead{Teff (K)} & \colhead{$\log g$} & \colhead{[M/H]} & \colhead{Mass} & \colhead{$\theta_{\rm PL}$ (mas)} & \colhead{$R/R_\odot$} \\
}
\startdata
1 & $\alpha$ CMi & HD 61421 & F5IV-V+DQZ & 6570 & 4.04 & -0.04 & 1.478 [1] & 5.417 $\pm$ 0.005 & 2.05 $\pm$ 0.01 \\
2 & $\alpha$ Per & HD 20902 & F5Ib & 6193 & 1.02 & 0.14 & 6.5 [2] & 3.226 $\pm$ 0.003 & 53.76 $\pm$ 1.39 \\
3 & $\eta$ Boo & HD 121370 & G0IV & 5901 & 3.78 & 0.25 & 1.71 [3] & 2.222 $\pm$ 0.006 & 2.72 $\pm$ 0.05 \\
4 & $\epsilon$ Leo & HD 84441 & G1II & 5576 & 2.34 & -0.09 & 3.71 [4] & 2.662 $\pm$ 0.004 & 19.94 $\pm$ 0.36 \\
5 & $\beta$ Dra & HD 159181 & G2Ib-IIa & 5184 & 1.66 & 0.06 & 6.0 [5] & 3.342 $\pm$ 0.005 & 42.82 $\pm$ 0.50 \\
6 & $\gamma$ Tau & HD 27371 & G9.5IIIabCN0.5 & 5124 & 2.63 & 0.12 & 2.7 [6] & 2.399 $\pm$ 0.006 & 11.94 $\pm$ 0.12 \\
7 & $\eta$ Her & HD 150997 & G7IIIFe-1 & 4994 & 2.53 & -0.22 & 2.01 [7] & 2.526 $\pm$ 0.006 & 9.28 $\pm$ 0.05 \\
8 & $\eta$ Ser & HD 168723 & K0III-IV & 4948 & 3.18 & -0.19 & 1.6 [4] & 2.968 $\pm$ 0.004 & 6.08 $\pm$ 0.02 \\
9 & $\epsilon$ Tau & HD 28305 & K0III & 4921 & 2.47 & 0.14 & 2.458 [8] & 2.596 $\pm$ 0.008 & 12.46 $\pm$ 0.10 \\
10 & $\delta$ Boo & HD 135722 & G8IIIFe & 4880 & 2.48 & -0.36 & 1.5 [4] & 2.756 $\pm$ 0.003 & 10.93 $\pm$ 0.06 \\
11 & $\phi^2$ Ori & HD 37160 & G9IV & 4823 & 2.61 & -0.54 & 1.07 [9] & 2.251 $\pm$ 0.008 & 8.45 $\pm$ 0.06 \\
12 & $\iota$ Cep & HD 216228 & K0III & 4779 & 2.61 & 0.01 & 2.15 [4] & 2.843 $\pm$ 0.007 & 11.20 $\pm$ 0.06 \\
13 & $\iota$ Gem & HD 58207 & G9IIIb & 4779 & 2.40 & -0.10 & 1.89 [4] & 2.472 $\pm$ 0.005 & 11.02 $\pm$ 0.09 \\
14 & $\eta$ Cyg & HD 188947 & K0III & 4770 & 2.64 & 0.02 & 0.9 [9] & 2.287 $\pm$ 0.005 & 10.43 $\pm$ 0.05 \\
15 & $\gamma$ Cep & HD 222404 & K1III-IVCN1 & 4684 & 3.20 & 0.15 & 1.27 [10] & 3.201 $\pm$ 0.004 & 4.75 $\pm$ 0.01 \\
16 & $\alpha$ Cas & HD 3712 & K0IIICN+1 & 4625 & 1.19 & -0.10 & 4.98 [11] & 5.664 $\pm$ 0.010 & 43.14 $\pm$ 1.50 \\
17 & $\xi$ Dra & HD 163588 & K2III & 4504 & 2.31 & -0.01 & 1.45 [4] & 3.078 $\pm$ 0.003 & 11.40 $\pm$ 0.05 \\
18 & $\mu$ Leo & HD 85503 & K2IIIb\_CN1\_Ca1 & 4433 & 2.50 & 0.25 & 1.5 [12] & 2.854 $\pm$ 0.004 & 11.74 $\pm$ 0.08 \\
19 & 109 Her & HD 169414 & K2III & 4387 & 2.24 & -0.06 & 1.05 [4] & 2.996 $\pm$ 0.003 & 11.93 $\pm$ 0.06 \\
20 & $\upsilon$ Per & HD 9927 & K3-III & 4374 & 2.06 & 0.05 & 1.75 [4] & 3.615 $\pm$ 0.003 & 20.91 $\pm$ 0.20 \\
21 & $\delta$ And & HD 3627 & K3III & 4347 & 2.04 & 0.21 & 1.3 [13] & 4.160 $\pm$ 0.008 & 14.45 $\pm$ 0.03 \\
22 & $\epsilon$ Gem & HD 48329 & G8Ib & 4315 & 0.76 & 0.11 & 5.29 [7] & 4.780 $\pm$ 0.009 & 136.39 $\pm$ 6.33 \\
23 & 11 Lac & HD 214868 & K2.5III & 4310 & 1.93 & -0.20 & 1.38 [4] & 2.633 $\pm$ 0.009 & 30.24 $\pm$ 0.34 \\
24 & $\sigma$ Per & HD 21552 & K3III & 4159 & 1.60 & -0.23 & 1.32 [4] & 3.125 $\pm$ 0.005 & 35.28 $\pm$ 0.77 \\
25 & $\zeta$ Cep & HD 210745 & K1.5Ib & 4151 & 0.77 & 0.07 & 10.1 [5] & 5.239 $\pm$ 0.013 & 171.55 $\pm$ 7.40 \\
26 & HD 220369 & HD 220369 & K3II & 4131 & 1.24 & 0.11 & 1.5 [9] & 2.542 $\pm$ 0.008 & 96.14 $\pm$ 67.13 \\
27 & $\lambda$ Her & HD 158899 & K3.5III & 4070 & 1.89 & -0.04 & 1.18 [4] & 3.109 $\pm$ 0.003 & 40.28 $\pm$ 0.48 \\
28 & V424 Lac & HD 216946 & K5Ib & 4013 & 0.32 & -0.02 & 6.8 [12] & 4.049 $\pm$ 0.008 & 304.30 $\pm$ 23.98 \\
29 & $\rho$ Ser & HD 141992 & K4.5III & 3920 & 1.68 & -0.17 & 1.0 & 3.371 $\pm$ 0.004 & 44.79 $\pm$ 0.57 \\
30 & $\upsilon$ Aur & HD 38944 & M0III & 3912 & 1.16 & -0.23 & 1.64 [7] & 4.316 $\pm$ 0.007 & 80.46 $\pm$ 2.66 \\
31 & $\alpha$ Vul & HD 183439 & M0III & 3690 & 1.30 & -0.38 & 0.97 [9] & 4.339 $\pm$ 0.004 & 41.48 $\pm$ 0.41 \\
\enddata
\tablecomments{Mass references: 1--\cite{alfCMi_Bond2015}; 2--\cite{alfPer_Wade2025}; 3--\cite{etaBoo_Guenther2005}; 4--\cite{Massarotti2008}; 5--\cite{Hohle2010}; 6--\cite{gamTau_daSilva2006}; 7--\cite{Baines2018}; 8--\cite{epsTau_Arentoft2019}; 9--\cite{AllendePrieto1999}; 10--\cite{gamCep_Knudstrup2023}; 11--\cite{alfCas_Rudrasingam2026}; 12--\cite{Lee2014}; 13--\cite{delAnd_Bottom2015}. For $\rho$ Ser, no suitable literature mass was found; we adopt $1.0\,M_\odot$ only for SATLAS grid selection.}
\end{deluxetable}
\twocolumngrid

\section{Observations and Data Reduction}
\label{sec:obs_data_redu}

To compare limb darkening in $H$ and $K$-bands, we obtained
simultaneous $H$- and $K$-band interferometric observations of each target
with MIRC-X and MYSTIC at the CHARA Array. 
The CHARA Array consists of six 1\,m telescopes (S1, S2, W1, W2, E1, and
E2) arranged in a Y-shaped configuration.
This yields an angular resolution of approximately $\lambda/2B$,
corresponding to spatial scales of $\sim0.5$~mas and $\sim0.7$~mas in the
near-infrared $H$ and $K$ bands, respectively. 
The long baselines up to 331~m and dense ($u,v$) coverage make CHARA well suited for this
program. With the six-telescope MIRC-X and MYSTIC beam combiners, each exposure
provides 15 $V^2$ measurements and 10 independent closure-phase ($\mathrm{T3}\phi$)
measurements per wavelength band. 

The observations were acquired simultaneously, with MYSTIC operating
as the fringe tracker while MIRC-X recorded science data simultaneously.
We used the standard observing procedures to reduce detector-correlated noise in MIRC-X,
including hot-pixel correction and frequent dark exposures taken approximately
every 5 minutes.
MIRC-X was used at spectral resolving power $\mathcal{R}\sim180$ in 2025, providing 31 spectral
channels across the $H$ band (1.5--1.75~$\mu$m), and at
$\mathcal{R}\sim102$ in 2026, providing 14 spectral channels. MYSTIC was used
at $\mathcal{R}\sim272$ in 2025, providing 61 spectral channels across the $K$
band (2.0--2.37~$\mu$m), and at $\mathcal{R}\sim102$ in 2026, providing 22
spectral channels. The lower spectral resolution mode in 2026 was adopted as an
observing-setup choice to increase per-channel sensitivity by concentrating the
flux into fewer spectral channels, improving the signal-to-noise ratio for
faint or low-contrast detections. The lower-resolution mode increases the
per-channel signal-to-noise ratio while preserving the bandpass-integrated
visibility information used for the limb-darkening fits. Depending on nightly conditions and array availability,
observations were conducted with five or six available telescopes.

Science targets and calibrators were observed in a repeated
SCI1--CAL--SCI2--CAL cadence (see Appendix~\ref{app:Appendix_obs_log}). Each observing sequence consisted of a 5-minute coherent (on-fringe) scan
sampled at 4~ms cadence, paired with a 10~ms coherent integration used to
estimate and subtract instrumental bias. These settings follow the standard
MIRC-X/MYSTIC observing procedure and were applied uniformly across the program
rather than adjusted target by target for magnitude or expected signal-to-noise
ratio.

The raw MIRC-X and MYSTIC data\footnote{Available at https://www.chara.gsu.edu/observers/database}
were reduced using the standard MIRC-X data-reduction pipeline (version~1.5.0;
\citealt{Anugu2020, lebouquin_2024})\footnote{Available at https://gitlab.chara.gsu.edu/lebouquj/mircx\_pipeline.git}, 
which produces calibrated $V^2$, closure phases ($\mathrm{T3}\phi$), and photometric fluxes. The final calibrated OIFITS files are available through the OIDB database\footnote{Available at https://oidb.jmmc.fr}.
For each target, we produced calibrated OIFITS files using 30~s averages and
10, 15, and 20~minute averages. To assess calibrator-related systematics, we also
calibrated the calibrators against other calibrators observed on the same
nights, allowing us to refine effective calibrator diameters and propagate
realistic calibrator-size uncertainties into the transfer function (see Appendix~\ref{app:Appendix_obs_log} and \mbox{Table~\ref{tab:calibrators}}). Some
calibrators have angular diameters larger than $\sim0.5$~mas and are therefore not
strictly unresolved on the longest baselines, particularly in the $H$ band; we
accordingly treat calibrator size as a contributor to the calibration error
budget rather than assuming all calibrators are point-like. As a final
calibration-stability check, we inspected the nightly transfer functions
separately for each baseline, calibrator, and spectral band (see Appendix~\ref{app:tf}). 
The transfer functions show the expected scatter from seeing, residual delay-line errors, and
vibrations, but no large discontinuities or systematic wavelength-dependent
offsets. We therefore find no evidence that the calibrated $V^2$ data are
dominated by a transfer-function artifact. 

We adopt instrument-specific wavelength-scale correction factors (see
Appendix~\ref{app:wavelength_calibration}). For MIRC-X in the $H$ band, we use
$s_\lambda = 0.999 \pm 0.001$ for the 2025 data and $1.0054 \pm 0.0006$ for the 2026 data. 
The difference between the 2025 and 2026 wavelength-scale factors arose because
the instrument focus was changed in 2026.
For MYSTIC in the $K$ band, we use $1.0067 \pm 0.0007$
\citep[J. Monnier, private communication;][]{Gardner2022AJ....164..184G}. In all cases, the corrected wavelength scale is computed as $\lambda_{\rm corr}=\lambda_{\rm pipeline}/s_\lambda.$

\section{Methods}
\label{sec:methods}

We present here our methodology for comparing CHARA limb-darkening measurements
with atmosphere-model predictions matched to the MIRC-X and MYSTIC $H$ and $K$
passbands. The main analysis steps are summarized in \mbox{Table~\ref{tab:methods}}.  Rather than adopting precomputed
coefficient tables, we use the underlying model intensity or CLV profiles,
integrate them over the individual $H$ and $K$-band passbands, fit the same analytic laws used for
the CHARA data, and propagate the adopted stellar-parameter uncertainties.

We fit four analytic limb-darkening laws (Section~\ref{sec:pmoired_laws_v2_fit})
and use the one-parameter power law as the reference prescription for
comparisons between CHARA and atmosphere-model predictions. This law provides a
stable coefficient summary across the sample, including targets with uneven
higher-lobe $V^2$ coverage. More flexible laws, especially the power--2 and
four-parameter forms, can reproduce realistic near-infrared CLVs more accurately
\citep{Verma2024,Claret2025A&A...699A..97C}; we therefore include those laws as consistency
checks, and their fitted coefficients are provided in the machine-readable table
in the Zenodo archive \citep{AnuguZenodo2026} and Appendix~\ref{app:laws_compare}.

We use $\theta_{\rm law}$ for empirical analytic-law angular diameters, with
$\theta_{\rm PL}$ denoting the power-law angular diameter from the joint
$H{+}K$ fit. The coefficients $\alpha_b$, with $b\in\{H,K\}$, denote the
empirical CHARA power-law limb-darkening coefficients measured independently
in the $H$ and $K$ bands, i.e., $\alpha_H$ and $\alpha_K$. For model
comparisons, $\alpha_{I,b}^g$ is the coefficient obtained by fitting the same
analytic law directly to the passband-integrated model intensity profile,
$I_b(\mu)$, where $g$ denotes the atmosphere grid. We use $I(\mu)$ for a generic
center-normalized CLV or analytic-law form. The coefficient
$\alpha_{{\rm SVAM},b}^g$ is computed from synthetic CHARA-sampled
visibilities generated from the same model CLV. Fixed-CLV angular diameters
are denoted by $\theta_{\rm CLV}^g$.

The analysis proceeds in three steps. First, we fit analytic laws directly to
the calibrated CHARA $V^2$ data using
PMOIRED\footnote{\url{https://github.com/amerand/PMOIRED}}
\citep{Merand2022SPIE12183E..1NM} (see Section~\ref{sec:pmoired_laws_v2_fit}). 
Second, we compute atmosphere-model
coefficients both from direct fits to $I_b(\mu)$ (Section~\ref{sec:exotic_laws_v2_fit}) and from synthetic CHARA
visibility sampling (Section~\ref{sec:exotic_svam}). Third, we fit the CHARA $V^2$ data directly with fixed
atmosphere-model CLV profiles to test the effect on angular diameter (Section~\ref{sec:diameters_fit_to_clv}).

\subsection{Analytic limb-darkening laws}
\label{sec:laws}

We describe the stellar CLV with the center-normalized specific intensity
$I(\mu)$, such that $I(1)=1$. For a circular disk of angular diameter
$\theta$,
$\mu~=~\sqrt{1-\left(2r/\theta\right)^2}$,
where $r$ is the projected angular distance from the disk center. We fit the
$V^2$ data with a uniform-disk (UD) baseline and four analytic
limb-darkening laws: linear \citep{Schwarzschild1906}, quadratic
\citep{Kopal1950}, power law \citep{Hestroffer1997}, and power--2
\citep{Maxted2018A&A...616A..39M}:

\begin{equation}
\label{eq:laws}
\begin{array}{@{}l@{\quad=\quad}l@{}}
I_{\rm lin}(\mu)  & 1 - a(1-\mu), \\
I_{\rm pow}(\mu)  & \mu^{\alpha}, \\
I_{\rm quad}(\mu) & 1 - u_1(1-\mu) - u_2(1-\mu)^2, \\
I_{\rm p2}(\mu)   & 1 - p_1\left(1-\mu^{\alpha_1}\right).
\end{array}
\end{equation}

The fitted coefficients are $a$, $\alpha$, $(u_1,u_2)$, and
$(p_1,\alpha_1)$; band-specific values use subscripts, e.g., $a_H$ and $a_K$.

\subsection{Analytic laws: visibility-domain fitting of CHARA $V^2$ data with PMOIRED}
\label{sec:pmoired_laws_v2_fit}

For a circularly symmetric CLV, the monochromatic complex visibility is the
normalized zeroth-order Hankel transform of the intensity profile
\citep[e.g.,][]{HanburyBrown1974,Merand2022SPIE12183E..1NM}:
\begin{equation}
V(q) =
\frac{\int_0^1 I(\mu)\,
J_0\!\left(\pi q\theta\sqrt{1-\mu^2}\right)\,\mu\,d\mu}
{\int_0^1 I(\mu)\,\mu\,d\mu},
\label{eq:clv_visibility}
\end{equation}
where $q=B_{\rm p}/\lambda$ is the projected spatial frequency,
$B_{\rm p}$ is the projected baseline length, $J_0$ is the zeroth-order
Bessel function of the first kind, and $\theta$ is the angular diameter.
Eq.~\ref{eq:clv_visibility} shows explicitly how the CLV enters the
visibility curve; the fitted observable is $|V|^2$ after bandpass integration
and sampling at the observed CHARA spatial frequencies.

We fit the calibrated CHARA $V^2$ data (Figure~\ref{fig:V2_fit}) using PMOIRED, a widely used Python package for parametric interferometric model fitting. For clarity, we describe the fitting procedure using the one-parameter power law; the other limb-darkening laws are treated analogously. For each
target, we jointly fit the $H$- and $K$-band $V^2$ data with one shared
power-law angular diameter, $\theta_{\rm PL}$, and one shared
visibility-normalization term, $V^2_0$. The power-law coefficient is allowed
to vary by band through independent $\alpha_H$ and $\alpha_K$. 
The band-specific uniform-disk diameters $\theta_{{\rm UD},H}$ and
$\theta_{{\rm UD},K}$ are reported as baseline references.

The fitted visibility model is scaled as
$V^2_{\rm fit}=V_0^2 V^2_{\rm LD}$, where the normalized limb-darkened model
satisfies $V^2_{\rm LD}(B=0)=1$. Thus $V_0^2$ represents a multiplicative
visibility-scale correction, with the ideal value equal to unity. Separate
$H$- and $K$-band fits yield consistent values of $V_0^2$, so we adopt a single
shared value for each target in the joint $H{+}K$ fit. This is appropriate
because MIRC-X and MYSTIC share the same calibrators, upstream optics, and
optical table. The normalization term accounts for baseline-independent
visibility offsets, while residual baseline- or wavelength-dependent structure
is treated as calibration variance.

The resulting fitting residuals are shown in Figure~\ref{fig:V2_fit}. In
$V^2$ space, different analytic laws can give similar residuals because they
share nearly the same null structure and overall visibility envelope. The
degree of discrimination depends on angular diameter, lobe coverage, and
measurement precision. Appendix~\ref{app:laws_compare} therefore also compares
the fitted laws in intensity space, where their differences are more apparent.

As a check on possible fit degeneracies, we compared the fitted coefficients
$\alpha_H$ and $\alpha_K$ against both $\theta_{\rm PL}$ and  $V^2_0$ (Appendix~\ref{app:diagnostic_plots}). No strong
monotonic dependence is evident.

Closure phases are not included in the limb-darkening fits but are used to vet epochs and targets for asymmetries or companions (Section~\ref{sec:companions_imaging}). All benchmark targets, including the evolved stars, are consistent with centro-symmetry within the uncertainties of the closure-phase measurements. Consequently, no targets were excluded on the basis of closure phases alone.

Parameter uncertainties are estimated from PMOIRED bootstrap covariances computed by resampling the 10-minute averaged scans within each target and band. \mbox{Table~\ref{tab:ld_summary}} presents the compact power-law summary, while the machine-readable version in Zenodo additionally includes the results for the other analytic limb-darkening laws and the direct fixed-CLV diameter fits (Section~\ref{sec:diameters_fit_to_clv}).

\begin{table*}[htbp]
\centering
\caption{Schematic summary of the main analysis steps in Section~\ref{sec:methods}. 
Representative results from these steps are shown in \mbox{Figures~\ref{fig:V2_fit}}, 
\ref{fig:h_vs_k_crossband}, 
\ref{fig:data_model_for_teff}, \ref{fig:spam_data_model_for_teff} and 
\ref{fig:v2_data_model_discrepancy}, with tabulated empirical power-law results 
in Table~\ref{tab:ld_summary}. The full set of CHARA visibility-domain analytic-law 
fits, model-grid limb-darkening coefficients, and fixed-CLV angular diameters is 
provided in the online machine-readable table.}
\label{tab:methods}
\small
\setlength{\tabcolsep}{6pt}
\renewcommand{\arraystretch}{1.15}
\begin{tabular}{
  p{0.21\textwidth}
  p{0.47\textwidth}
  p{0.24\textwidth}}
\toprule
Method & What is fitted & Main output \\
\midrule
Empirical analytic-law fits 
(Section~\ref{sec:pmoired_laws_v2_fit}) &
The calibrated CHARA $V^2$ data are fit with a chosen analytic limb-darkening law (Section~\ref{sec:laws}).
For the joint power-law fit, the free parameters are $\theta_{\rm PL}$, $V_0^2$,
$\alpha_H$, and $\alpha_K$. &
Empirical angular diameters and limb-darkening coefficients 
(Tables~\ref{tab:ld_summary} and \ref{tab:hk_model_summary}; 
\mbox{Figures~\ref{fig:V2_fit}}  and 
\ref{fig:h_vs_k_crossband}). \\

Intensity-domain model limb-darkening coefficients 
(Section~\ref{sec:exotic_laws_v2_fit}) &
Bandpass-integrated model CLV profiles, $I_b(\mu)$, are computed from the
atmosphere grids and fit directly with the same analytic laws in intensity space. &
Model intensity-domain coefficients, $\alpha_{I,b}^g$ 
(Figure~\ref{fig:data_model_for_teff}). \\

Synthetic-visibility atmosphere-model coefficients 
(Section~\ref{sec:exotic_svam}) &
The same model CLV profiles are converted to synthetic $V^2$ values, sampled at
the observed CHARA $(u,v,\lambda)$ points, and refit with the same analytic laws
in the visibility domain. &
Model visibility-domain coefficients, $\alpha_{{\rm SVAM},b}^g$ 
(Figure~\ref{fig:spam_data_model_for_teff}). \\

Fixed-grid CLV fitting 
(Section~\ref{sec:diameters_fit_to_clv}) &
The CHARA $V^2$ data are fit directly with fixed bandpass-integrated model CLV
profiles from the atmosphere grids, solving only for $\theta_{\rm CLV}^g$ and
$V_0^2$ while keeping the model CLV shape fixed. &
Model-dependent CLV angular diameters 
(Table~\ref{tab:ld_summary}). \\
\bottomrule
\end{tabular}
\end{table*}

\begin{figure*}
\centering
\includegraphics[width=0.49\textwidth]{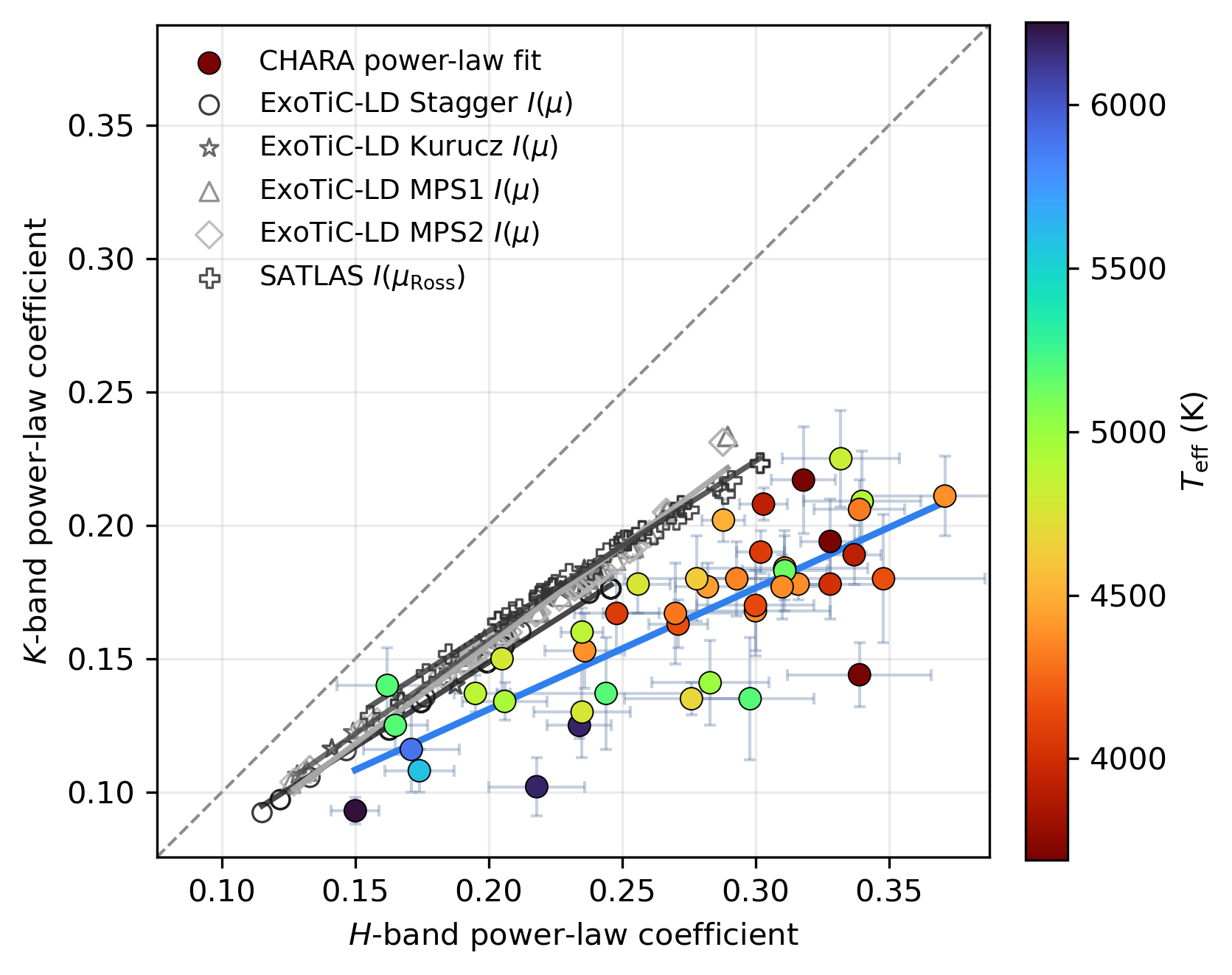}
\includegraphics[width=0.49\textwidth]{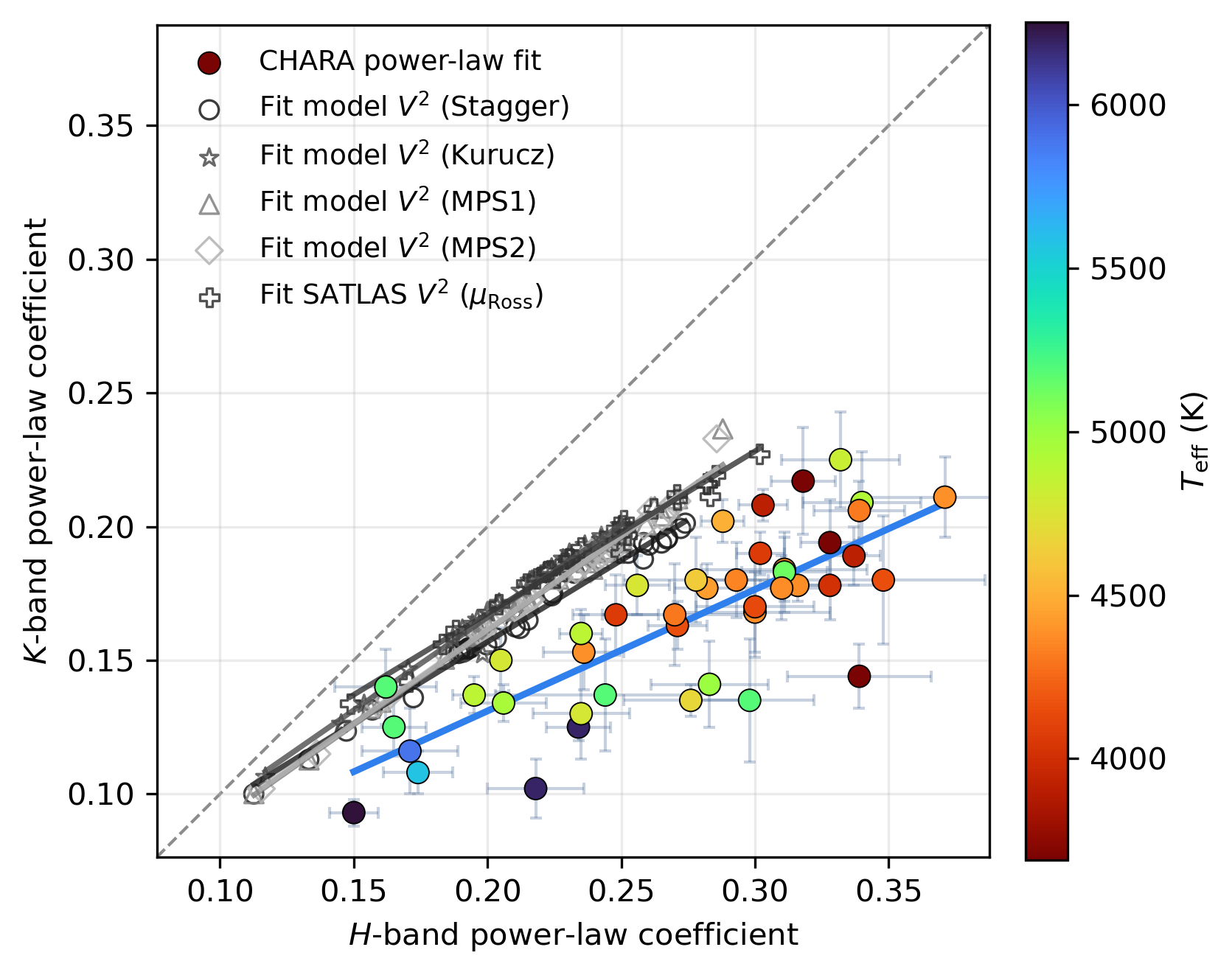}
\caption{Comparison of $H$- and $K$-band power-law coefficients from CHARA data
and atmosphere-model predictions. The filled colored points show the empirical
CHARA coefficients, $(\alpha_H,\alpha_K)$, measured by fitting the CHARA
$V^2$ data and color-coded by $T_{\rm eff}$. The gray sequences show the
corresponding model-grid predictions. Left: direct intensity-domain model
coefficients, $(\alpha^g_{I,H},\alpha^g_{I,K})$, obtained by fitting the
analytic law to the passband-integrated model intensity profiles $I_b(\mu)$
(Section~\ref{sec:exotic_laws_v2_fit}). Right: SVAM coefficients,
$(\alpha^g_{{\rm SVAM},H},\alpha^g_{{\rm SVAM},K})$, recovered by fitting the
same law to synthetic CHARA-sampled visibilities generated from each model CLV
(Section~\ref{sec:exotic_svam}). The dashed line marks equality between the
$H$- and $K$-band coefficients. The systematic separation between the colored
CHARA and the gray model points shows the model--data discrepancy
discussed in the text.}
\label{fig:h_vs_k_crossband}
\end{figure*}

\begin{figure*}[htbp]
\centering
\includegraphics[width=0.9\textwidth]{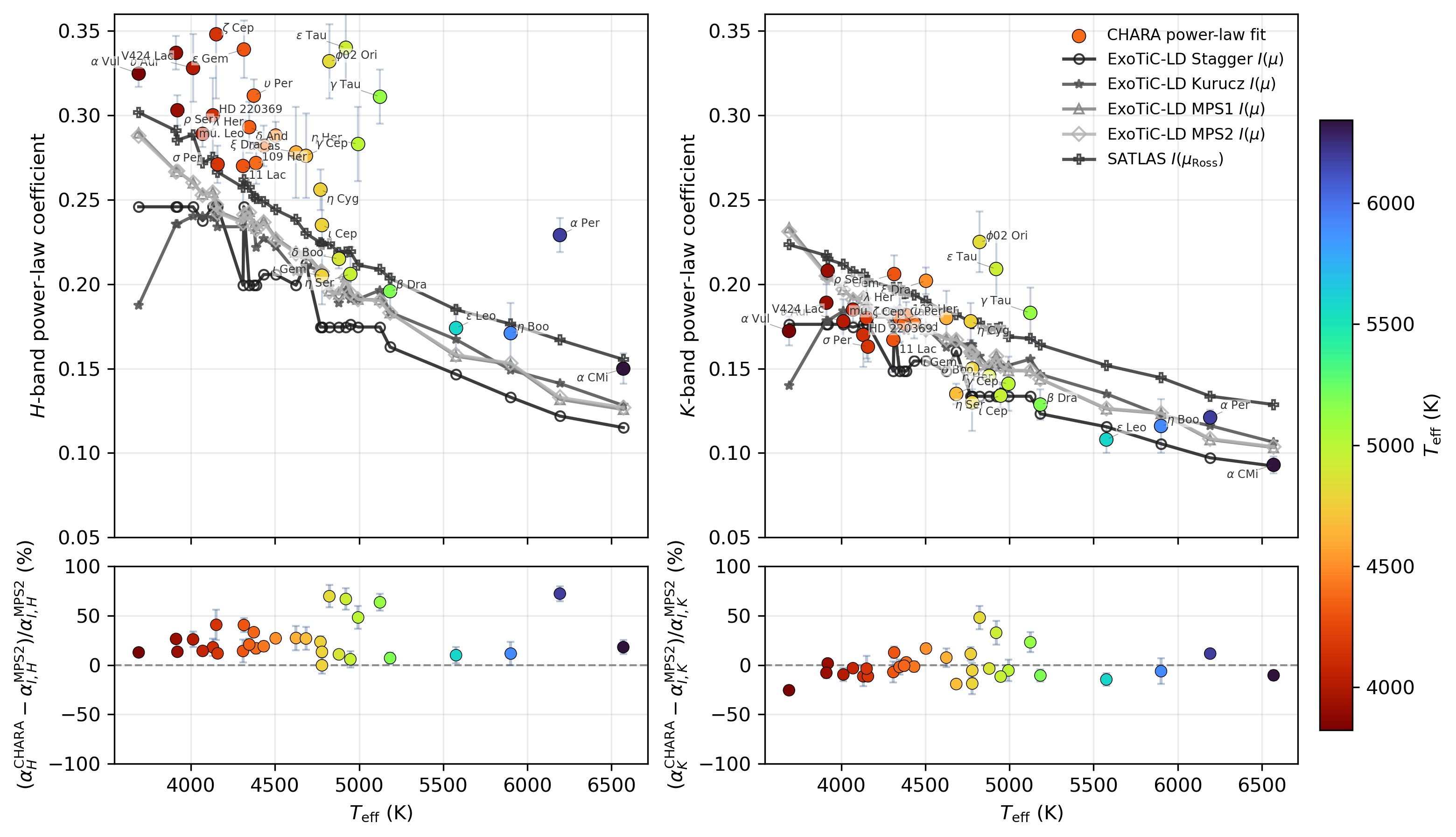}\\
\caption{CHARA power-law coefficients compared with intensity-domain
coefficients predicted by the atmosphere-model grids. Top: $H$-band
coefficients, $\alpha_H$ and $\alpha^g_{I,H}$, as a function of
$T_{\rm eff}$ (left), and $K$-band coefficients, $\alpha_K$ and
$\alpha^g_{I,K}$, as a function of $T_{\rm eff}$ (right). Filled colored
circles show the empirical CHARA $V^2$ fit coefficients, color-coded by
$T_{\rm eff}$, while open symbols and gray curves show the corresponding
model predictions from Stagger, Kurucz, MPS1, MPS2, and SATLAS. Bottom:
fractional offsets of the empirical CHARA coefficients relative to the MPS2
intensity-domain predictions, shown as
$100(\alpha_b-\alpha^{\rm MPS2}_{I,b})/
\alpha^{\rm MPS2}_{I,b}$, with $b\in\{H,K\}$ and CHARA $1\sigma$ error bars.
Apparent discontinuities in some model tracks reflect edge-limited grid
coverage for parts of the target sample. The limitations differ by grid:
Stagger is limited for the coolest targets and some low-gravity/metal-rich
stars, Kurucz and MPS1/MPS2 are mainly limited by their high-\(\log g\)
coverage for evolved stars, and SATLAS assumes solar composition
(Table~\ref{tab:model_grids}).}
\label{fig:data_model_for_teff}
\end{figure*}

\begin{figure*}[htbp]
\centering
\includegraphics[width=0.9\textwidth]{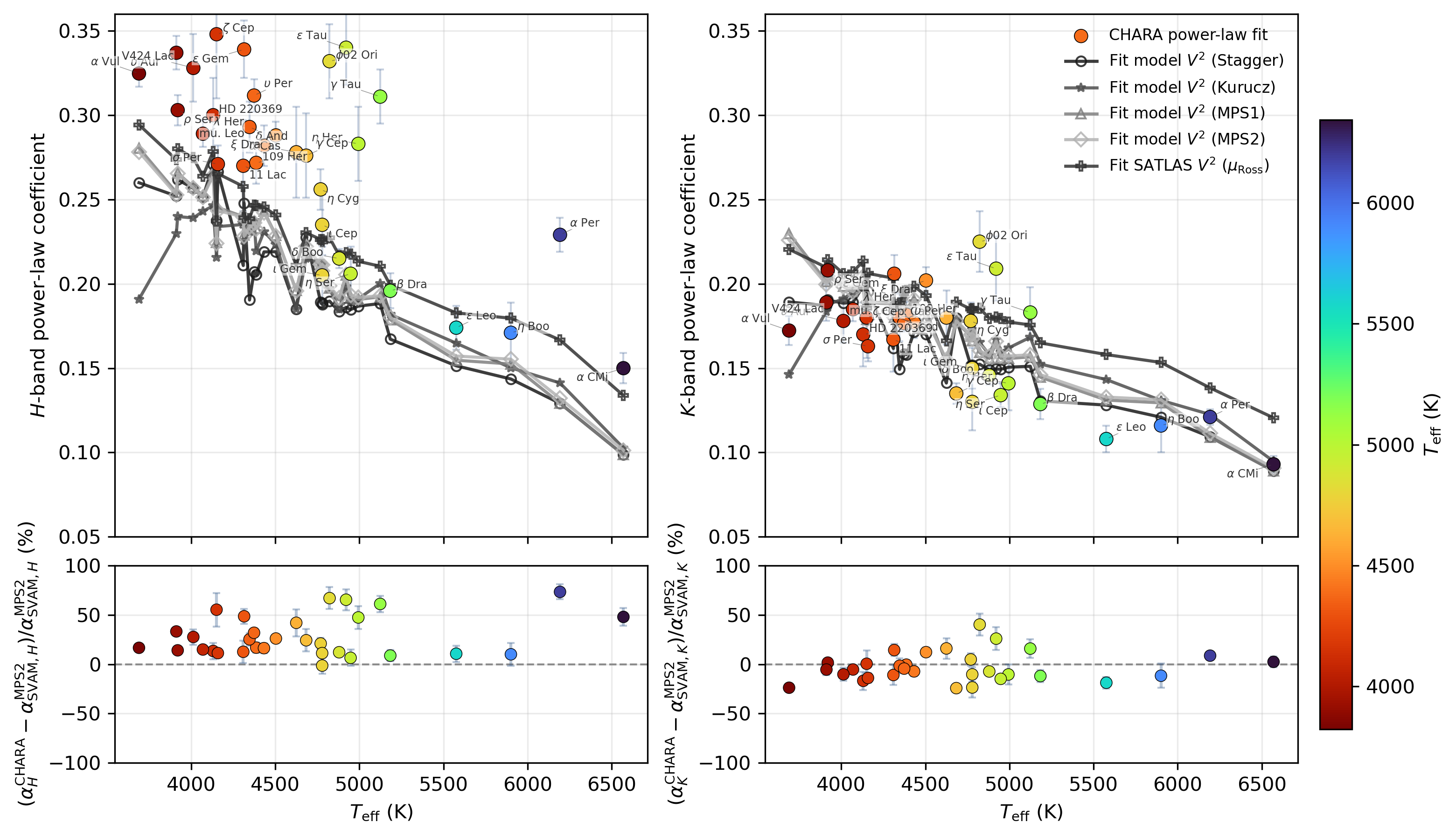}\\
\caption{Same format as Figure~\ref{fig:data_model_for_teff}, but comparing
the empirical CHARA coefficients with SVAM coefficients,
$\alpha^g_{{\rm SVAM},b}$, computed from synthetic CHARA-sampled
visibilities generated from each model CLV (see Section~\ref{sec:exotic_svam}).  Target labels are shown on the upper CHARA panels.}
\label{fig:spam_data_model_for_teff}
\end{figure*}

\begin{deluxetable*}{llll}
\tabletypesize{\scriptsize}
\tablecaption{Atmosphere-model grids used for model limb-darkening comparisons.
The first row gives the parameter range of the observed CHARA sample for
reference.
\label{tab:model_grids}}
\tablehead{
\colhead{Model set} &
\colhead{Type} &
\colhead{Inputs} &
\colhead{Coverage}
}
\startdata
CHARA sample (31 targets) &
Observed sample &
$T_{\rm eff},\log g,[{\rm M/H}]$ &
$T=3690$--6570 K; $\log g=0.32$--4.04; $Z=-0.54$ to $+0.25$ \\
Kurucz ATLAS (741) &
1D ATLAS &
$T_{\rm eff},\log g,[{\rm M/H}]$ &
$T=3500$--6500 K; $\log g=4.0$--5.0; $Z=-5.0$ to $+1.0$ \\
MPS1 (34160) &
1D MPS-ATLAS &
$T_{\rm eff},\log g,[{\rm M/H}]$ &
$T=3500$--9000 K; $\log g=3.0$--5.0; $Z=-5.0$ to $+1.5$ \\
MPS2 (34160) &
1D MPS-ATLAS &
$T_{\rm eff},\log g,[{\rm M/H}]$ &
$T=3500$--9000 K; $\log g=3.0$--5.0; $Z=-5.0$ to $+1.5$ \\
Stagger (99) &
3D RHD &
$T_{\rm eff},\log g,[{\rm M/H}]$ &
$T=4000$--7000 K; $\log g=1.5$--5.0; $Z=-3.0$ to $0.0$ \\
SATLAS &
1D spherical &
$T_{\rm eff},\log g,M$ &
$T=3000$--8000 K; $\log g=-1$ to $3$; $M=0.5,1,2.5$--$20\,M_\odot$
\enddata
\tablecomments{
Here $T\equiv T_{\rm eff}$ and $Z$ denotes $[{\rm M/H}]$ for compactness.
The CHARA-sample row gives the range spanned by the targets in
Table~\ref{tab:target_properties}.
Kurucz, MPS1, MPS2, and Stagger are accessed through ExoTiC-LD; SATLAS CLV
profiles are accessed through PMOIRED and placed on the Rosseland-radius scale. SATLAS assumes solar composition; therefore no metallicity dimension is varied for that grid.
The listed ranges refer to the grid coverage used for the model comparisons. 
}
\end{deluxetable*}

\subsection{Analytic laws: intensity-domain fitting of stellar-atmosphere predictions}
\label{sec:exotic_laws_v2_fit}

We compute bandpass-matched model limb-darkening coefficients from five
stellar-atmosphere model sets: Kurucz ATLAS, MPS1, MPS2, Stagger, and
spherical ATLAS/SATLAS. These model sets differ in physical assumptions,
geometry, and stellar-parameter coverage. Kurucz \citep{Kurucz1993} provides a
classical one-dimensional ATLAS grid. MPS1 and MPS2 are newer
one-dimensional MPS-ATLAS grids \citep{Kostogryz2022,Kostogryz2023}, which
differ in their adopted abundance mixture and mixing-length treatment. Stagger
provides the three-dimensional radiation--hydrodynamic comparison, with
spatially inhomogeneous convection and velocity fields
\citep{Magic2013,Magic2015}. SATLAS provides the spherical low-gravity
comparison, for which atmospheric extension can affect both the CLV and the
interferometric angular-diameter correction \citep{Neilson2013}. The grid sizes
and parameter ranges are summarized in \mbox{Table~\ref{tab:model_grids}}.

For each model set, we fit the same analytic limb-darkening laws to the
bandpass-integrated model intensity profiles, $I_b(\mu)$, so that the model coefficients can be compared consistently with the empirical
CHARA visibility-domain coefficients.

The Kurucz, MPS1, MPS2, and Stagger grids are accessed through ExoTiC-LD
\citep{Grant2024JOSS}, which distributes the underlying atmosphere-grid files
through the ExoTiC-LD stellar-grid archive.\footnote{\url{https://www.star.bris.ac.uk/exotic-ld-data/}}
For these four grids, the input stellar parameters are
$(T_{\rm eff},\log g,[{\rm M/H}])$. ExoTiC-LD interpolates the model
specific-intensity profiles to the adopted stellar parameters, integrates them
over the MIRC-X and MYSTIC passbands, and fits the resulting
bandpass-matched $I_b(\mu)$ profiles with the specified analytic
limb-darkening laws. ExoTiC-LD natively supports several common analytic laws,
including the linear, quadratic, square-root, Kipping, three-parameter,
four-parameter, and power--2 laws; for this work, we additionally implemented
the one-parameter power law used in the CHARA $V^2$ fits.

The SATLAS models are treated separately as they take $(T_{\rm eff},\log g,M)$  as input parameters. The SATLAS CLV
profiles are those of \citet{Neilson2013}, distributed through the CDS/VizieR
archive,\footnote{\url{http://cdsarc.u-strasbg.fr/viz-bin/qcat?J/A+A/554/A98}}
and accessed in this work through the PMOIRED implementation. The grid assumes
solar composition and spans $T_{\rm eff}=3000$--8000~K, $\log g=-1$ to $3$,
and stellar masses from $0.5$ to $20\,M_\odot$. For each target, we adopt a
literature-based mass as listed in \mbox{Table~\ref{tab:target_properties}}.
These masses are used only to select the spherical CLV profile and are not
fitted to the CHARA data.

For SATLAS, all coefficients and angular diameters reported in this paper are
placed on the Rosseland-radius scale \citep{Neilson2013}. The native SATLAS
CLV profiles are tabulated relative to the model outer radius, $R_{\rm out}$.
We convert the native SATLAS coordinate from the outer-radius scale to the Rosseland-radius scale using the tabulated $R_{\rm Ross}/R_{\rm out}$
ratio. For the analytic-law coefficient fits, only rays with projected impact
parameter $p \leq R_{\rm Ross}$ are used after rescaling to the
Rosseland-radius coordinate. In the fixed-CLV visibility fits, we retain the
full spherical SATLAS intensity profile, including the layers outside
$R_{\rm Ross}$, but report the fitted angular scale as the Rosseland angular
diameter. Thus $\theta_{\rm CLV}^{\rm SATLAS}$ denotes the Rosseland angular
diameter.

For each target, we propagate the uncertainties in the adopted stellar
parameters into the model limb-darkening coefficients using a Monte Carlo
procedure. For the ExoTiC-LD grids, we draw $10^4$ realizations of
$(T_{\rm eff},\log g,[{\rm M/H}])$ around the adopted values in
\mbox{Table~\ref{tab:target_properties}}, assuming independent Gaussian uncertainties
of $\sigma_T=100$~K, $\sigma_{\log g}=0.20$~dex, and
$\sigma_{[{\rm M/H}]}=0.20$~dex. These perturbations are used only to propagate
stellar-parameter uncertainties into the model predictions and are not fitted
to the CHARA data. For each star and band, we adopt the median coefficient as
the model prediction and the 16th--84th percentile range as the model-based
uncertainty. For SATLAS, the adopted mass is used only to select the spherical CLV profile;
mass uncertainties are not propagated into the coefficient uncertainties.

We fit each analytic law over the full tabulated CLV range down to the native
grid limit, $\mu_{\rm min}$, which is typically $\sim 0.01$. We adopt this
convention because the CHARA $V^2$ measurements are sensitive to the full
center-to-limb profile, including the treatment of the outer stellar limb. We
therefore retain the native model limit rather than truncating the fit at a
critical value such as $\mu_{\rm crit}$ to regularize the analytic-law
coefficients \citep{claret_mucrit_2018}. Truncated-$\mu$ coefficient tables
can be useful for other applications, but they are not strictly like-for-like
with the CHARA visibility-domain measurements considered here.

Our benchmark sample includes evolved stars with $\log g$ values below the
formal gravity coverage of the Kurucz and MPS grids, so predictions from those
grids for the lowest-gravity targets should be interpreted as edge-limited
comparisons rather than fully matched atmosphere predictions. Stagger reaches
lower gravities but is sparse and non-uniform in $T_{\rm eff}$, $\log g$, and
$[{\rm M/H}]$. SATLAS provides the most directly matched low-gravity spherical
comparison, but the grid assumes solar composition and uses stellar mass rather
than metallicity as an input parameter. We therefore interpret inter-grid
differences as model sensitivity rather than as a strict ranking of
atmosphere-grid accuracy.

Although PHOENIX models are available through ExoTiC-LD, we exclude them from
the final comparison set because the full-$\mu$ analytic-law fits were not
stable for the present visibility-domain comparison. Models like MARCS \citep{Gustafsson2008} and NewEra \citep{Hauschildt2025A&A...698A..47H} would be a
natural future extension for cool evolved stars.

For the main residual plots, we use MPS2 as a homogeneous reference grid
because it provides dense and regular parameter-space coverage across the
sample. \mbox{Figures~\ref{fig:h_vs_k_crossband}} and
\ref{fig:data_model_for_teff} compare the coefficients computed from the five
model sets.

\subsection{Analytic laws: like-for-like model comparison of intensity-domain and synthetic-visibility coefficients}
\label{sec:exotic_svam}

Limb-darkening coefficients depend on the adopted law, fitting weights, and
fitting domain \citep{Howarth2011MNRAS.418.1165H,Espinoza2015MNRAS.450.1879E}. A coefficient fit directly
to $I_b(\mu)$ need not match one measured from interferometric observables,
because visibility fitting weights the CLV after transformation into Fourier
space and sampling at the observed spatial frequencies.

We therefore compute two model coefficient sets. The first,
$\alpha_{I,b}^g$, is obtained by fitting the analytic law directly to
$I_b(\mu)$. The second, $\alpha_{{\rm SVAM},b}^g$, is obtained by transforming
the same model CLV into synthetic $V^2$, sampling it at the observed CHARA
$(u,v,\lambda)$ points with the empirical OIFITS uncertainties and flags, and evaluated at the empirical
power-law angular diameter $\theta_{\rm PL}$. We then
refit those synthetic visibilities with PMOIRED. 

Figures~\ref{fig:h_vs_k_crossband}, \ref{fig:data_model_for_teff} and
\ref{fig:spam_data_model_for_teff} show the corresponding comparisons.
Unless stated otherwise, we interpret model--data differences using the SVAM
coefficients and retain the direct $I_b(\mu)$ coefficients as context.

For each target, band, law, and grid we define
\begin{equation}
\sigma_{{\rm comb},g} =
\sqrt{\sigma_{\rm CHARA}^2 + \sigma_g^2},
\end{equation}
where $\sigma_{\rm CHARA}$ is the visibility-fit bootstrap uncertainty and
$\sigma_g$ is the corresponding model-coefficient uncertainty from the
Monte Carlo propagation of the adopted stellar-parameter uncertainties, where
available. Multi-epoch measurements are averaged.

For each band $b\in\{H,K\}$ and atmosphere grid $g$, we quantify the empirical $H$--$K$ coefficient contrast as
\begin{equation}
C^\alpha_{\rm HK} =
100\,\frac{\alpha_H-\alpha_K}
{\alpha_H},
\label{eq:chara_hk_contrast}
\end{equation}
and the corresponding model contrast as
\begin{equation}
C^{\alpha,g}_{\rm HK} =
100\,\frac{\alpha^g_{\rm SVAM,H}-\alpha^g_{\rm SVAM,K}}
{\alpha^g_{\rm SVAM,H}} .
\label{eq:model_hk_contrast}
\end{equation}
When needed below, we distinguish the corresponding direct intensity-domain
contrast, $C^{\alpha,g}_{\rm HK}(I)$, from the synthetic-visibility contrast,
$C^{\alpha,g}_{\rm HK}({\rm SVAM})$.

We define the absolute
CHARA--model coefficient offset as
\begin{equation}
\delta\alpha^g_b =
\alpha_b-\alpha^g_{\rm SVAM,b},
\label{eq:absolute_alpha_offset}
\end{equation}
and the corresponding fractional offset as
\begin{equation}
\Delta^g_b =
100\,\frac{\alpha_b-\alpha^g_{\rm SVAM,b}}
{\alpha^g_{\rm SVAM,b}} .
\label{eq:fractional_alpha_offset}
\end{equation}

To compare coefficients obtained directly from the model intensity profile with
those recovered from synthetic CHARA-sampled visibilities, we define
\begin{equation}
\Delta^g_{I-\mathrm{SVAM},b} =
100\,\frac{\alpha^g_{I,b}-\alpha^g_{\rm SVAM,b}}
{\alpha^g_{I,b}} .
\label{eq:intensity_svam_offset}
\end{equation}

\begin{figure*}[htbp]
\centering
\includegraphics[width=0.9\textwidth]{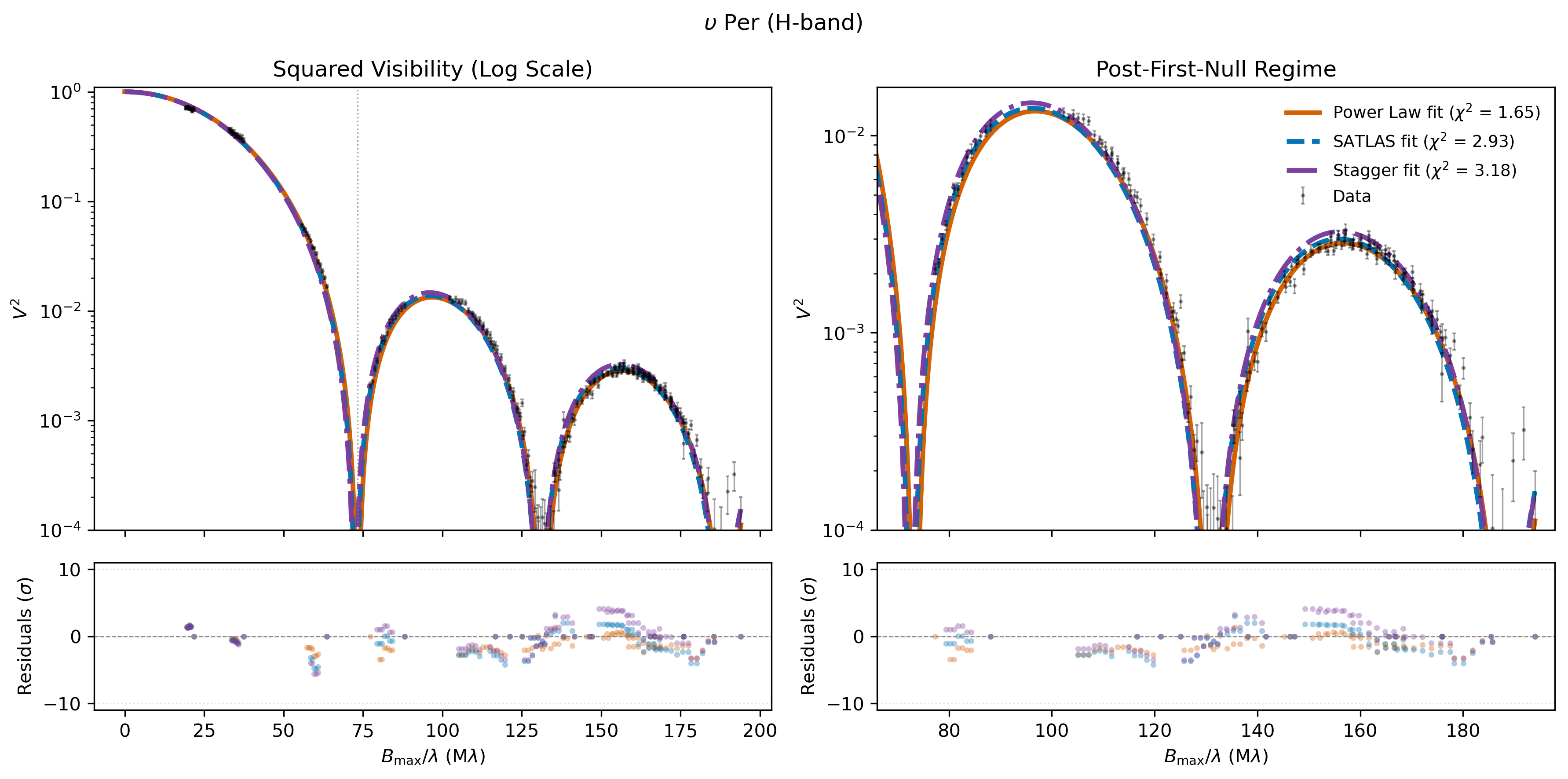}
\includegraphics[width=0.9\textwidth]{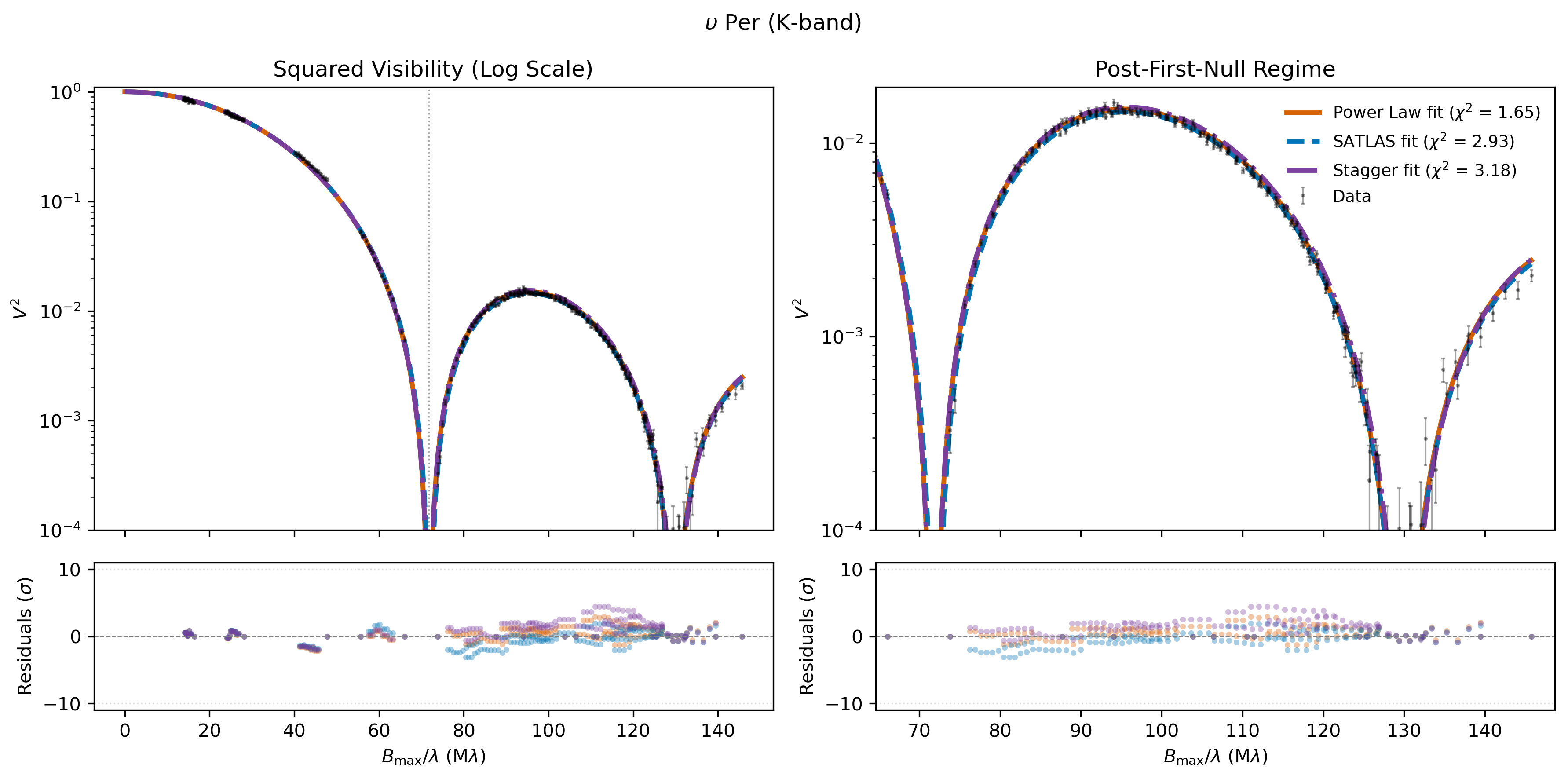}
    \caption{
    Illustrative data-vs-model comparison for the well-resolved star
    $\upsilon$~Per (HD 9927, $\theta_{\rm PL} = 3.615 \pm 0.003$ mas, $T_{\rm eff} = 4374 \pm 17$), shown in
    the $H$ band (top) and $K$ band (bottom). This target is shown because it
    is a cool giant with clear multi-lobe coverage, so the visibility curvature beyond the first null
    and the model--data discrepancy are both easy to see visually. The
    empirical CHARA power-law fit and model-based CLV predictions projected
    into the same observed $(u,v,\lambda)$ sampling are over-plotted on the
    data. The Stagger CLV predictions come from the ExoTiC-LD interpolated
    $I(\mu)$ profiles and the SATLAS predictions from \cite{Neilson2011,Neilson2013};
    in both cases the CLV shape is held fixed while the angular diameter and
    $V^2_0$ are fitted. The right panels zoom into
    the regime beyond the first null, where the visibility curvature is most sensitive
    to limb darkening. For this target, the empirical power-law fit better
    reproduces the higher-lobe visibility structure than the model-based CLV
    predictions. The complete figure set (42 target/epoch pairs; 84 band-specific plots) is available in the online version of the article and in the Zenodo figure-set archive \citep{AnuguZenodo2026}.}
    \label{fig:v2_data_model_discrepancy}
\end{figure*}

\subsection{Fixed Model Grid CLV Fitting: Angular Diameters}
\label{sec:diameters_fit_to_clv}
After comparing empirical and model-based limb-darkening coefficients, we
implement a complementary diameter measurement by fitting the CHARA $V^2$ data directly
with fixed atmosphere-model CLV profiles. Here ``fixed-CLV'' means that the
bandpass-integrated model $I_b(\mu)$ profile is held fixed, while only the
angular diameter, $\theta_{\rm CLV}^g$, and visibility normalization, $V_0^2$,
are fitted. The Kurucz, MPS1, MPS2, and Stagger CLVs are generated with
ExoTiC-LD as described in Section~\ref{sec:exotic_laws_v2_fit}; the SATLAS
CLVs are obtained through PMOIRED. 

For the fixed-CLV visibility fits, we retain the full spherical SATLAS CLV,
including its outer atmospheric extension, but report the fitted angular scale
as the Rosseland angular diameter.

These fits therefore yield model-dependent diameters, $\theta_{\rm CLV}^g$.
Because each grid predicts a different CLV shape, the fitted diameter is the
angular scale required for that specific model profile to reproduce the same
observed visibility curve. Inter-grid diameter differences therefore measure the systematic effect of the
adopted CLV shape on the inferred angular diameter. 
We compare $\theta_{\rm CLV}^g$ with $\theta_{\rm PL}$ to quantify
the effect of using analytic laws instead of full model CLVs.

Figure~\ref{fig:v2_data_model_discrepancy} illustrates the model--data comparison
directly in the visibility domain for $\upsilon$~Per. We use $\upsilon$~Per as an illustrative example because it has clear
sampling of the visibility curvature beyond the first null, observations at
multiple epochs, and $T_{\rm eff}<5000$~K. It also shows the main wavelength
dependent behavior discussed in this work: a larger CHARA--model discrepancy in
$H$ and closer agreement in $K$.


\startlongtable
\begin{deluxetable}{r l r r r r r r r}
\tablecaption{Uniform-disk ($\theta_{{\rm UD},H}$ and $\theta_{{\rm UD},K}$) angular diameters in the $H$ (MIRC-X) and $K$ (MYSTIC) bands, the combined H+K power-law limb-darkened angular diameter ($\theta_{\rm PL}$), the MPS2 fixed-CLV angular diameter ($\theta_{\rm CLV}^{\rm MPS2}$), the fitted visibility scale factor ($V^2_0$), and the corresponding empirical CHARA band-dependent power-law coefficients ($\alpha_H^{\rm CHARA}$ and $\alpha_K^{\rm CHARA}$). Same $ID$ and order as Table~\ref{tab:target_properties}.}\label{tab:ld_summary}
\tablewidth{0pt}
\tablehead{\colhead{$ID$} & \colhead{Target} & \colhead{$\theta_{{\rm UD},H}$ (mas)} & \colhead{$\theta_{{\rm UD},K}$ (mas)} & \colhead{$\theta_{\rm PL}$ (mas)} & \colhead{$\theta_{\rm CLV}^{\rm MPS2}$ (mas)} & \colhead{$V^2_0$} & \colhead{$\alpha_H^{\rm CHARA}$} & \colhead{$\alpha_K^{\rm CHARA}$} \\}
\startdata
1 & $\alpha$ CMi & 5.337 $\pm$ 0.006 & 5.333 $\pm$ 0.006 & 5.417 $\pm$ 0.005 & 5.422 $\pm$ 0.004 & 1.023 & 0.150 $\pm$ 0.009 & 0.093 $\pm$ 0.005 \\
2 & $\alpha$ Per & 3.148 $\pm$ 0.004 & 3.143 $\pm$ 0.003 & 3.226 $\pm$ 0.003 & 3.216 $\pm$ 0.003 & 1.025 & 0.229 $\pm$ 0.010 & 0.121 $\pm$ 0.005 \\
3 & $\eta$ Boo & 2.167 $\pm$ 0.006 & 2.164 $\pm$ 0.002 & 2.222 $\pm$ 0.006 & 2.226 $\pm$ 0.002 & 1.024 & 0.171 $\pm$ 0.018 & 0.116 $\pm$ 0.016 \\
4 & $\epsilon$ Leo & 2.596 $\pm$ 0.003 & 2.603 $\pm$ 0.003 & 2.662 $\pm$ 0.004 & 2.672 $\pm$ 0.002 & 1.111 & 0.174 $\pm$ 0.013 & 0.108 $\pm$ 0.008 \\
5 & $\beta$ Dra & 3.255 $\pm$ 0.006 & 3.263 $\pm$ 0.004 & 3.342 $\pm$ 0.005 & 3.354 $\pm$ 0.002 & 1.067 & 0.196 $\pm$ 0.010 & 0.129 $\pm$ 0.009 \\
6 & $\gamma$ Tau & 2.317 $\pm$ 0.011 & 2.305 $\pm$ 0.003 & 2.399 $\pm$ 0.006 & 2.386 $\pm$ 0.003 & 1.047 & 0.311 $\pm$ 0.016 & 0.183 $\pm$ 0.015 \\
7 & $\eta$ Her & 2.426 $\pm$ 0.006 & 2.445 $\pm$ 0.002 & 2.526 $\pm$ 0.006 & 2.519 $\pm$ 0.005 & 1.161 & 0.283 $\pm$ 0.022 & 0.141 $\pm$ 0.016 \\
8 & $\eta$ Ser & 2.868 $\pm$ 0.010 & 2.890 $\pm$ 0.004 & 2.968 $\pm$ 0.004 & 2.979 $\pm$ 0.003 & 1.136 & 0.206 $\pm$ 0.016 & 0.134 $\pm$ 0.007 \\
9 & $\epsilon$ Tau & 2.491 $\pm$ 0.007 & 2.484 $\pm$ 0.003 & 2.596 $\pm$ 0.008 & 2.573 $\pm$ 0.003 & 1.102 & 0.340 $\pm$ 0.022 & 0.209 $\pm$ 0.019 \\
10 & $\delta$ Boo & 2.671 $\pm$ 0.004 & 2.675 $\pm$ 0.002 & 2.756 $\pm$ 0.003 & 2.761 $\pm$ 0.002 & 1.075 & 0.215 $\pm$ 0.006 & 0.146 $\pm$ 0.006 \\
11 & $\phi^2$ Ori & 2.161 $\pm$ 0.013 & 2.143 $\pm$ 0.002 & 2.251 $\pm$ 0.008 & 2.224 $\pm$ 0.003 & 0.981 & 0.332 $\pm$ 0.022 & 0.225 $\pm$ 0.018 \\
12 & $\iota$ Cep & 2.765 $\pm$ 0.008 & 2.756 $\pm$ 0.006 & 2.843 $\pm$ 0.007 & 2.853 $\pm$ 0.005 & 1.105 & 0.235 $\pm$ 0.018 & 0.130 $\pm$ 0.017 \\
13 & $\iota$ Gem & 2.398 $\pm$ 0.009 & 2.365 $\pm$ 0.002 & 2.472 $\pm$ 0.005 & 2.480 $\pm$ 0.002 & 1.099 & 0.205 $\pm$ 0.017 & 0.150 $\pm$ 0.012 \\
14 & $\eta$ Cyg & 2.193 $\pm$ 0.004 & 2.205 $\pm$ 0.002 & 2.287 $\pm$ 0.005 & 2.283 $\pm$ 0.002 & 0.988 & 0.256 $\pm$ 0.012 & 0.178 $\pm$ 0.011 \\
15 & $\gamma$ Cep & 3.078 $\pm$ 0.012 & 3.108 $\pm$ 0.009 & 3.201 $\pm$ 0.004 & 3.213 $\pm$ 0.006 & 1.128 & 0.276 $\pm$ 0.025 & 0.135 $\pm$ 0.006 \\
16 & $\alpha$ Cas & 5.502 $\pm$ 0.023 & 5.529 $\pm$ 0.007 & 5.664 $\pm$ 0.010 & 5.659 $\pm$ 0.006 & 1.094 & 0.278 $\pm$ 0.027 & 0.180 $\pm$ 0.016 \\
17 & $\xi$ Dra & 2.969 $\pm$ 0.011 & 2.983 $\pm$ 0.007 & 3.078 $\pm$ 0.003 & 3.068 $\pm$ 0.003 & 1.109 & 0.288 $\pm$ 0.008 & 0.202 $\pm$ 0.008 \\
18 & $\mu$ Leo & 2.756 $\pm$ 0.006 & 2.765 $\pm$ 0.005 & 2.854 $\pm$ 0.004 & 2.860 $\pm$ 0.003 & 1.052 & 0.282 $\pm$ 0.012 & 0.177 $\pm$ 0.009 \\
19 & 109 Her & 2.906 $\pm$ 0.009 & 2.895 $\pm$ 0.003 & 2.996 $\pm$ 0.003 & 2.998 $\pm$ 0.002 & 1.024 & 0.272 $\pm$ 0.012 & 0.182 $\pm$ 0.008 \\
20 & $\upsilon$ Per & 3.495 $\pm$ 0.007 & 3.508 $\pm$ 0.004 & 3.615 $\pm$ 0.003 & 3.616 $\pm$ 0.002 & 1.011 & 0.312 $\pm$ 0.010 & 0.177 $\pm$ 0.005 \\
21 & $\delta$ And & 4.035 $\pm$ 0.014 & 4.043 $\pm$ 0.013 & 4.160 $\pm$ 0.008 & 4.165 $\pm$ 0.005 & 1.032 & 0.293 $\pm$ 0.015 & 0.180 $\pm$ 0.014 \\
22 & $\epsilon$ Gem & 4.625 $\pm$ 0.015 & 4.633 $\pm$ 0.012 & 4.780 $\pm$ 0.009 & 4.767 $\pm$ 0.006 & 1.083 & 0.339 $\pm$ 0.017 & 0.206 $\pm$ 0.011 \\
23 & 11 Lac & 2.547 $\pm$ 0.007 & 2.537 $\pm$ 0.003 & 2.633 $\pm$ 0.009 & 2.641 $\pm$ 0.003 & 0.978 & 0.270 $\pm$ 0.027 & 0.167 $\pm$ 0.019 \\
24 & $\sigma$ Per & 3.029 $\pm$ 0.007 & 3.024 $\pm$ 0.006 & 3.125 $\pm$ 0.005 & 3.137 $\pm$ 0.005 & 1.123 & 0.271 $\pm$ 0.011 & 0.163 $\pm$ 0.009 \\
25 & $\zeta$ Cep & 5.066 $\pm$ 0.017 & 5.099 $\pm$ 0.011 & 5.239 $\pm$ 0.013 & 5.244 $\pm$ 0.009 & 1.028 & 0.348 $\pm$ 0.038 & 0.180 $\pm$ 0.024 \\
26 & HD 220369 & 2.453 $\pm$ 0.011 & 2.448 $\pm$ 0.003 & 2.542 $\pm$ 0.008 & 2.556 $\pm$ 0.003 & 0.966 & 0.300 $\pm$ 0.022 & 0.170 $\pm$ 0.019 \\
27 & $\lambda$ Her & 3.004 $\pm$ 0.013 & 3.006 $\pm$ 0.004 & 3.109 $\pm$ 0.003 & 3.114 $\pm$ 0.002 & 1.049 & 0.289 $\pm$ 0.008 & 0.185 $\pm$ 0.007 \\
28 & V424 Lac & 3.922 $\pm$ 0.022 & 3.929 $\pm$ 0.007 & 4.049 $\pm$ 0.008 & 4.066 $\pm$ 0.005 & 0.928 & 0.328 $\pm$ 0.020 & 0.178 $\pm$ 0.013 \\
29 & $\rho$ Ser & 3.253 $\pm$ 0.011 & 3.253 $\pm$ 0.009 & 3.371 $\pm$ 0.004 & 3.374 $\pm$ 0.002 & 1.132 & 0.303 $\pm$ 0.009 & 0.208 $\pm$ 0.006 \\
30 & $\upsilon$ Aur & 4.140 $\pm$ 0.029 & 4.181 $\pm$ 0.014 & 4.316 $\pm$ 0.007 & 4.314 $\pm$ 0.007 & 1.067 & 0.337 $\pm$ 0.010 & 0.189 $\pm$ 0.011 \\
31 & $\alpha$ Vul & 4.193 $\pm$ 0.016 & 4.230 $\pm$ 0.004 & 4.339 $\pm$ 0.004 & 4.356 $\pm$ 0.003 & 1.068 & 0.325 $\pm$ 0.008 & 0.172 $\pm$ 0.009 \\
\enddata
\end{deluxetable}
\twocolumngrid

\section{Results}
\label{sec:results}

We report the fitting results and address three questions: how the CHARA
$V^2$ data constrain limb darkening, how the empirical coefficients compare
with bandpass-matched atmosphere-model predictions, and how the adopted CLV
prescription affects angular diameters. Unless stated otherwise, model--data
offsets are computed using the like-for-like SVAM coefficients defined in
Section~\ref{sec:exotic_svam}.

\subsection{CHARA visibility-domain limb-darkening fits}
\label{sec:pmoired_results}

We fitted the calibrated $V^2$ with a uniform-disk baseline and
four analytic limb-darkening laws: linear, power law, quadratic, and power--2
(Section~\ref{sec:pmoired_laws_v2_fit}). Figure~\ref{fig:V2_fit} shows the
well-resolved target $\alpha$~CMi as an illustrative example. Across the
sample, the limb-darkened-law fits give lower $\chi^2$ values than the
uniform-disk approximation, confirming that the data constrain CLV curvature in
addition to angular diameter.

The angular diameter is primarily anchored by the visibility nulls, whereas the
visibility curvature beyond the first null provides the strongest constraint on
the CLV shape. Higher-lobe $V^2$ coverage therefore helps in two related ways:
it reduces the covariance between the fitted angular diameter and
limb-darkening coefficient, and it provides the main leverage for
distinguishing among analytic laws. Targets with extended higher-lobe coverage
consequently yield more precise and robust limb-darkening coefficients.

Because different analytic laws can still produce similar $V^2$ residuals, we
also compare the fitted laws in intensity space in
Appendix~\ref{app:laws_compare}. Those comparisons show where different
coefficient prescriptions correspond to visibly different CLV shapes despite
similar visibility-domain fits.

\subsection{Checks for companions and non-centrosymmetric structure}\label{sec:companions_imaging}

Companions or surface asymmetries could perturb the visibility curvature and
bias the inferred coefficients, so we used the 10 independent closure phases, $T3\phi$, from each six-telescope observation to vet
the sample. With PMOIRED, we searched for the full interferometric field of view,
approximately $\lambda \mathcal{R}/B \simeq 100$--$270$~mas, in both bands for
close companions and non-centrosymmetric structure
\citep[e.g.,][]{Monnier2007,roettenbacher_no_2016,Anugu2024}. The search
jointly fit $V^2$ and $\mathrm{T3}\phi$ so that spurious solutions driven by
only one observable were rejected.
Target-by-target closure-phase residuals are provided in the Zenodo archive
\citep{AnuguZenodo2026}. Figure~\ref{fig:closure_phase_example} shows one
example for $\upsilon$~Per.

No companion candidate was detected in either the $H$- or $K$-band data with
consistent astrometry and contrast at a significance of $\geq 3\sigma$ after
accounting for systematics. For this search, we adopted a power-law
limb-darkening model for the primary and added a companion at position
$(c_x,c_y)$, exploring separations from $\theta_{\rm PL}/2$ to
the maximum field of view. The band-averaged detection limit is
$\Delta m \simeq 5$~mag, implying that any companion in the searched region
must be fainter than about $1\%$ of the primary flux. 

We also fit a single-spot model on the stellar surface, again using a
power-law limb-darkened primary. With the present data, we would detect an
isolated spot-like asymmetry if it contributed at the $\gtrsim1\%$ level of the
stellar flux and were larger than the angular resolution limit,
$\lambda/(2B_{\max})$. We did not detect any such feature. Thus, within these contrast
and angular-scale limits, we find no evidence that companions or large isolated
spots dominate the sample-level trends. Fainter, smaller-scale, or nearly
centro-symmetric surface structure cannot be excluded.

\begin{figure}[htbp]
\centering
\includegraphics[width=0.48\textwidth]{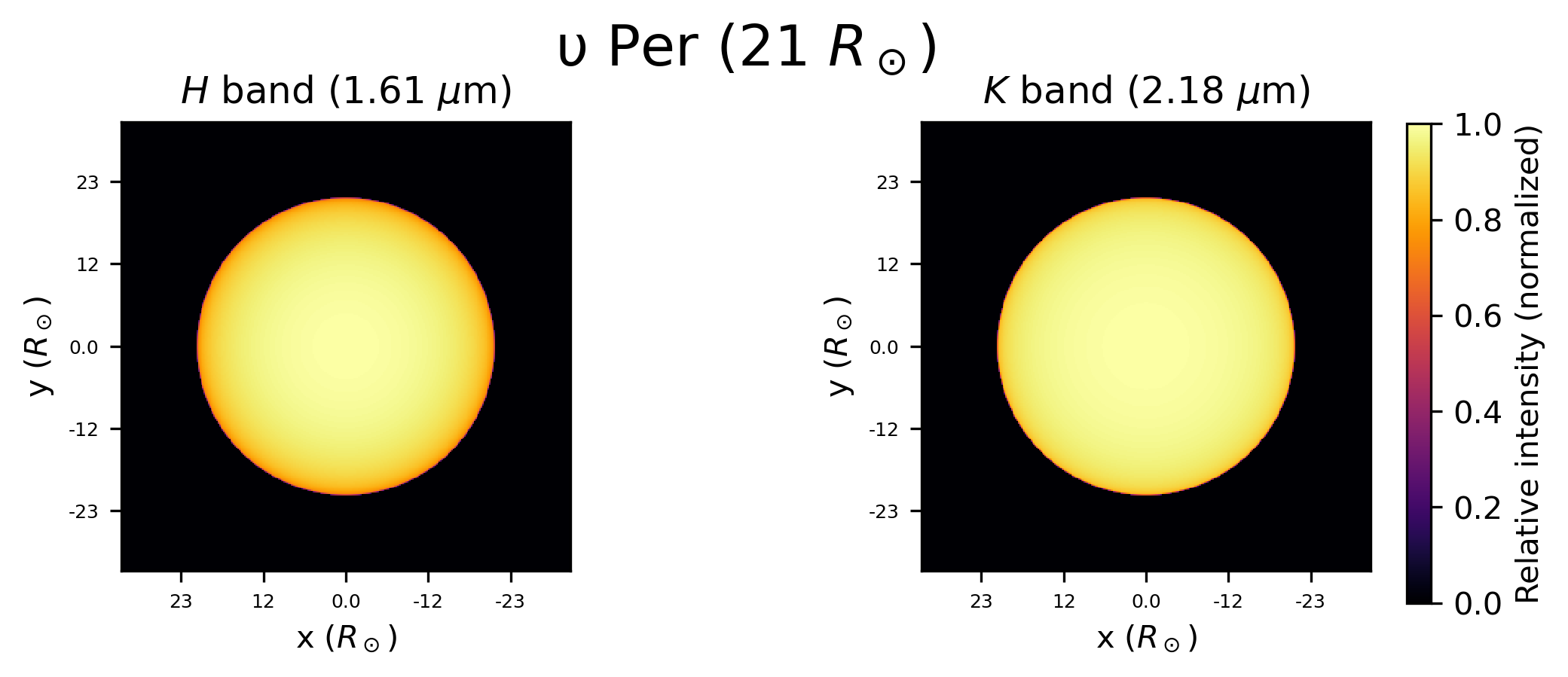}
\caption{
Illustrative power-law model images fitted with PMOIRED for $\upsilon$~Per,
shown on a physical-radius scale in the $H$ and $K$ bands. The weaker limb
darkening in $K$ produces a slightly flatter center-to-limb intensity profile
than in $H$. The complete figure set (31 images) is available in the online
version of the article and in the Zenodo figure-set archive
\citep{AnuguZenodo2026}.}
\label{fig:ups_per_model_images}
\end{figure}

As a complementary visual check, we generated PMOIRED model images of the best-fit limb-darkened stellar disks. These images provide a direct visualization of the symmetric photospheric models used in the companion and spot searches, and help illustrate the angular scales probed by the data. Figure~\ref{fig:ups_per_model_images} shows an example for $\upsilon$~Per.

\subsection{$H$- and $K$-band empirical behavior}
\label{sec:crossband}

Before comparing the empirical coefficients with atmosphere-model predictions,
we first check whether the measured $H$--$K$ behavior is driven by parameters
of the joint visibility fit. 
The fitted power-law coefficients, $\alpha_H$ and
$\alpha_K$, show no strong dependence on either the fitted angular diameter
$\theta_{\rm PL}$ or the visibility normalization $V_0^2$
(Appendix~\ref{app:diagnostic_plots}). 
Thus the systematic separation between
$\alpha_H$ and $\alpha_K$ is not primarily a by-product of covariance with the
angular scale or the visibility-scale factor.

Figure~\ref{fig:h_vs_k_crossband} compares the empirical and model-predicted
power-law coefficients in the $\alpha_H$--$\alpha_K$ plane. The colored CHARA points lie systematically below the one-to-one line, showing
that the fitted limb darkening is weaker in $K$ than in $H$, as expected for
near-infrared stellar atmospheres. The gray model sequences show the
corresponding atmosphere-grid predictions. However, the empirical $H$-to-$K$ change is substantially larger
than predicted by the atmosphere grids. 

Using Equation~\ref{eq:chara_hk_contrast}, the CHARA sample has a median
$H$--$K$ coefficient contrast of
$C^\alpha_{\rm HK}=38.15\%$, with a 16--84\% range of $33$--$46\%$.

The gray model sequences in Figure~\ref{fig:h_vs_k_crossband} show the
corresponding atmosphere-grid predictions. For the SVAM coefficients, the
ExoTiC-LD grids give $C^{\alpha,g}_{\rm HK}({\rm SVAM})=17.12$--$21.68\%$, while SATLAS gives
$19.10\%$
(\mbox{Table~\ref{tab:hk_model_summary}}). The MPS2 reference grid predicts
$C_{\rm HK}^{\alpha,{\rm MPS2}}({\rm SVAM})=19.78\%$, roughly half of the empirical CHARA value.

\begin{table*}
\centering
\caption{Median $H$-to-$K$ power-law coefficient chromaticity and
CHARA--model coefficient offsets. For the empirical CHARA coefficients, the
median over the sample is
$\alpha_H^{\rm CHARA}-\alpha_K^{\rm CHARA}=0.1050$, corresponding to
$C_{\rm HK}^{\alpha}=38.15\%$
(Equation~\ref{eq:chara_hk_contrast}). The second and third columns give the
model $H$--$K$ coefficient contrast for coefficients fit directly to the
passband-integrated model profiles, $\alpha^g_{I,b}$, and for SVAM
coefficients, $\alpha^g_{\rm SVAM,b}$, respectively. The final two columns give
the CHARA--SVAM fractional coefficient offsets, $\Delta^g_b$
(Equation~\ref{eq:fractional_alpha_offset}), for $b\in\{H,K\}$.}
\label{tab:hk_model_summary}
\begin{tabular}{lcccc}
\toprule
Grid &
$C_{\rm HK}^{\alpha,g}(I)$ (\%) &
$C_{\rm HK}^{\alpha,g}({\rm SVAM})$ (\%) &
$\Delta^g_H$ (\%) &
$\Delta^g_K$ (\%) \\
\midrule
Stagger & 24.90 & 21.68 & +28.81 &  +1.04 \\
Kurucz  & 21.58 & 17.12 & +26.89 &  -5.22 \\
MPS1    & 23.00 & 20.02 & +21.23 &  -6.09 \\
MPS2    & 23.07 & 19.78 & +21.04 &  -5.38 \\
SATLAS  & 22.19 & 19.10 & +13.11 & -12.46 \\
\bottomrule
\end{tabular}
\end{table*}

\subsection{Band-dependent empirical--model coefficient offsets}
\label{sec:model-data}

Figures~\ref{fig:data_model_for_teff} and
\ref{fig:spam_data_model_for_teff} compare the empirical CHARA power-law
coefficients with atmosphere-model predictions as a function of
$T_{\rm eff}$. The direct intensity-domain comparison uses
$\alpha^g_{I,b}$, while the preferred like-for-like comparison uses the SVAM
coefficients, $\alpha^g_{{\rm SVAM},b}$, recovered from synthetic
CHARA-sampled visibilities. We quantify the fractional CHARA--model
coefficient offsets in \mbox{Table~\ref{tab:hk_model_summary}}.

Across the sample, the CHARA--model offsets exceed the combined empirical and
model-propagated uncertainties at the sample level. To test whether this
behavior is driven by different $H$- and $K$-band spatial-frequency coverage,
we carried out a matched-resolution robustness test. As discussed in
Section~\ref{sec:Tests_systematic_biases}, the sample-level $H$-band trends
are unchanged within uncertainties when the $H$-band data are restricted to
the spatial-frequency range sampled in $K$.

The largest effective offset occurs in $H$, where the CHARA coefficients are
systematically larger than all five atmosphere-model predictions. Using
Equation~\ref{eq:absolute_alpha_offset}, the median SVAM absolute offsets are
$\delta\alpha_H^g=+0.0648$, $+0.0609$, $+0.0448$, $+0.0465$, and $+0.0304$
for Stagger, Kurucz, MPS1, MPS2, and SATLAS, respectively. The corresponding
fractional offsets, defined by Equation~\ref{eq:fractional_alpha_offset}, are
$\Delta_H^g=+28.81\%$, $+26.89\%$, $+21.23\%$, $+21.04\%$, and $+13.11\%$.

In the $K$ band, the SVAM offsets are smaller and more grid-dependent. The
median absolute offsets are
$\delta\alpha_K^g=+0.0019$, $-0.0102$, $-0.0107$, $-0.0107$, and $-0.0222$
for Stagger, Kurucz, MPS1, MPS2, and SATLAS, respectively, with corresponding
fractional offsets of
$\Delta_K^g=+1.04\%$, $-5.22\%$, $-6.09\%$, $-5.38\%$, and $-12.46\%$.
Thus, relative to the MPS2 SVAM reference, CHARA is higher in $H$ by
$21.04\%$ but lower in $K$ by $5.38\%$.

SATLAS gives the smallest absolute $H$-band offset among the model families
considered, but it overshoots in $K$ and still does not reproduce the observed
$H$--$K$ chromaticity
(\mbox{Figures~\ref{fig:h_vs_k_crossband}} and
\ref{fig:spam_data_model_for_teff}). The $H$-band discrepancy is systematic
across model families, whereas in $K$ the inter-grid differences are comparable
to the empirical--model coefficient offsets. These offsets are therefore
effective, law-dependent diagnostics of model--data mismatch, not direct
measurements of the underlying atmosphere structure.

The corresponding diameter-level effect is discussed separately in
Section~\ref{sec:results_fixed_clv_fits}.

\begin{figure}[htbp]
\centering
\includegraphics[width=0.48\textwidth]{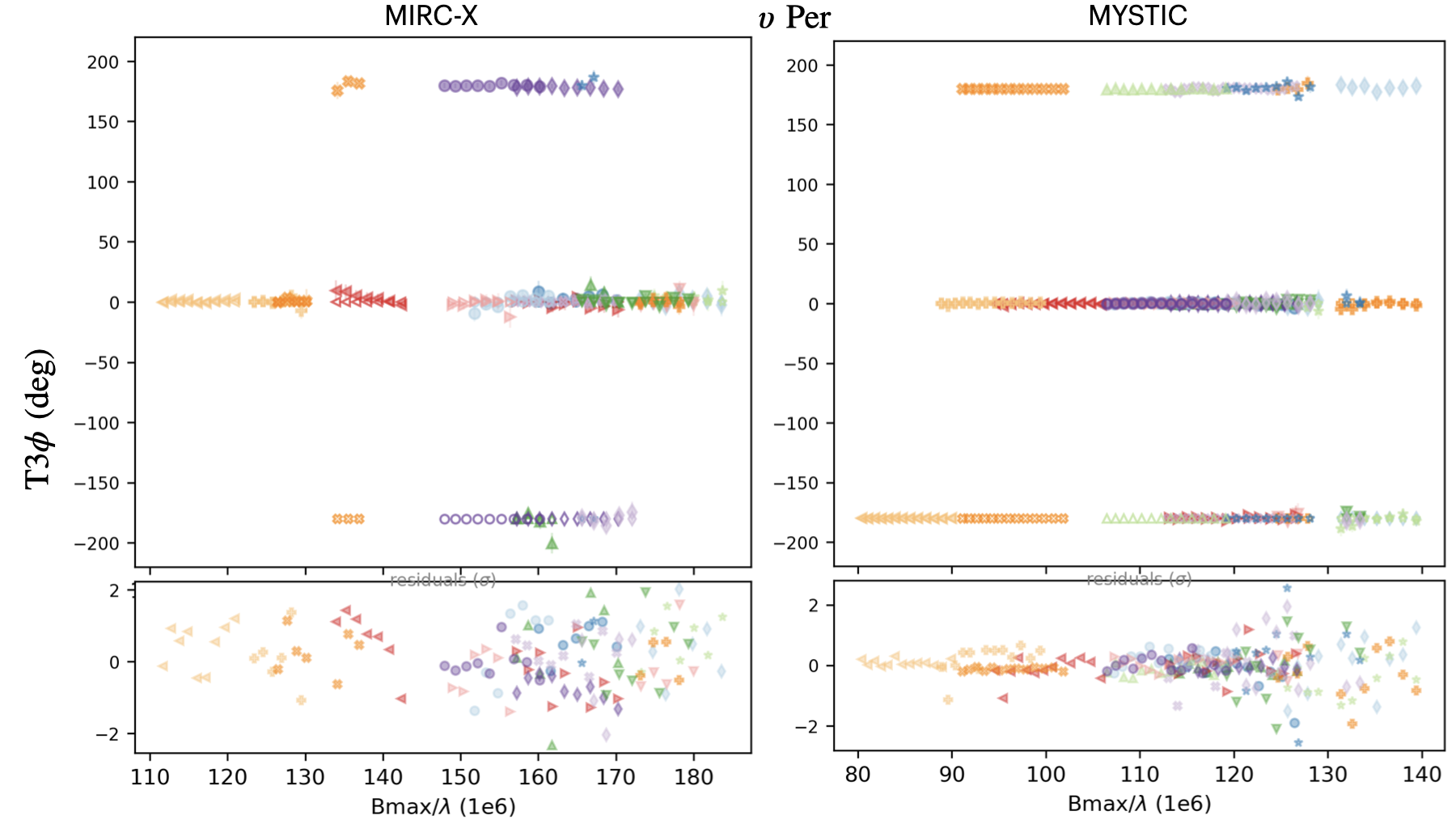}
\caption{
Illustrative closure-phase example for $\upsilon$~Per, shown for MIRC-X in
the $H$ band (left) and MYSTIC in the $K$ band (right). The closure phases are wrapped modulo $180^\circ$ so that a centro-symmetric source maps to $0^\circ$. Within the measurement uncertainties, the data are consistent with
centro-symmetry in both bands. The complete figure set (82 images) is available
in the online version of the article and in the Zenodo figure-set
archive \citep{AnuguZenodo2026}.}
\label{fig:closure_phase_example}
\end{figure}

\begin{figure}[htbp]
\centering
\includegraphics[width=0.48\textwidth]{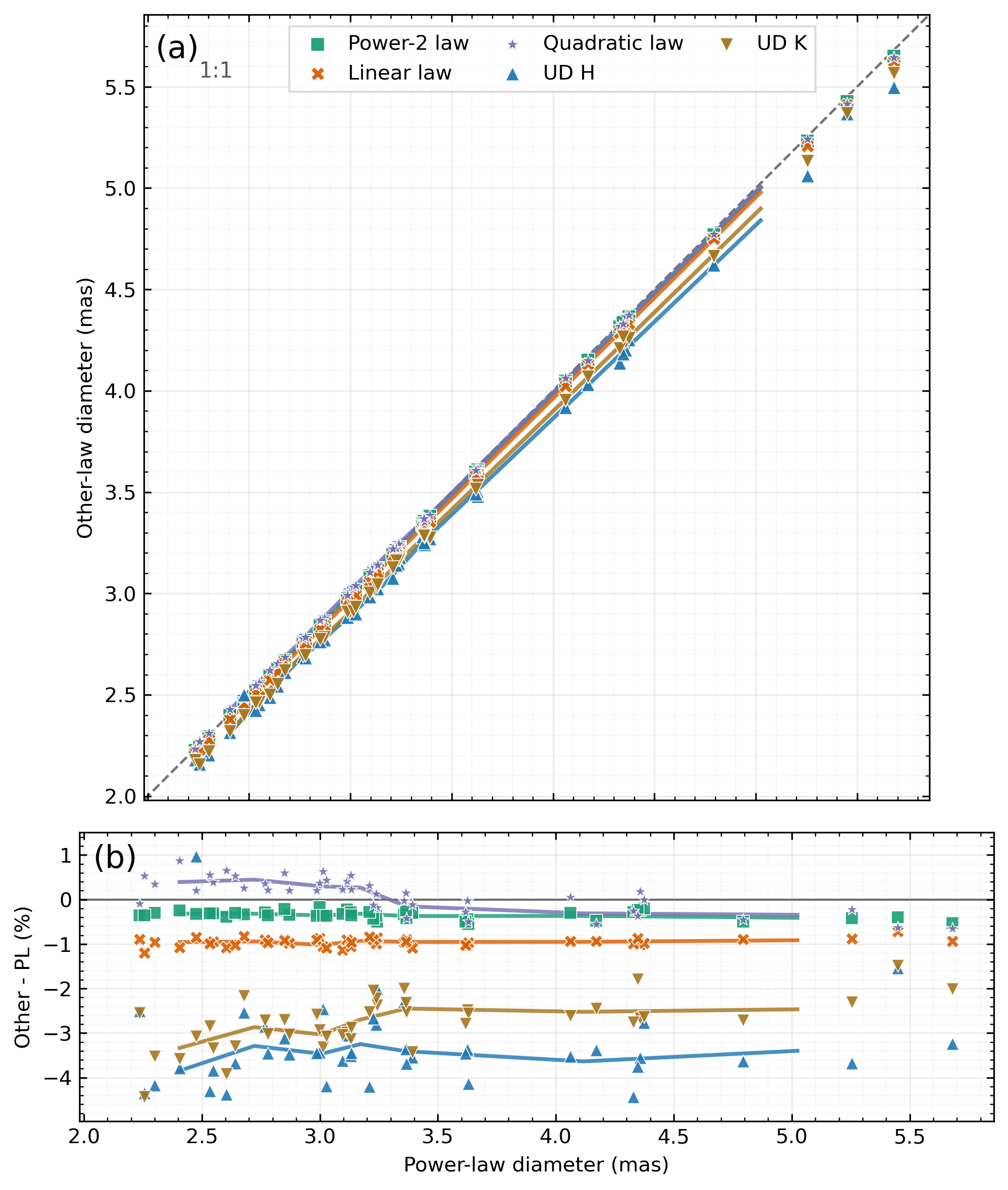}
\caption{
Angular diameters from the power-law fit compared with other laws.
Panel (a) shows diameter from other fitting methods ($\theta_{\rm other}$) versus $\theta_{\rm PL}$; the black dashed diagonal is the one-to-one relation
$\theta_{\rm other}=\theta_{\rm PL}$. Panel (b) shows the fractional
difference, $(\theta_{\rm other}-\theta_{\rm PL})/\theta_{\rm PL}$, in
percent. The colored solid curves in panel (b) are running-median trends for
each law, computed from the same point sets shown in panel (a). Power--2
and quadratic diameters remain closest to $\theta_{\rm PL}$, typically within
$0.5\%$. The quadratic law often gives higher $\chi^2$ when fewer than two
visibility nulls are sampled, reflecting parameter degeneracy. Linear-law and
uniform-disk diameters are smaller because they underrepresent or ignore
limb-darkening curvature, with the largest uniform-disk offsets in $H$.
}
\label{fig:diameters_ld_laws}
\end{figure}

\begin{figure}[htbp]
\centering
\includegraphics[width=0.499\textwidth]{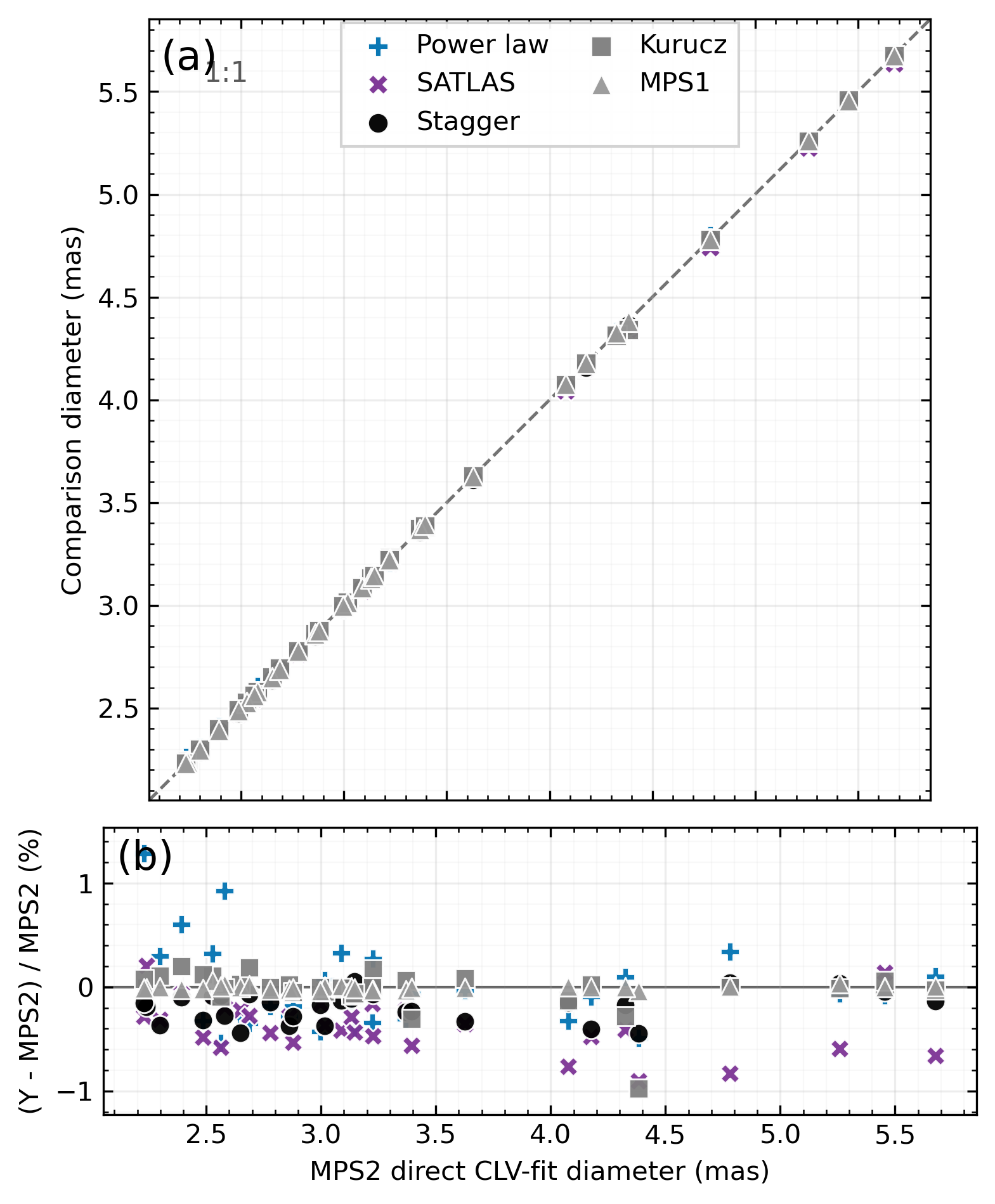}
\caption{
Comparison of the MPS2 fixed-CLV fit diameter, $\theta_{\rm CLV}^{\rm MPS2}$,
with $\theta_{\rm PL}$ and with $\theta_{\rm CLV}^g$ from Stagger, Kurucz, and
MPS1. The upper panel shows the one-to-one comparison; the lower panel shows
fractional residuals relative to MPS2. Fixed-CLV fit diameters from different
grids agree closely, while $\theta_{\rm PL}$ shows larger target-to-target
scatter relative to MPS2.
}
\label{fig:diameters_direct_clv_law}
\end{figure}

\begin{figure}[htbp]
\centering
\includegraphics[width=0.49\textwidth]{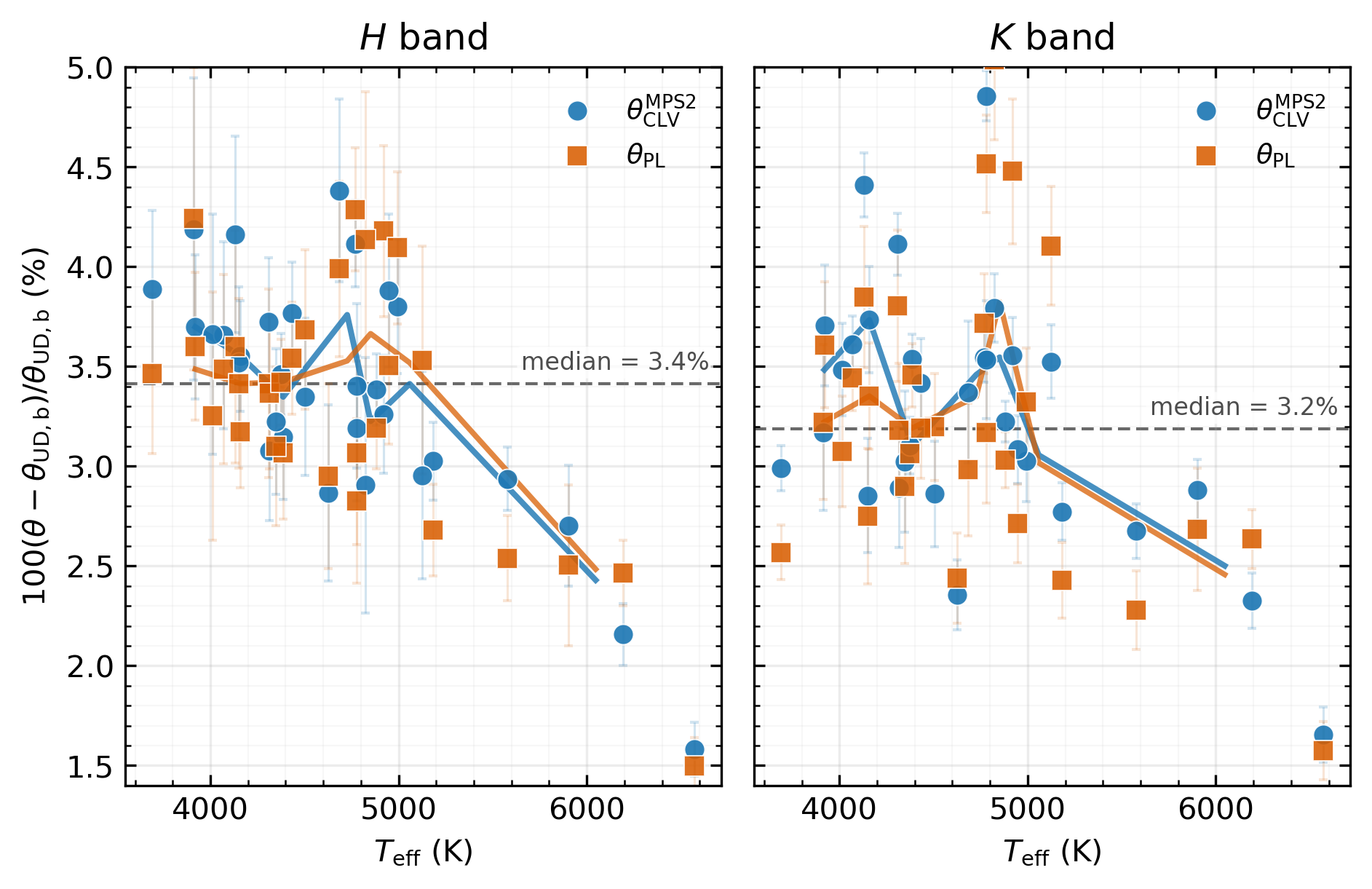}
\caption{
Uniform-disk to limb-darkened diameter corrections in the $H$ and
$K$ bands. For each target and band ($H$ or $K$), the ordinate gives
$100\,(\theta-\theta_{\rm UD,b})/\theta_{\rm UD,b}$, where $\theta_{\rm UD,b}$ is the band-specific uniform-disk diameter and
$\theta$ is either the empirical joint-fit power-law diameter, $\theta_{\rm PL}$,
or the MPS2 fixed-CLV diameter, $\theta_{\rm CLV}^{\rm MPS2}$. The correction is
larger in $H$ than in $K$, consistent with stronger limb darkening at shorter
wavelengths. The empirical power-law and MPS2 fixed-CLV corrections are similar
in amplitude, showing that the overall diameter correction is stable even where
the limb-darkening coefficients differ between CHARA and the models.
}
\label{fig:ud_to_ld_diameter_correction}
\end{figure}

\begin{figure}[htbp]
\centering
\includegraphics[width=0.48\textwidth]{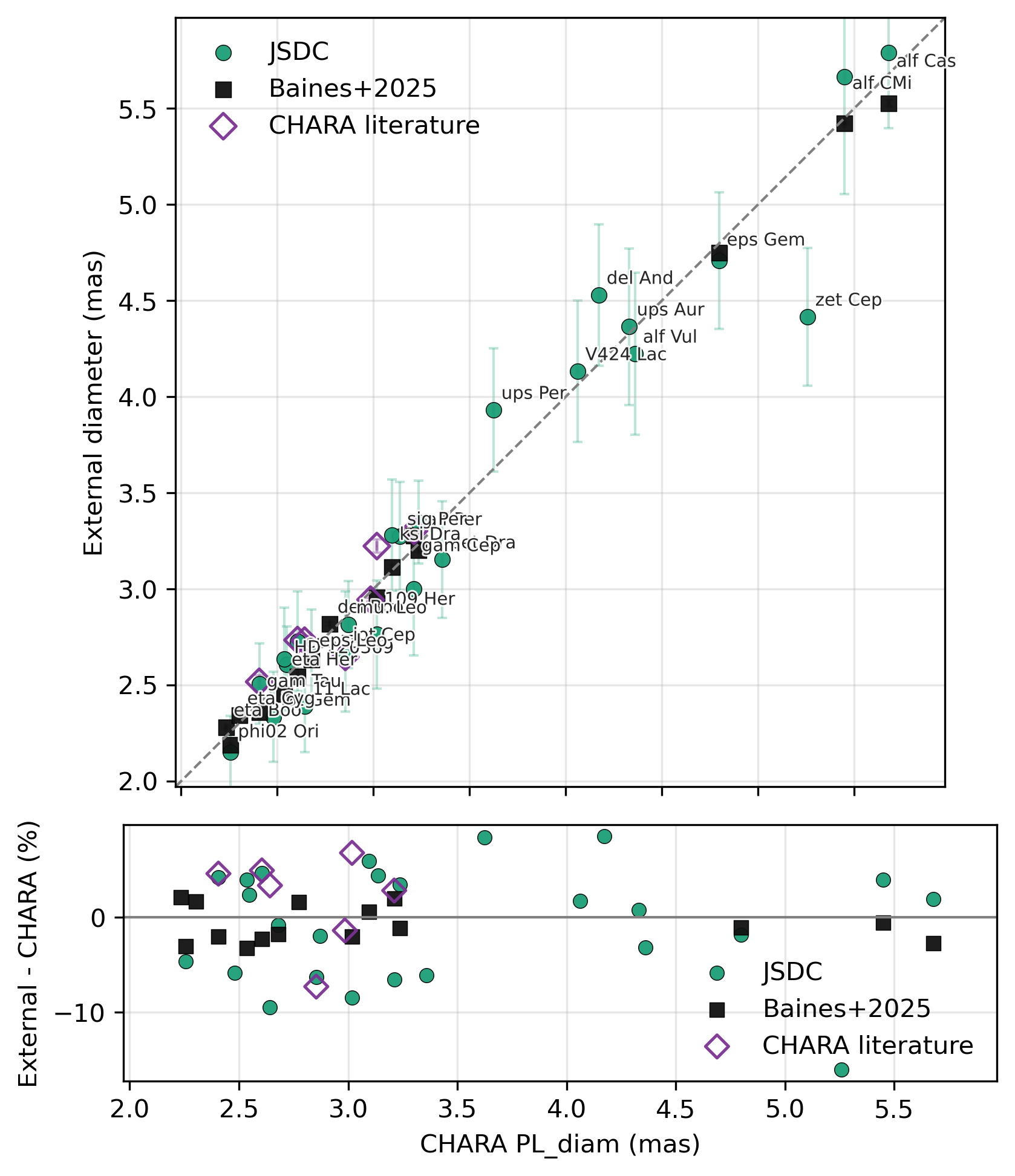}
\caption{
CHARA power-law limb-darkened diameters compared with literature. The
upper panel compares $\theta_{\rm PL}$ with JMMC catalog diameters
\citep{Chelli2016,Bourges2017}, NPOI $R$-band diameters from
\citet{Baines2025}, and literature CHARA diameters
\citep{Boyajian2009ApJ6911243,Baines2009ApJ701154,Baines2010ApJ7101365,Berio2011AA53559,Merand2010A&A...517A..64M};
the dashed line marks equality. The lower panel shows
$(\theta_{\rm ext}-\theta_{\rm PL})/\theta_{\rm PL}$ in percent. The comparison
is generally broadly consistent with $\theta_{\rm PL}$, with scatter dominated by
source-to-source literature differences.
}
\label{fig:comapre_diameters_Baines2025}
\end{figure}

\subsection{Intensity-domain versus visibility-domain model coefficients}
\label{sec:synthetic_validation}

Figures~\ref{fig:h_vs_k_crossband} and
\ref{fig:spam_data_model_for_teff} compare model coefficients obtained from
direct fits to the passband-integrated intensity profiles, $I_b(\mu)$, with
those recovered from synthetic CHARA-sampled visibilities.

Thus, transforming the model CLVs into the visibility domain and sampling them
at the observed CHARA $(u,v,\lambda)$ points changes the recovered coefficients
at a measurable $10^{-2}$ level. This effect is smaller than the
CHARA--model coefficient offsets reported in Section~\ref{sec:model-data}.
Finite spatial-frequency sampling and visibility-domain fitting therefore do
not by themselves explain the dominant discrepancies.

\subsection{Angular Diameters from Analytic-Law and Fixed-CLV Fits}
\label{sec:results_fixed_clv_fits}

Figure~\ref{fig:diameters_ld_laws} compares angular diameters obtained with
different analytic laws. The power--2 and quadratic diameters remain close to
the power-law values, typically within $0.5\%$. Linear-law diameters are
systematically smaller by $\sim1.1$\%, consistent with underestimated CLV curvature, and
uniform-disk diameters are smaller still, especially in $H$, where limb
darkening is stronger. The corresponding law-to-law comparison in intensity space is shown in
Appendix~\ref{app:laws_compare}.

Figure~\ref{fig:diameters_direct_clv_law} compares the empirical analytic-law
diameters with fixed-CLV diameters from the atmosphere grids. In the fixed-CLV
fits, the model CLV shape is held fixed while the common $H{+}K$ angular
diameter and visibility normalization $V_0^2$ are fitted
(Section~\ref{sec:diameters_fit_to_clv}). We compare the MPS2 fixed-CLV
diameter, $\theta_{\rm CLV}^{\rm MPS2}$, with fixed-CLV diameters from
Stagger, Kurucz, MPS1, and SATLAS, and with the empirical analytic-law
diameters.

The main result is that the limb-darkened angular diameters are much more
stable than the limb-darkening coefficients. Replacing the empirical power-law
profile with the MPS2 fixed CLV changes the diameter by only a few tenths of a
percent:
$(\theta_{\rm CLV}^{\rm MPS2}-\theta_{\rm PL})/
\theta_{\rm CLV}^{\rm MPS2}$ has a median of $0.08\%$ and an RMS scatter of
$0.61\%$. Other analytic laws
produce larger law-dependent shifts, especially the linear law, whose median
offset is $1.11\%$. The quadratic and power--2 laws remain closer to the
power-law diameters, with median offsets of $0.30\%$ and $0.48\%$,
respectively. These differences are therefore best interpreted as the
diameter response to the assumed CLV shape, not as evidence for a large bias
in $\theta_{\rm PL}$.

Similar small differences are seen when comparing fixed-CLV fits across
atmosphere grids. Using
$100(\theta-\theta_{\rm CLV}^{\rm MPS2})/\theta_{\rm CLV}^{\rm MPS2}$, the
median offsets are $-0.365\%$ for SATLAS, $-0.153\%$ for Stagger,
$-0.003\%$ for Kurucz, and $-0.004\%$ for MPS1; the corresponding RMS scatters
are $0.446\%$, $0.235\%$, $0.207\%$, and $0.022\%$. Thus, the adopted
atmosphere grid changes the fitted diameter only weakly. However, agreement in
the fitted angular diameters does not mean that the model CLV profiles
reproduce the observed visibility curvature equally well. The fixed-CLV fits
use one common $H{+}K$ diameter, and the weaker limb darkening in $K$ helps
anchor the diameter scale. The stronger model--data differences in $H$,
especially for the cooler stars, therefore appear mainly in the fitted
limb-darkening coefficients and in the residual visibility curvature rather
than as a large shift in the combined angular diameter. The fixed-CLV fits are
therefore a useful diameter consistency check, but they do not remove the
larger coefficient-level discrepancy.

Figure~\ref{fig:ud_to_ld_diameter_correction} shows the wavelength dependence
from the angular-diameter perspective. Both $\theta_{\rm PL}$ and
$\theta_{\rm CLV}^{\rm MPS2}$ are larger than the band-specific uniform-disk
diameters, as expected when limb darkening is included. The median correction is
slightly larger in $H$ than in $K$, consistent with stronger near-infrared limb
darkening at shorter wavelengths. The corrections also tend to decrease toward
higher $T_{\rm eff}$, following the same qualitative temperature dependence
seen in the fitted power-law coefficients.
The close agreement between the empirical power-law and MPS2 fixed-CLV diameter
corrections shows that the gross angular-diameter correction behaves normally.
Thus, the stronger CHARA--model tension appears mainly in the fitted
coefficients and residual higher-lobe visibility curvature, rather than as a
large shift in the combined angular diameter.

\subsection{Assessment of systematic effects}
\label{sec:Tests_systematic_biases}

We examined whether the observed $H$--$K$ chromaticity and CHARA--model
coefficient offsets could be produced by observational or methodological
systematics rather than by differences between the empirical and model CLVs.

\begin{itemize}

\item The closure-phase checks do not support contamination from companions or
strong non-centrosymmetric surface structure
(Section~\ref{sec:companions_imaging}).

\item We tested whether the coefficient-fitting framework itself could produce
the observed offsets. For the CHARA data, the fitted $\alpha_H$ and $\alpha_K$
coefficients are reproducible within uncertainties when using both 10-minute
and 30-second averaged datasets. Synthetic datasets generated with PMOIRED
using CHARA-like $(u,v,\lambda)$ coverage and MPS2-based input CLVs recover the
input power-law coefficients within the quoted uncertainties
(Appendix~\ref{app:alpha_diameter_recovery_test}). In addition, the
intensity-domain model coefficients,
$(\alpha^{g}_{I,H},\alpha^{g}_{I,K})$, and the synthetic-visibility
coefficients,
$(\alpha^{g}_{\mathrm{SVAM},H},\alpha^{g}_{\mathrm{SVAM},K})$, are generally
consistent within uncertainties (\mbox{Figures~\ref{fig:h_vs_k_crossband}}, \ref{fig:data_model_for_teff} and \ref{fig:spam_data_model_for_teff}). These tests indicate that numerical
instability, averaging choices, and incomplete spatial-frequency sampling are
unlikely to explain the observed star-to-star scatter or the mean
CHARA--model coefficient offsets.

\item Calibration checks do not indicate that calibrator-size errors drive the
chromatic discrepancy. Nearly all 42 calibrators are unresolved in $K$, while
about half are unresolved in $H$, where calibrator-size uncertainties are more
important (Appendix~\ref{app:Appendix_obs_log}); we therefore used same-night cross-calibration to refine the
effective calibrator diameters. A subset of eight targets were observed on 2026
April 4. This night used calibrators with diameters $<0.4$ mas, and those data reproduce
the same $H$--$K$ behavior. Restricting the MIRC-X data to the MYSTIC
spatial-frequency range ($B<240$ m) and refitting the power-law coefficients
also leaves the sample-level trends unchanged. Thus, neither partially resolved
calibrators nor the longest $H$-band baselines explain the chromatic
discrepancy.

\item MIRC-X and MYSTIC operate simultaneously on the same CHARA beams, share
the upstream optical train and science--calibrator cadence, and are reduced
within the same calibration framework. We found no night-dependent differential
offset that would reproduce the sample-wide $H$--$K$ behavior.

\item We studied the impact of delay line tracking errors. \citet{Anugu2026} showed that delay-line tracking errors can produce
systematic $V^2$ losses. Our observations were obtained in 2025--2026
(Appendix~\ref{app:Appendix_obs_log}), after the delay-line time-jitter problem
was corrected in early 2025. During these observations, the residual delay-line
tracking errors were below 20~nm, corresponding to instrumental $V^2$ losses
of $<1\%$ in both $H$ and $K$. Delay-line tracking errors are therefore
unlikely to drive the observed $H$--$K$ discrepancy.

\item Multi-epoch observations of $\alpha$~Per, $\upsilon$~Per, $\alpha$~Vul,
$\lambda$~Her, 109~Her, $\delta$~Boo, and $\beta$~Dra show that power-law
diameters are repeatable at the few-tenths-of-a-percent level, while the
coefficients are repeatable within their quoted uncertainties. The individual
epoch-by-epoch fitted values and corresponding visibility-fit comparisons are
provided in the machine-readable table and the Zenodo archive
\citep{AnuguZenodo2026}. 

\end{itemize}

\section{Discussion}
\label{sec:discussion}

\subsection{Physical interpretation of the model--data power-law coefficient offsets}

The results in Section~\ref{sec:results} show that the main tension with the
atmosphere grids is the wavelength dependence of the limb darkening. The
empirical $H$--$K$ change is larger than predicted by all five grids, while
the single-band empirical--model coefficient offsets are largest in $H$ and weaker or more
model-dependent in $K$.

The empirical--model coefficient offsets in Figure~\ref{fig:spam_data_model_for_teff} suggest
that the largest $H$-band discrepancies occur preferentially among cooler
and/or more evolved stars. Several of the largest offsets are found for stars
with $T_{\rm eff}\lesssim5000$~K, qualitatively similar to the behavior found
in TESS transit analyses, where empirical quadratic limb-darkening coefficients
for stars cooler than $\sim5000$~K deviate systematically from PHOENIX/ATLAS
predictions \citep{Patel2022AJ}. However, the prominent offset of hot and F-type
supergiant $\alpha$~Per ($T_{\rm eff} = 6193$ K; see \mbox{Figures~\ref{fig:data_model_for_teff}} and ~\ref{fig:spam_data_model_for_teff}) shows that effective temperature alone cannot explain
the discrepancy. Luminosity class, surface gravity, atmospheric extension, and
surface inhomogeneity are also likely relevant.

A pure analytic-law effect is unlikely to explain the observed pattern. Changing
the adopted law can shift the absolute coefficient scale, but such a shift would
be expected to act more uniformly across bands and stellar types. Instead, the
largest discrepancies are concentrated in $H$ and among the cooler and/or more
evolved targets, while the $K$-band agreement is generally better. 
The persistence of the same qualitative trend in the alternative-law and SVAM comparisons suggests that the discrepancy is not driven solely by the one-parameter power-law parameterization.

SATLAS provides a useful limiting comparison because it is the only spherical,
low-gravity atmosphere grid in our model set. It gives the smallest absolute
$H$-band coefficient offset among the model sets considered, but it does not
correct the $H$--$K$ chromaticity: the predicted coefficient contrast remains
far smaller than the empirical CHARA value. Moreover,  SATLAS gives the most negative $K$-band offset. Thus, spherical
low-gravity geometry affects the absolute coefficient scale but does not solve
the wavelength-dependent discrepancy.

For cooler and more evolved stars, empirical--model coefficient offsets are expected to be more
sensitive to atmospheric extension, convection, molecular opacity, magnetic
surface structure, and the breakdown of simple one-dimensional or
plane-parallel assumptions
\citep[e.g.,][]{Neilson2011,Neilson2013,Magic2015,Kostogryz2024,Hauschildt2025A&A...698A..47H}.
The concentration of larger offsets among cool and/or highly evolved stars is
therefore suggestive, but the present sample does not isolate a single
controlling stellar parameter.

Independent empirical studies point to the same broad conclusion that model
CLVs can differ measurably from direct limb-darkening constraints. The
VLTI/PIONIER study of alf~Cen~A and B found weaker $H$-band power-law
limb darkening than predicted by both 1D and 3D atmosphere models
\citep{Kervella2017}. Recent CHARA/PAVO measurements of $\kappa$~Cyg likewise
found that Stagger predictions overestimated the $R$-band limb darkening
relative to direct interferometric fits \citep{2026MNRAS.548ag719C}. Similar
empirical--model coefficient offsets have also been reported in high-precision Kepler, TESS,
and JWST transit analyses \citep[e.g.,][]{Maxted2023,Rustamkulov2023,
Kostogryz2024}. The present CHARA $H$- and $K$-band measurements therefore add
a near-infrared, evolved-star benchmark to these empirical tests of
stellar-atmosphere CLVs.

\subsection{Interpreting empirical diameters and limb-darkening coefficients}

Although the coefficient-level CHARA--model coefficient offsets are significant, the
fixed-CLV fits show no corresponding large systematic bias in the empirical
power-law diameters relative to full atmosphere-model CLVs
(Figure~\ref{fig:diameters_direct_clv_law}). The remaining target-to-target
scatter between $\theta_{\rm PL}$ and $\theta_{\rm CLV}^{\rm MPS2}$ shows that
individual diameters retain some sensitivity to whether the visibility profile
is represented by a fixed model CLV or by an empirical power-law fit. The spread
among $\theta_{\rm CLV}^g$ values from different atmosphere grids is also larger
than the offset between $\theta_{\rm PL}$ and any single reference grid (see Figure~\ref{fig:diameters_direct_clv_law}). We
therefore treat the fixed-CLV diameters as a model-dependent systematic check,
rather than as a replacement for the empirical $\theta_{\rm PL}$ scale.

We interpret the fitted power-law coefficients, $\alpha_H$ and $\alpha_K$, as
compact empirical summaries of the visibility-curve curvature constrained by
the data, not as unique physical atmosphere descriptions. The corresponding
power-law diameters provide a practical internal diameter scale and remain
broadly consistent with previous interferometric and catalog estimates after
accounting for wavelength band, spatial-frequency coverage, and
limb-darkening prescription differences
(Figure~\ref{fig:comapre_diameters_Baines2025}). Thus, the coefficient-level
CHARA--model discrepancy should not be interpreted as a comparably large
diameter bias. The angular diameter is mainly anchored by the visibility nulls,
whereas the limb-darkening coefficient is more sensitive to the higher-lobe
curvature and therefore to the detailed CLV shape.

\subsection{Scope of the benchmark and future tests}

The present benchmark is complementary to transit limb-darkening studies,
which usually constrain compact main-sequence stars with short-period planets.
The CHARA sample instead provides direct near-infrared CLV constraints for
evolved stars with luminosity classes IV--I. For subgiants and giants, larger
radii, lower transit probabilities, and shallower transit depths make limb
darkening more difficult to constrain from transits
\citep[e.g.,][]{Howarth2011MNRAS.418.1165H,Maxted2023}. Eclipsing binaries can
also constrain limb darkening, but evolved-star systems often require additional
modeling for tidal distortion, gravity darkening, reflection, and non-spherical
geometry. To our knowledge, there are no published transit-based
limb-darkening constraints from TESS or JWST for the targets in this sample.
CHARA therefore provides direct CLV constraints for evolved stars that are
difficult to test by occultation methods. The same fits also provide a
homogeneous $H$/$K$ diameter scale for bright giants and supergiants, which can
be combined with parallaxes and bolometric fluxes for radius and effective
temperature work.

A subset of the targets was observed at multiple epochs. For these stars, the
power-law angular diameters are repeatable at the few-tenths-of-a-percent level,
and the fitted power-law coefficients remain consistent within their quoted
uncertainties across nights (Section~\ref{sec:Tests_systematic_biases}). This
repeatability, seen for targets such as $\alpha$~Per, $\upsilon$~Per,
$\alpha$~Vul, $\lambda$~Her, 109~Her, $\delta$~Boo, and $\beta$~Dra, supports
the internal consistency of the measurements and is much smaller than the
sample-wide star-to-star dispersion discussed in this paper.

An important limitation is that we do not yet have contemporaneous photometric
or spectroscopic monitoring for the full sample to quantify surface activity or
time variability on timescales relevant for low-contrast convection, spots, or
pulsation. While our closure-phase checks reject strong non-centrosymmetric
structure and companions (Section~\ref{sec:companions_imaging}), they do not
rule out low-contrast or nearly axisymmetric surface structure that could
influence the inferred CLV and contribute to the observed star-to-star
dispersion. A useful next step is targeted time-domain analysis of available
TESS light curves for these bright targets, following the approach of
\citet{alfCas_Rudrasingam2026}, to search for signatures of convection, spots, or
pulsation that can be compared with the interferometric limb-darkening
measurements. Combined with additional multi-epoch interferometric monitoring,
such data would help separate intrinsic target-to-target differences from
time-variable behavior.

Future work should also extend the comparison to cooler, lower-gravity evolved
stars and to broader wavelength coverage. Cool red giants occupy the regime
where spherical extension, molecular opacity, and convective surface structure
can make the CLV especially sensitive to the adopted atmosphere model; recent
spherical model grids emphasize that center-to-limb differences from
plane-parallel models are especially large for giants
\citep{Hauschildt2025A&A...698A..47H}. Simultaneous $J{+}H{+}K$ coverage with
MIRC-X and MYSTIC would test whether the empirical--model mismatch strengthens
toward shorter near-infrared wavelengths. A split-band test within the existing
$H$ and $K$ coverage already suggests the expected sense of this wavelength
dependence, with the shorter-wavelength sub-bands generally yielding stronger
fitted limb darkening (Appendix~\ref{app:split_band_ld_trends}). Visible-band
interferometry \citep[SPICA,][]{Mourard2022}, where limb darkening is
intrinsically stronger, would provide an even more sensitive test of these
CHARA--model differences.

\section{Conclusions}\label{sec:conclusion}

We have presented a dual-band ($H$ and $K$) interferometric
study of stellar limb darkening for 31 bright stars observed with the CHARA
Array. We fitted the combined $H+K$ squared
visibilities with four analytic laws. Among them, the one-parameter power-law 
provides the most uniform coefficient-level summary for this dataset, with
higher-order laws used as consistency checks.

We used closure-phase data to search for companions or isolated
non-centrosymmetric surface structures and found no detections brighter than
$1\%$ of the primary stellar flux (Figure~\ref{fig:closure_phase_example}). 
The closure-phase non-detections do not indicate companions or large isolated
non-centrosymmetric structures at the tested contrast level, which is
consistent with interpreting the dominant visibility-curvature signal as limb
darkening.

We compared the resulting CHARA fitted coefficients with bandpass-matched
model-grid predictions. The strongest discrepancy lies in the wavelength
dependence: the empirical $H$--$K$ decrease in the power-law coefficient is
nearly twice the model-predicted decrease. Relative to MPS2 in the preferred
SVAM comparison, the empirical CHARA coefficients are higher by
$\delta\alpha_H^{\rm MPS2}=+0.0465$ ($\Delta_H^{\rm MPS2}=+21.04\%$) in the
$H$ band and lower by $\delta\alpha_K^{\rm MPS2}=-0.0107$
($\Delta_K^{\rm MPS2}=-5.38\%$) in the $K$ band. Among the model families
considered, SATLAS gives the smallest absolute $H$-band offset,
$\delta\alpha_H^{\rm SATLAS}=+0.0304$
($\Delta_H^{\rm SATLAS}=+13.11\%$), but overpredicts the $K$-band coefficient,
leaving $\delta\alpha_K^{\rm SATLAS}=-0.0222$
($\Delta_K^{\rm SATLAS}=-12.46\%$) and still failing to reproduce the observed
chromatic contrast.

The target-to-target scatter in the empirical--model coefficient offsets exceeds the formally propagated
uncertainties. Synthetic recovery tests, closure-phase companion checks,
and matched $H$- and $K$-band spatial-frequency tests indicate that the excess scatter is unlikely to be
explained primarily by numerical instability, unresolved companions, or simple
cross-band resolution mismatch. We therefore interpret the coefficient offsets
as evidence that the observed and model CLV shapes differ in the adopted
analytic-law representation, rather than as direct measurements of a specific
atmospheric parameter. 

In addition to the limb-darkening comparison, this work provides a homogeneous set of $H$/$K$ limb-darkened angular diameters for 31 bright evolved stars. The formal diameter uncertainties are typically below 0.3\%, and the coefficients and diameters are provided in machine-readable form to support future atmosphere-model and calibration tests.

Future CHARA measurements extending this visibility-domain framework to cooler
red giants, additional epochs, and broader wavelength coverage from $R$ through
$K$ will test whether the empirical--model discrepancy strengthens toward lower
$T_{\rm eff}$, lower $\log g$, and shorter wavelengths.

\begin{acknowledgments}
We thank the anonymous referee for the careful reading and constructive suggestions, which improved the clarity and presentation of the paper.
We thank the CHARA support staff for their invaluable support during observations. 
We thank E. K. Baines and S. Mahadevan for helpful discussions about this project. 
This work is based upon observations obtained with the Georgia State University Center for High Angular Resolution Astronomy Array at Mount Wilson Observatory.  
The CHARA Array is supported by the National Science Foundation under Grant No. AST-2034336 and AST-2407956. Institutional support has been provided from the GSU College of Arts and Sciences, Office of the Provost, and Office of the Vice President for Research and Economic Development.
SK acknowledges funding for MIRC-X from the European Research Council (ERC) under the European Union's Horizon 2020 research and innovation programme (Starting Grant No. 639889 and Consolidated Grant No. 101003096). 
JDM acknowledges funding for the development of MIRC-X (NASA-XRP NNX16AD43G, NSF-AST 2009489) and MYSTIC (NSF-ATI 1506540, NSF-AST 1909165). 
AG acknowledges support from the Agencia Nacional de Investigaci\'on y Desarrollo (ANID) through FONDECYT Regular grant 1241073.
This research has made use of the Jean-Marie Mariotti Center Aspro and SearchCal services.
ChatGPT (GPT-5.1) was used to assist with language editing and readability improvements in parts of the manuscript text.
\end{acknowledgments}

\begin{contribution}
Anugu developed the project, obtained the CHARA observations, led the data analysis and interpretation, and wrote the manuscript. All authors discussed the results, reviewed the manuscript, and provided feedback.
\end{contribution}

\facilities{CHARA (MIRC-X \& MYSTIC)}

\software{PMOIRED \citep{Merand2022SPIE12183E..1NM},
        ExoTiC-LD \citep{Grant2024JOSS},
        MIRC-X pipeline \citep{Anugu2020},
        JMMC OIFITS and Aspro \citep{Tallon-Bosc2024}}

\FloatBarrier
\appendix

\begin{figure*}[htbp]
\centering
\includegraphics[width=\textwidth]{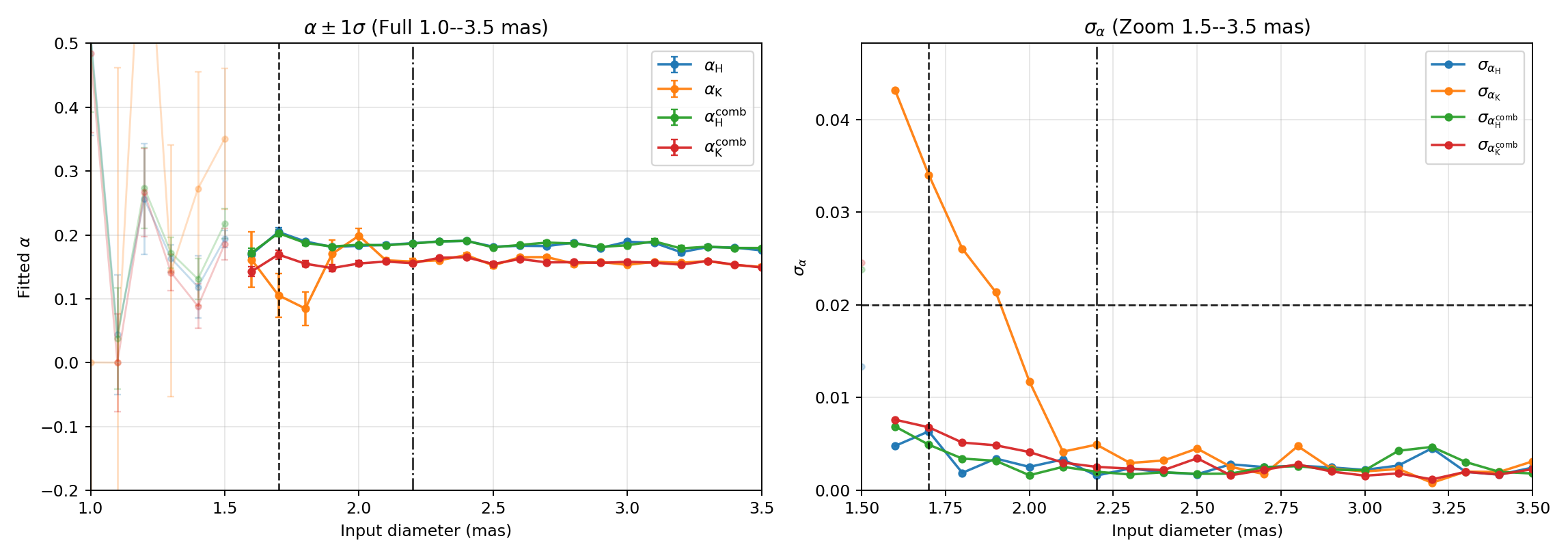}
\caption{Synthetic recovery test for the power-law limb-darkening coefficient as
a function of input angular diameter. The left panel shows the recovered
$H$- and $K$-band coefficients for independent and joint $H{+}K$ fits, with
$1\sigma$ uncertainties; points at $\theta<1.5$~mas are de-emphasized because
the recovery is unstable. The right panel shows the corresponding coefficient
uncertainties, with $\sigma_\alpha=0.02$ marked for reference. Vertical lines at
$\theta=1.8$ and $2.3$~mas indicate the approximate transition to robust
$H$-band and independent $K$-band recovery, respectively.}
\label{fig:diam_alpha_scan}
\end{figure*}

\begin{figure}[htbp]
\centering
\includegraphics[width=0.48\textwidth]{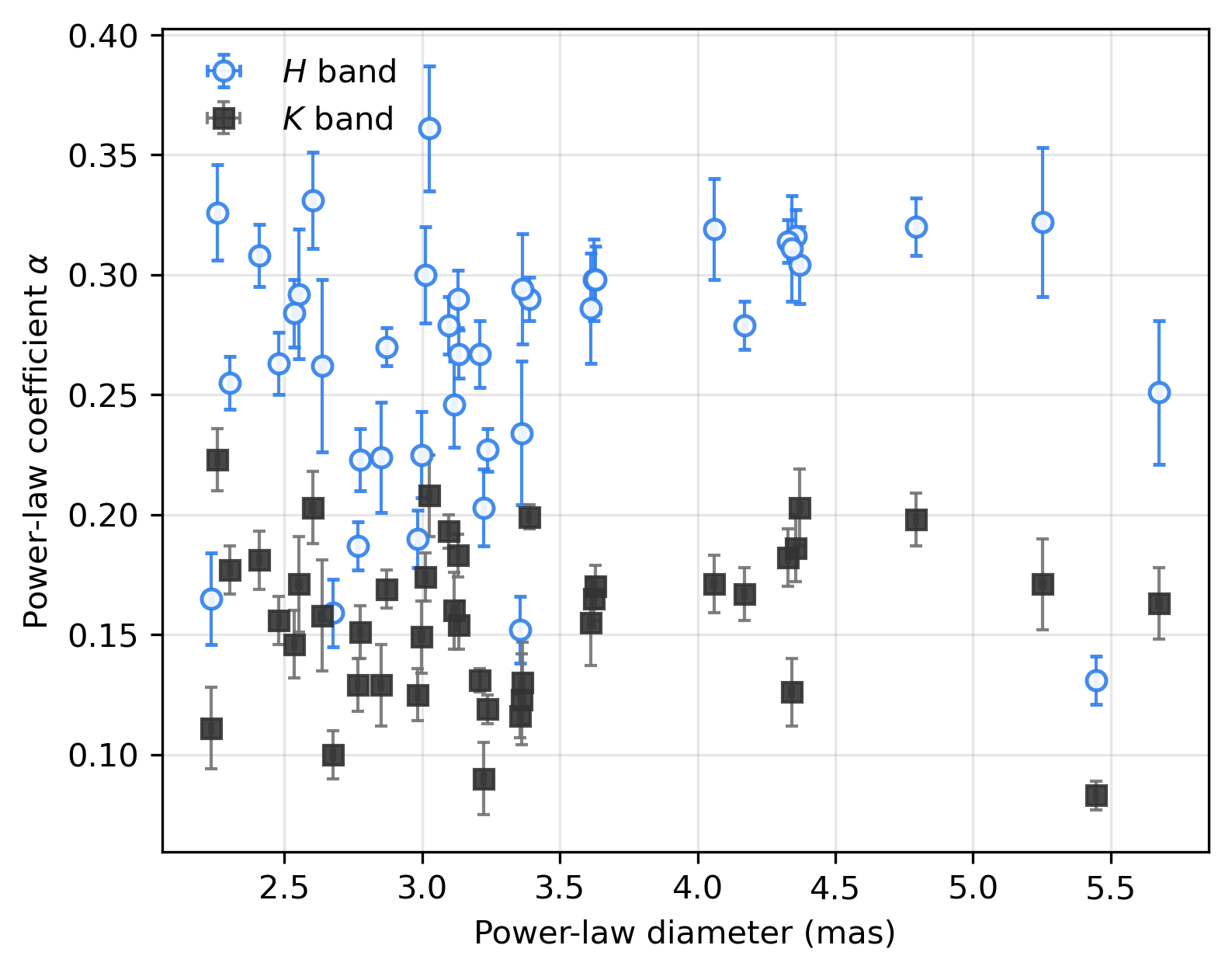}
\includegraphics[width=0.48\textwidth]{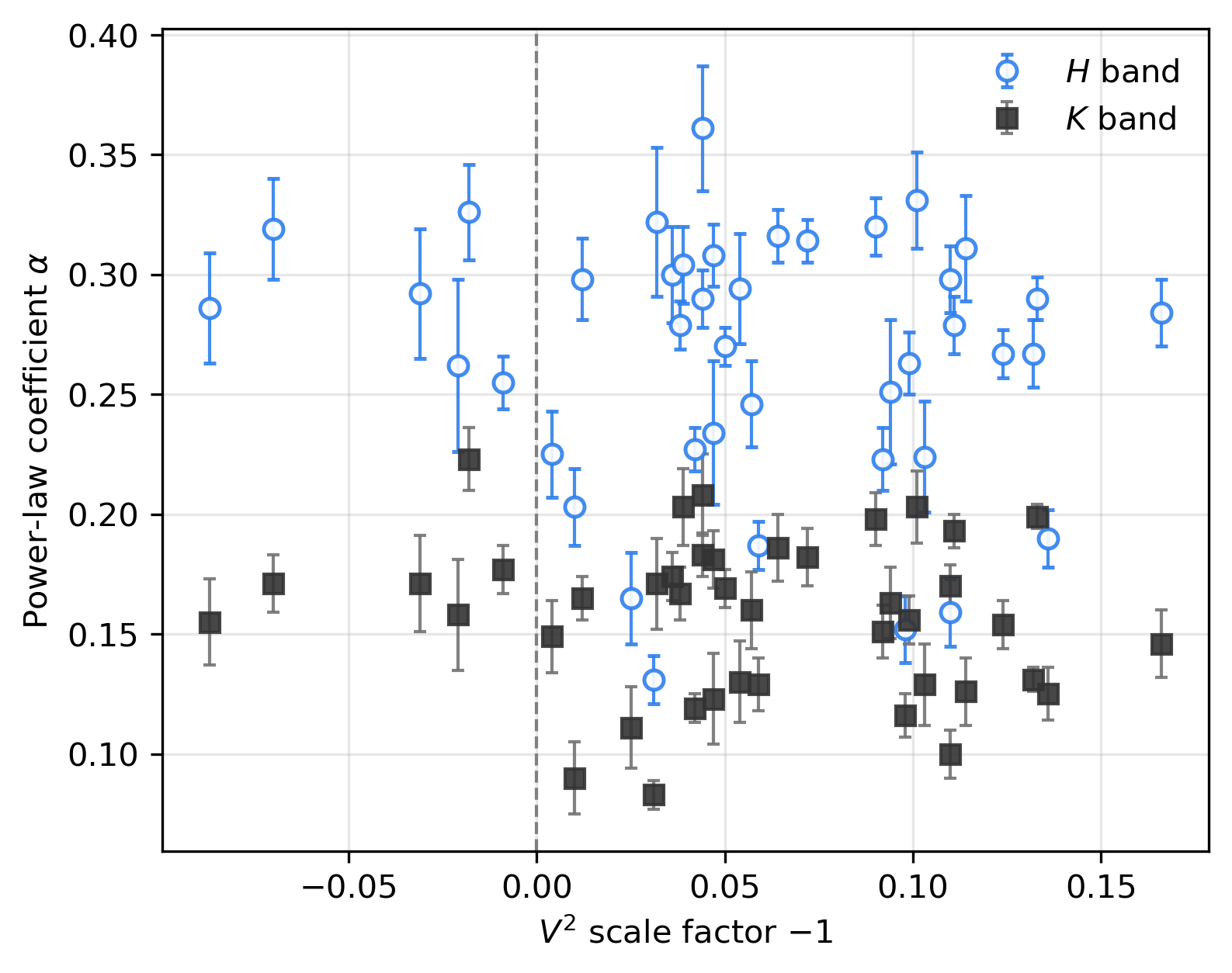}
\caption{
Checks on the fitted power-law coefficients. Left: fitted $H$- and $K$-band
coefficients, $\alpha_H$ and $\alpha_K$, plotted against the fitted power-law
angular diameter, $\theta_{\rm PL}$. Right: the same coefficients plotted
against the fitted visibility scale factor, shown as $V^2_0 - 1$. In both
panels, the $H$-band coefficients are systematically larger than the $K$-band
coefficients, consistent with stronger limb darkening at shorter wavelengths.
No strong monotonic dependence of $\alpha_H$ and $\alpha_K$ on either
$\theta_{\rm PL}$ or $V^2_0$ is evident over the sample, indicating that the
measured limb-darkening coefficients are not primarily driven by the overall
angular scale or by the visibility normalization term.
}
\label{fig:diagnostic_plots}
\end{figure}

\begin{figure*}[htbp]
\centering
\includegraphics[width=0.49\textwidth]{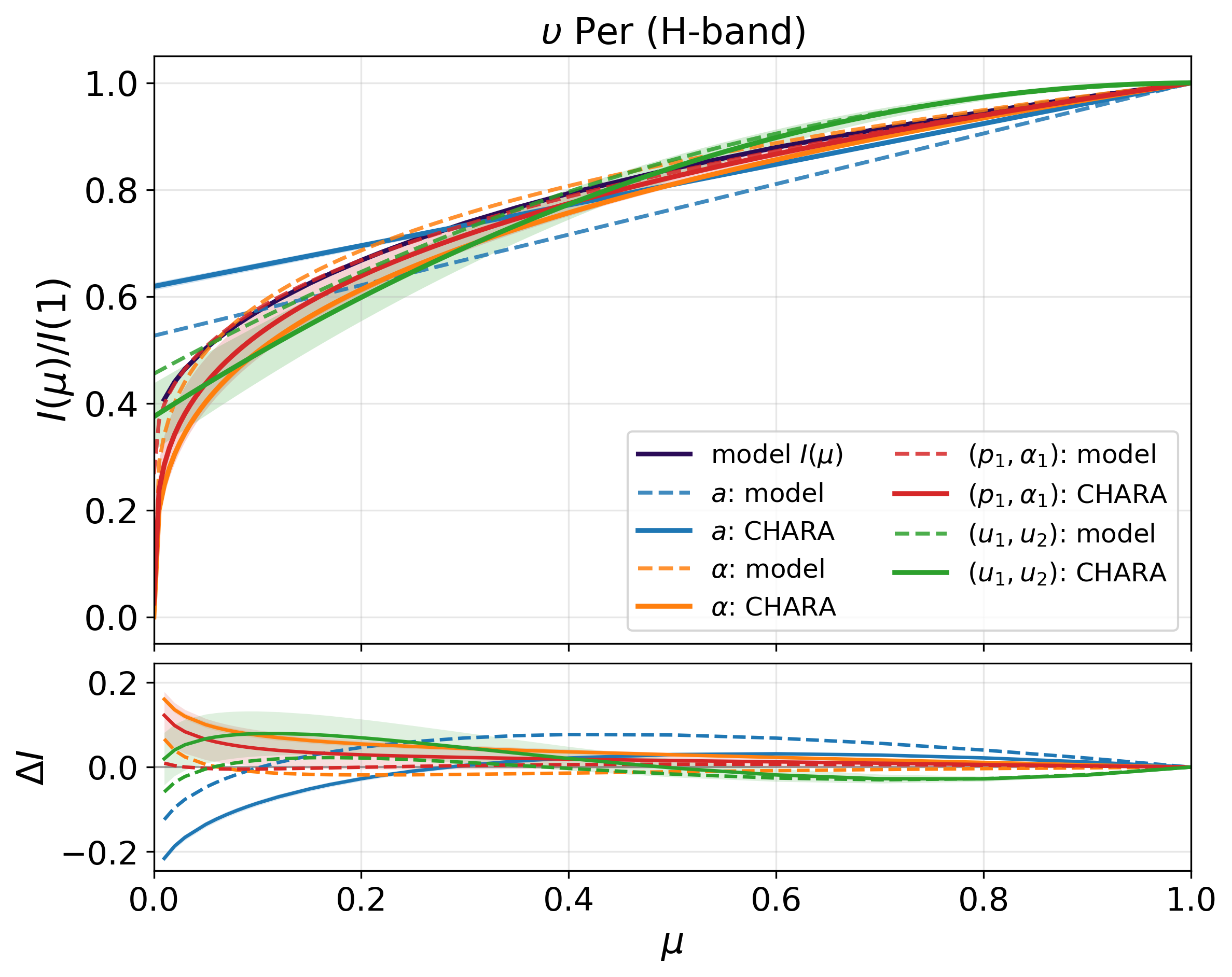}
\includegraphics[width=0.49\textwidth]{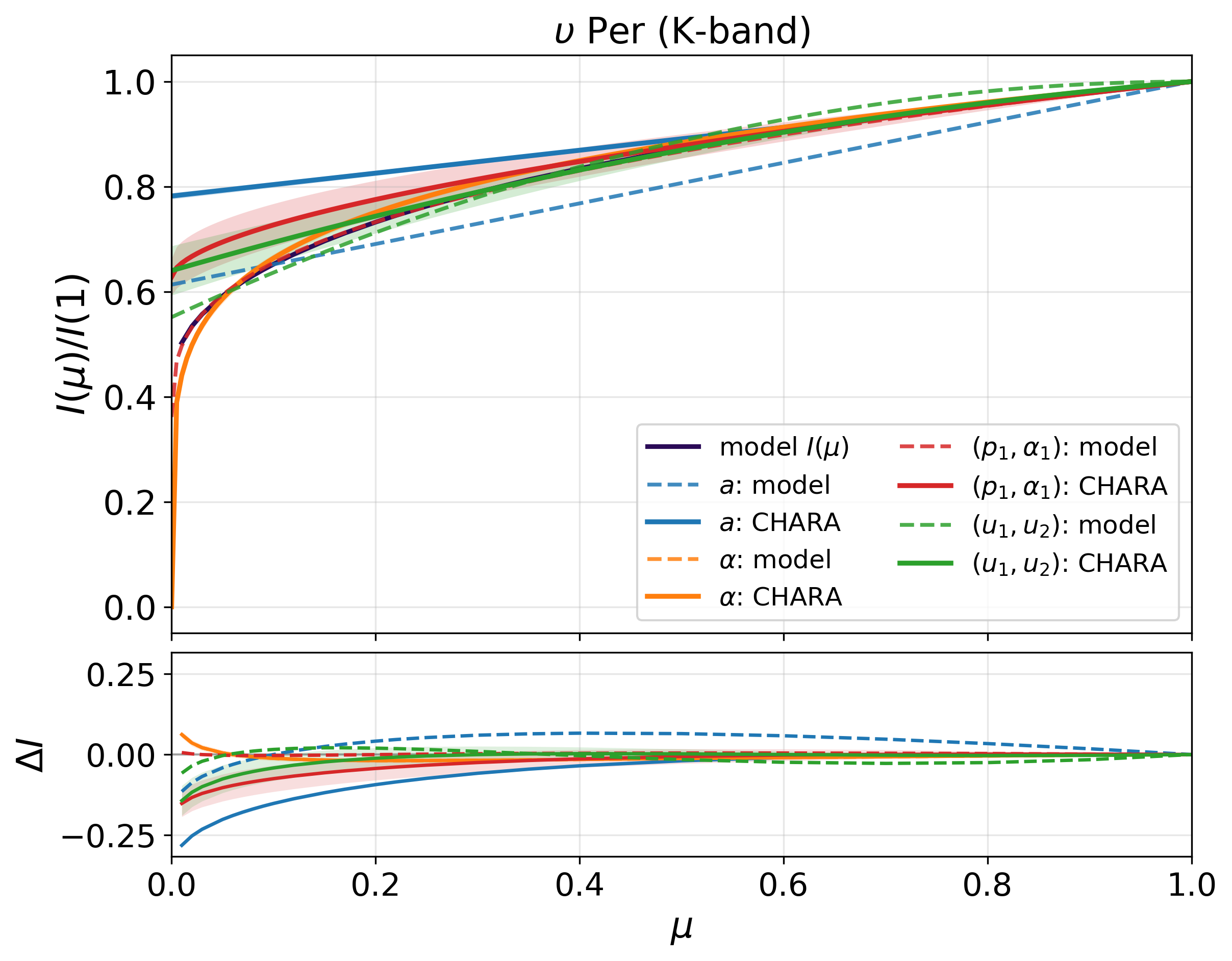}
\caption{Illustrative intensity-space comparison for $\upsilon$~Per in the
$H$ band (left) and $K$ band (right). The parent MPS2 model CLV is shown
along with the analytic-law profiles reconstructed from the model coefficients
and from the CHARA visibility-domain fits. This view makes the law-to-law
differences more apparent than in the fitted $V^2$ curves alone. The complete figure set (42 target/epoch pairs; 84 band-specific plots) is available in the online version of the article and in the Zenodo figure-set archive \citep{AnuguZenodo2026}.}
\label{fig:appendix_laws_compare}
\end{figure*}

\begin{figure*}[htbp]
\centering
\includegraphics[width=\textwidth]{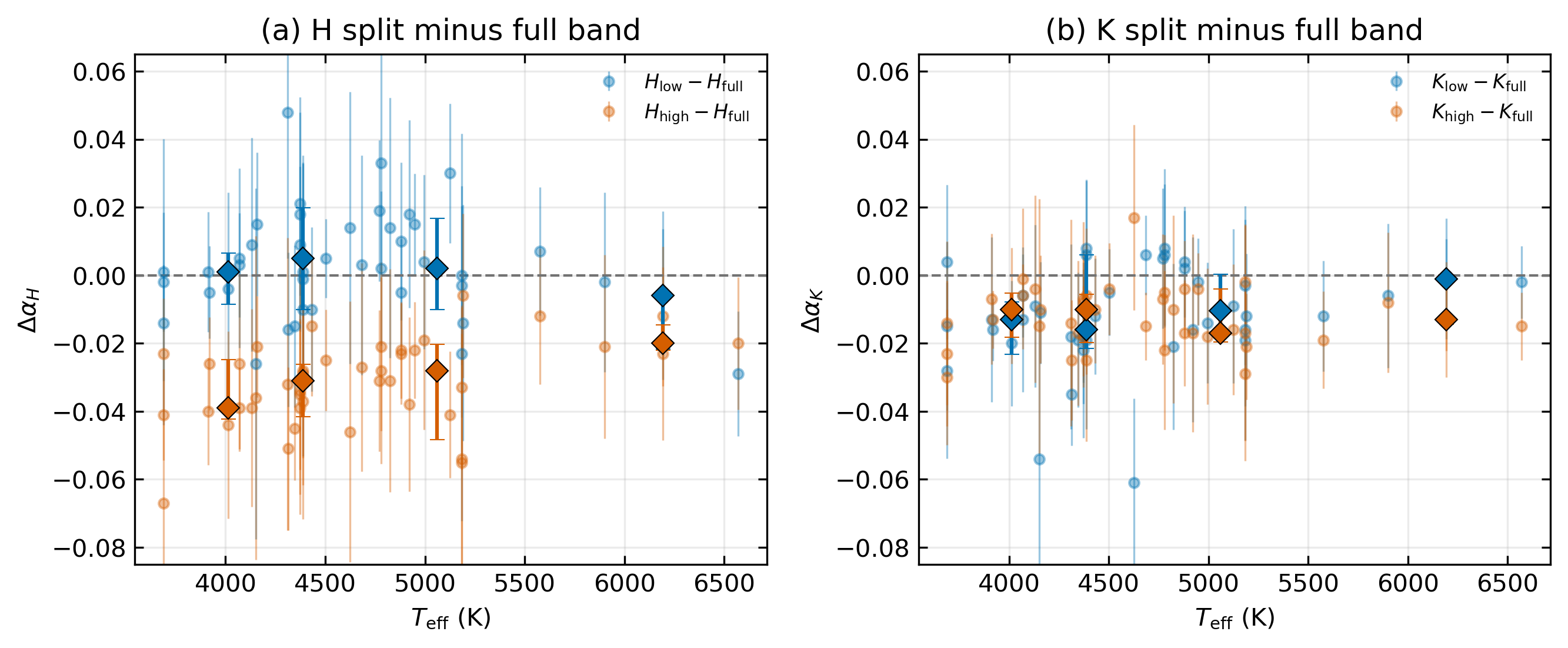}
\caption{Empirical power-law limb-darkening coefficients as a function of
wavelength within the broad $H$ and $K$ bands. Panel (a) shows the offsets
between the lower- and higher-wavelength $H$-band fits and the full-$H$ fit;
panel (b) shows the analogous offsets for $K$. Circles show individual targets,
and diamonds show median residuals in broad $T_{\rm eff}$ bins, with error bars
spanning the 16th--84th percentile range. The split-band fits provide an
internal check on wavelength sensitivity within each passband.}
\label{fig:hk_split_teff}
\end{figure*}

The appendices provide supporting material for the main results, including a
test of the minimum angular diameter required to recover the limb-darkening
coefficient reliably (Appendix~\ref{app:alpha_diameter_recovery_test}), the full observing log
and calibration steps (Appendix~\ref{app:Appendix_obs_log}), a direct
intensity-space comparison of the analytic-law profiles
(Appendix~\ref{app:laws_compare}), and an
internal split-band wavelength diagnostic within the existing $H$ and $K$
coverage (Appendix~\ref{app:split_band_ld_trends}).

\section{Star diameter dependence of limb darkening power-law coefficient recovery accuracy}
\label{app:alpha_diameter_recovery_test}

To directly test the practical meaning of the 1.5-$V^2$ null requirement, we
performed an additional CHARA-like synthetic data simulations using the same
six-telescope configuration. We adopted a fixed limb-darkening setup
($T_{\rm eff}=5772$\,K, $\log g=4.438$, $M=1.0\,M_\odot$) based on a spherical
SATLAS atmosphere model \citep{Neilson2011,Neilson2013}, implemented in PMOIRED. We varied only the
input angular diameter. Synthetic-data generation and coefficient recovery were
carried out with PMOIRED.
For each input diameter, we generated
synthetic visibilities on the same representative $(u,v,\lambda)$ coverage and
fit the power-law prescription in $H$, $K$, and in a combined fit.

Figure~\ref{fig:diam_alpha_scan} summarizes the behavior of the recovered
power-law coefficients and their bootstrap uncertainties. At small diameters
($\lesssim 1.0$~mas), the recovered coefficients are unstable and the
uncertainties are large, consistent with insufficient sampling of visibility
curvature. The synthetic scan indicates band-dependent thresholds:
$H$-band recovery is robust from $\theta\gtrsim1.8$~mas (about 1.5 $V^2$
nulls), whereas independent $K$-only recovery becomes robust at
$\theta\gtrsim2.3$~mas. The higher $K$ threshold is expected because, at
longer wavelength, a fixed projected baseline samples lower spatial frequency and
therefore covers less of the visibility curvature. In joint $H{+}K$ fits, the recovered
$K$-band coefficient stabilizes near $\theta\sim1.8$~mas because the
shared diameter is constrained by the stronger $H$-band curvature signal,
reducing the $K$-band $\alpha$--diameter degeneracy. We therefore adopt
$\theta\gtrsim2.3$~mas as a conservative threshold for independent $K$-band
limb-darkening constraints, while joint $H{+}K$ inference remains effective at
smaller diameter at 1.75 mas.

\section{CHARA Observing Log and Calibrations}\label{app:Appendix_obs_log}
Table~\ref{tab:calibrators} lists the calibrators used for each
target and epoch. Calibrator selection is particularly important because
unrecognized multiplicity or partially resolved calibrators can imprint
baseline-dependent transfer-function errors. Before the observations, we screened
candidate calibrators for binarity using Gaia EDR3 astrometric indicators as
defined by \citet{Kervella2022A&A...657A...7K} (RUWE $< 1.4$ and
$\texttt{snrPMaH2EG2b} < 2$, $\texttt{snrPMaH2EG3b} < 2$). As an additional
empirical safeguard, we verified that calibrator closure phases are consistent
with zero (i.e., a centro-symmetric brightness distribution) within our
measurement precision.

All observations analyzed here were obtained in 2025 and 2026, after correcting timing-jitter-related delay-line tracking errors ($<20$~nm)  and subsequent system improvements
that enabled stable fringe tracking \citep{Anugu2026}. 
We verified stable instrument performance across the observing runs via standard pipeline quality diagnostics and conservative calibrator propagation, as described below.

\subsection{Transfer-function calibration and calibrator sizes}\label{app:tf}

The calibration of $V^2$ followed the standard MIRC-X/MYSTIC pipeline. Initial estimates of the calibrator diameters
were adopted from the JMMC Stellar Diameters Catalogue
\citep[JSDC;][]{Chelli2016,Bourges2017}. We then refined these values by
calibrating calibrators against other calibrators observed on the same nights
and fitting power-law limb-darkened diameters (LDDs) for the calibrators. The
calibrator--calibrator fits were carried out separately by each night so that the
diameter refinement used only calibrators sharing the same instrumental
transfer function. For calibrators
observed in multiple epochs, the adopted value is the mean of the fitted
epoch-level LDDs. The calibrator diameters listed in \mbox{Table~\ref{tab:calibrators}}
are these final adopted recalibrated LDDs, not the original JSDC starting
values.

The standard MIRC-X/MYSTIC calibration scripts normally read and use JSDC uniform-disk
calibrator diameters, but for this analysis we manually supplied the fitted
calibrator LDDs as the angular-size inputs used in the calibration. The LDDs
were fitted in the $H$ band and adopted for both the $H$- and $K$-band
calibration, because the limb-darkened angular diameter represents the physical
stellar disk scale. The center-to-limb profile is wavelength dependent, but for
these mostly unresolved calibrators the resulting difference in the
diameter-to-visibility correction between $H$ and $K$ is negligible compared
with the adopted diameter uncertainties. Calibrator diameter uncertainties are
propagated into the transfer function so that the calibrated $V^2$
uncertainties capture both statistical and calibration terms. This propagation
is especially important in the $H$ band, where a subset of calibrators are
partially resolved on the longest baselines. For each
calibrator and scan, the contribution of diameter uncertainty is computed from
the derivative of the calibrator visibility model,
\begin{equation}
\left.\frac{\partial V^2_{\rm cal}}{\partial \theta}\right|_{B,\lambda}
= 2 V_{\rm cal}
\left.\frac{\partial V_{\rm cal}}{\partial \theta}\right|_{B,\lambda},
\end{equation}
and added in quadrature to the statistical terms.

\subsection{Wavelength calibration}\label{app:wavelength_calibration}

Accurate wavelength calibration is important because the spatial frequency
$(B/\lambda)$ sets the location of the visibility lobes and nulls and thus
propagates directly into angular-diameter and limb-darkening inferences. For
MIRC-X, we took standard datasets for wavelength-monitoring approach described by
\citet{Gardner2022AJ....164..184G}. In that scheme, a custom six-beam optical
etalon module (named ARMADA) is inserted into the CHARA beams to generate interference packets
that mimic a binary-star signal. The etalon data are reduced with the same
pipeline and spectral-channel definition as the science data, and a per-night
multiplicative correction is derived for the channel wavelengths by fitting the
etalon signal and bringing each night to a common wavelength scale. We apply the
resulting wavelength correction to the channel wavelengths used to compute
spatial frequencies and to perform bandpass integration in the forward modeling.

\subsection{Diagnostic checks on fitted power-law parameters}
\label{app:diagnostic_plots}

Figure~\ref{fig:diagnostic_plots} checks whether the fitted power-law
coefficients are driven by either the fitted angular diameter or the
visibility-normalization term. The $H$-band coefficients remain systematically
larger than the $K$-band coefficients, while no strong monotonic dependence on
$\theta_{\rm PL}$ or $V^2_0$ is apparent.

\section{Intensity-Space Comparison of Analytic Laws}
\label{app:laws_compare}

As an additional diagnostic, we compare the analytic-law fits directly in
intensity space using the same coefficients derived from the visibility-domain
fits. This complements the $V^2$ residual and reduced-$\chi^2$ comparisons in
the main text by showing how each fitted law maps onto the corresponding
center-to-limb intensity profile.

Figure~\ref{fig:appendix_laws_compare} shows an illustrative example for
$\upsilon$~Per in the $H$ and $K$ bands. The figure compares the model CLV with
analytic-law profiles reconstructed from both the model-law coefficients and
the empirical CHARA visibility-domain coefficients. In this representation, the
linear law is more visibly offset from the model CLV, whereas the power-law and
power--2 prescriptions track the model profile more closely over most of the
disk. For targets with $V^2$ sampling beyond the second null, the two-parameters of the power--2 law  matches better than the
one-parameter power law. The two-parameter quadratic law is unstable. The corresponding plots for the remaining targets are
provided in the online Zenodo archive.

\section{Split-band wavelength trends within $H$ and $K$}
\label{app:split_band_ld_trends}

The simultaneous MIRC-X/MYSTIC data allow a finer check of how the empirical
limb-darkening coefficients vary with wavelength inside the broad $H$ and $K$
bands. We subdivided each passband into lower- and higher-wavelength sub-bands
and refit the same power-law limb-darkening prescription in each case.
Figure~\ref{fig:hk_split_teff} shows the difference between each split-band
coefficient and the corresponding full-band coefficient. The effect is clearest
in the $H$ band, where the
lower-wavelength subset tends to give larger $\alpha$ than the
higher-wavelength subset. In $K$, the split-band differences are smaller and
consistent with weak or undetected wavelength dependence within the measurement
scatter. This is expected: near-infrared limb darkening is weaker at longer
wavelengths. We use this split-band exercise only as an
internal test of wavelength sensitivity; a full multi-band analysis will require
broader simultaneous coverage.

\startlongtable
\begin{deluxetable*}{r l c l}
\digitalasset
\tablecaption{Calibrators used for each target and epoch. The entries in parentheses are the final adopted recalibrated limb-darkened calibrator diameters and uncertainties used in the visibility calibration, in mas. Initial diameter estimates were taken from the JMMC/JSDC catalog \citep{Chelli2016,Bourges2017} and then refined as described in Appendix~\ref{app:Appendix_obs_log}. }\label{tab:calibrators}
\tablewidth{0pt}
\tablehead{
\colhead{$ID$} & \colhead{Target} & \colhead{Date} & \colhead{Calibrators, $\theta_{{\rm PL},{\rm cal}}$ (mas)} \\
}
\startdata
1 & $\alpha$ CMi & 2026-04-04 & $\xi$ Vir (${0.442} \pm {0.032}$), HD 57006 (${0.473} \pm {0.016}$) \\
1 &  &  & $\beta$ Leo (${0.447} \pm {0.035}$) \\
2 & $\alpha$ Per & 2025-07-25 & HD 13137 (${0.669} \pm {0.048}$), 7 And (${0.693} \pm {0.068}$) \\
2 &  & 2025-08-27 & HD 25056 (${0.626} \pm {0.050}$), 36 Per (${0.529} \pm {0.043}$) \\
2 &  &  & HD 50384 (${0.645} \pm {0.054}$) \\
3 & $\eta$ Boo & 2026-04-04 & $\pi$ Ser (${0.381} \pm {0.035}$), HD 159332 (${0.517} \pm {0.044}$) \\
3 &  &  & 101 Her (${0.429} \pm {0.032}$) \\
4 & $\epsilon$ Leo & 2026-04-04 & b Leo (${0.447} \pm {0.035}$), 16 Com (${0.401} \pm {0.027}$) \\
4 &  &  & 9 Leo (${0.364} \pm {0.009}$) \\
5 & $\beta$ Dra & 2025-06-05 & HD 132254 (${0.539} \pm {0.013}$), 53 Her (${0.481} \pm {0.013}$) \\
5 &  & 2025-07-20 & HD 173920 (${0.611} \pm {0.049}$), HD 222932 (${0.656} \pm {0.017}$) \\
5 &  &  & 7 And (${0.693} \pm {0.068}$), HD 212136 (${0.631} \pm {0.047}$) \\
5 &  & 2026-03-23 & HD 161053 (${0.545} \pm {0.014}$), HD 191195 (${0.426} \pm {0.038}$) \\
5 &  &  & HD 152306 (${0.478} \pm {0.049}$) \\
6 & $\gamma$ Tau & 2025-09-07 & HD 31294 (${0.681} \pm {0.016}$), HD 29038 (${0.670} \pm {0.016}$) \\
6 &  &  & HD 19903 (${0.565} \pm {0.018}$), HD 28424 (${0.595} \pm {0.014}$) \\
7 & $\eta$ Her & 2025-06-05 & 53 Her (${0.481} \pm {0.013}$), HD 132254 (${0.539} \pm {0.013}$) \\
7 &  &  & HD 167472 (${0.647} \pm {0.053}$) \\
8 & $\eta$ Ser & 2026-05-01 & 10 Ser (${0.424} \pm {0.012}$), HD 115539 (${0.501} \pm {0.012}$) \\
8 &  &  & 60 Her (${0.376} \pm {0.030}$) \\
9 & $\epsilon$ Tau & 2025-09-07 & HD 28424 (${0.595} \pm {0.014}$), HD 29038 (${0.670} \pm {0.016}$) \\
9 &  &  & HD 19903 (${0.565} \pm {0.018}$), HD 31294 (${0.681} \pm {0.016}$) \\
10 & $\delta$ Boo & 2026-04-04 & HD 185999 (${0.553} \pm {0.046}$), HD 173417 (${0.433} \pm {0.011}$) \\
10 &  &  & 53 Her (${0.481} \pm {0.013}$) \\
10 &  & 2026-05-01 & HD 152306 (${0.478} \pm {0.049}$), pi. Ser (${0.381} \pm {0.035}$) \\
10 &  &  & 101 Her (${0.429} \pm {0.032}$) \\
11 & $\phi^2$ Ori & 2025-09-07 & HD 29038 (${0.670} \pm {0.016}$), HD 19903 (${0.565} \pm {0.018}$) \\
11 &  &  & HD 28424 (${0.595} \pm {0.014}$) \\
12 & $\iota$ Cep & 2025-06-04 & HD 212136 (${0.631} \pm {0.047}$), HD 2589 (${0.685} \pm {0.060}$) \\
12 &  &  & HD 193556 (${0.643} \pm {0.056}$), HD 189671 (${0.681} \pm {0.052}$) \\
13 & $\iota$ Gem & 2025-09-09 & HD 59684 (${0.604} \pm {0.014}$), HD 46277 (${0.564} \pm {0.013}$) \\
13 &  &  & HD 26972 (${0.508} \pm {0.011}$) \\
14 & $\eta$ Cyg & 2026-04-04 & HD 189108 (${0.543} \pm {0.013}$), 53 Her (${0.481} \pm {0.013}$) \\
14 &  &  & HD 173417 (${0.433} \pm {0.011}$) \\
15 & $\gamma$ Cep & 2025-06-04 & HD 2589 (${0.685} \pm {0.060}$), HD 212136 (${0.631} \pm {0.047}$) \\
15 &  &  & HD 193556 (${0.643} \pm {0.056}$), HD 189671 (${0.681} \pm {0.052}$) \\
16 & $\alpha$ Cas & 2025-07-20 & HD 13137 (${0.669} \pm {0.048}$), HD 212136 (${0.631} \pm {0.047}$) \\
16 &  &  & HD 222932 (${0.656} \pm {0.017}$), 7 And (${0.693} \pm {0.068}$) \\
17 & $\xi$ Dra & 2026-03-23 & HD 161053 (${0.545} \pm {0.014}$), HD 191195 (${0.426} \pm {0.038}$) \\
17 &  &  & HD 152306 (${0.478} \pm {0.049}$) \\
18 & $\mu$ Leo & 2026-04-04 & 9 Leo (${0.364} \pm {0.009}$), 16 Com (${0.401} \pm {0.027}$) \\
18 &  &  & b Leo (${0.447} \pm {0.035}$) \\
19 & 109 Her & 2025-06-05 & HD 167472 (${0.647} \pm {0.053}$), 53 Her (${0.481} \pm {0.013}$) \\
19 &  &  & HD 132254 (${0.539} \pm {0.013}$) \\
19 &  & 2026-04-04 & 101 Her (${0.429} \pm {0.032}$), HD 189108 (${0.543} \pm {0.013}$) \\
19 &  &  & HD 159332 (${0.517} \pm {0.044}$) \\
19 &  & 2026-05-01 & 101 Her (${0.429} \pm {0.032}$), pi. Ser (${0.381} \pm {0.035}$) \\
19 &  &  & 60 Her (${0.376} \pm {0.030}$) \\
20 & $\upsilon$ Per & 2025-07-20 & HD 13137 (${0.669} \pm {0.048}$), 7 And (${0.693} \pm {0.068}$) \\
20 &  &  & HD 173920 (${0.611} \pm {0.049}$), HD 222932 (${0.656} \pm {0.017}$) \\
20 &  & 2025-07-22 & HD 212136 (${0.631} \pm {0.047}$), HD 13137 (${0.669} \pm {0.048}$) \\
20 &  &  & 7 And (${0.693} \pm {0.068}$) \\
20 &  & 2025-10-20 & HD 20675 (${0.415} \pm {0.040}$), HD 11151 (${0.420} \pm {0.011}$) \\
20 &  &  & 7 And (${0.693} \pm {0.068}$) \\
21 & $\delta$ And & 2025-09-07 & HD 29038 (${0.670} \pm {0.016}$), 7 And (${0.693} \pm {0.068}$) \\
21 &  &  & HD 28424 (${0.595} \pm {0.014}$), HD 31294 (${0.681} \pm {0.016}$) \\
22 & $\epsilon$ Gem & 2025-09-09 & HD 59684 (${0.604} \pm {0.014}$), HD 46277 (${0.564} \pm {0.013}$) \\
22 &  &  & HD 26972 (${0.508} \pm {0.011}$) \\
23 & 11 Lac & 2025-07-25 & 7 And (${0.693} \pm {0.068}$), HD 13137 (${0.669} \pm {0.048}$) \\
24 & $\sigma$ Per & 2025-08-27 & 36 Per (${0.529} \pm {0.043}$), HD 25056 (${0.626} \pm {0.050}$) \\
25 & $\zeta$ Cep & 2025-07-20 & HD 212136 (${0.631} \pm {0.047}$), HD 222932 (${0.656} \pm {0.017}$) \\
25 &  &  & HD 173920 (${0.611} \pm {0.049}$) \\
26 & HD 220369 & 2025-07-25 & 7 And (${0.693} \pm {0.068}$), HD 13137 (${0.669} \pm {0.048}$) \\
27 & $\lambda$ Her & 2026-04-04 & pi. Ser (${0.381} \pm {0.035}$), HD 185999 (${0.553} \pm {0.046}$) \\
27 &  &  & HD 173417 (${0.433} \pm {0.011}$) \\
27 &  & 2026-05-01 & pi. Ser (${0.381} \pm {0.035}$), 101 Her (${0.429} \pm {0.032}$) \\
27 &  &  & HD 152306 (${0.478} \pm {0.049}$) \\
28 & V424 Lac & 2025-07-25 & 7 And (${0.693} \pm {0.068}$), HD 13137 (${0.669} \pm {0.048}$) \\
29 & $\rho$ Ser & 2026-05-01 & HD 115539 (${0.501} \pm {0.012}$), HD 152306 (${0.478} \pm {0.049}$) \\
29 &  &  & pi. Ser (${0.381} \pm {0.035}$) \\
30 & $\upsilon$ Aur & 2025-08-27 & 36 Per (${0.529} \pm {0.043}$), HD 50384 (${0.645} \pm {0.054}$) \\
31 & $\alpha$ Vul & 2025-06-04 & HD 189671 (${0.681} \pm {0.052}$), HD 193556 (${0.643} \pm {0.056}$) \\
31 &  &  & HD 212136 (${0.631} \pm {0.047}$) \\
31 &  & 2026-04-04 & pi. Ser (${0.381} \pm {0.035}$), HD 159332 (${0.517} \pm {0.044}$) \\
31 &  &  & 101 Her (${0.429} \pm {0.032}$) \\
31 &  & 2026-05-01 & HD 138039 (${0.546} \pm {0.080}$), HD 187923 (${0.480} \pm {0.046}$) \\
31 &  &  & 101 Her (${0.429} \pm {0.032}$) \\
\enddata
\end{deluxetable*}


\bibliographystyle{aasjournalv7}
\bibliography{sample631}

\begin{thebibliography}{}
\expandafter\ifx\csname natexlab\endcsname\relax\def\natexlab#1{#1}\fi
\providecommand{\url}[1]{\href{#1}{#1}}
\providecommand{\dodoi}[1]{doi:~\href{http://doi.org/#1}{\nolinkurl{#1}}}
\providecommand{\doeprint}[1]{\href{http://ascl.net/#1}{\nolinkurl{http://ascl.net/#1}}}
\providecommand{\doarXiv}[1]{\href{https://arxiv.org/abs/#1}{\nolinkurl{https://arxiv.org/abs/#1}}}

\bibitem[{C. Afonso {et~al.}(2000)Afonso {et~al.}}]{Afonso2000}
Afonso, C., {et~al.} 2000, \bibinfo{title}{Limb Darkening of a K Giant in the Galactic Bulge: PLANET Photometry of Microlensing Event MACHO 97-BLG-28,} The Astrophysical Journal, 532, 340, \dodoi{10.1086/308537}

\bibitem[{C. {Allende Prieto} \& D.~L. {Lambert}(1999){Allende Prieto} \& {Lambert}}]{AllendePrieto1999}
{Allende Prieto}, C., \& {Lambert}, D.~L. 1999, \bibinfo{title}{{Fundamental parameters of nearby stars from the comparison with evolutionary calculations: masses, radii and effective temperatures},} \aap, 352, 555

\bibitem[{N. {Anugu}(2026){Anugu}}]{AnuguZenodo2026}
{Anugu}, N. 2026, Software and Figures for ``Empirical {H}- and {K}-band Limb Darkening for 31 {CHARA} Stars: A Near-Infrared Benchmark for Stellar-Atmosphere Models'', v1.0.0 Zenodo, \dodoi{10.5281/zenodo.21285841}

\bibitem[{N. {Anugu} {et~al.}(2020){Anugu}, {Le Bouquin}, {Monnier}, {Kraus}, {Setterholm}, {Labdon}, {Davies}, {Lanthermann}, {Gardner}, {Ennis}, {Johnson}, {Ten Brummelaar}, {Schaefer}, \& {Sturmann}}]{Anugu2020}
{Anugu}, N., {Le Bouquin}, J.-B., {Monnier}, J.~D., {et~al.} 2020, \bibinfo{title}{{MIRC-X: A Highly Sensitive Six-telescope Interferometric Imager at the CHARA Array},} \aj, 160, 158, \dodoi{10.3847/1538-3881/aba957}

\bibitem[{N. {Anugu} {et~al.}(2024){Anugu}, {Gies}, {Roettenbacher}, {Monnier}, {Montarg{\'e}s}, {M{\'e}rand}, {Baron}, {Schaefer}, {Shepard}, {Kraus}, {Anderson}, {Codron}, {Gardner}, {Gutierrez}, {K{\"o}hler}, {Kubiak}, {Lanthermann}, {Majoinen}, {Scott}, \& {Vollmann}}]{Anugu2024}
{Anugu}, N., {Gies}, D.~R., {Roettenbacher}, R.~M., {et~al.} 2024, \bibinfo{title}{{Time Evolution Images of the Hypergiant RW Cephei during the Rebrightening Phase Following the Great Dimming},} \apjl, 973, L5, \dodoi{10.3847/2041-8213/ad736c}

\bibitem[{N. {Anugu} {et~al.}(2026){Anugu}, {Turner}, {ten Brummelaar}, {Schaefer}, {B{\'e}rio}, {Farrington}, {Flores}, {Gies}, {Kraus}, {Ligon}, {Majoinen}, {Monnier}, {Mourard}, {Scott}, \& {Vargas}}]{Anugu2026}
{Anugu}, N., {Turner}, N.~H., {ten Brummelaar}, T.~A., {et~al.} 2026, \bibinfo{title}{{CHARA Array delay lines: upgrades, performance, and future directions},} Journal of Astronomical Telescopes, Instruments, and Systems, 12, 015008, \dodoi{10.1117/1.JATIS.12.1.015008}

\bibitem[{T. {Arentoft} {et~al.}(2019){Arentoft}, {Grundahl}, {White}, {Slumstrup}, {Handberg}, {Lund}, {Brogaard}, \& {others}}]{epsTau_Arentoft2019}
{Arentoft}, T., {Grundahl}, F., {White}, T.~R., {et~al.} 2019, \bibinfo{title}{{Asteroseismology of the Hyades red giant and planet host epsilon Tauri},} \aap, 622, A190, \dodoi{10.1051/0004-6361/201834690}

\bibitem[{J.~P. {Aufdenberg} {et~al.}(2005){Aufdenberg}, {Ludwig}, \& {Kervella}}]{Aufdenberg2005}
{Aufdenberg}, J.~P., {Ludwig}, H.~G., \& {Kervella}, P. 2005, \bibinfo{title}{{On the Limb Darkening, Spectral Energy Distribution, and Temperature Structure of Procyon},} \apj, 633, 424, \dodoi{10.1086/452622}

\bibitem[{C.~A.~L. {Bailer-Jones} {et~al.}(2021){Bailer-Jones}, {Rybizki}, {Fouesneau}, {Demleitner}, \& {Andrae}}]{BailerJones2021}
{Bailer-Jones}, C.~A.~L., {Rybizki}, J., {Fouesneau}, M., {Demleitner}, M., \& {Andrae}, R. 2021, \bibinfo{title}{{Estimating Distances from Parallaxes. V. Geometric and Photogeometric Distances to 1.47 Billion Stars in Gaia Early Data Release 3},} \aj, 161, 147, \dodoi{10.3847/1538-3881/abd806}

\bibitem[{E.~K. {Baines} {et~al.}(2018){Baines}, {Armstrong}, {Schmitt}, {Zavala}, {Benson}, {Hutter}, {Tycner}, \& {van Belle}}]{Baines2018}
{Baines}, E.~K., {Armstrong}, J.~T., {Schmitt}, H.~R., {et~al.} 2018, \bibinfo{title}{{Fundamental Parameters of 87 Stars from the Navy Precision Optical Interferometer},} \aj, 155, 30, \dodoi{10.3847/1538-3881/aa9d8b}

\bibitem[{E.~K. {Baines} {et~al.}(2025){Baines}, {Clark}, {Kingsley}, {Schmitt}, \& {Stone}}]{Baines2025}
{Baines}, E.~K., {Clark}, J.~H., {Kingsley}, B.~I., {Schmitt}, H.~R., \& {Stone}, J.~M. 2025, \bibinfo{title}{{Vintage NPOI: New and Updated Angular Diameters for 145 Stars},} \aj, 169, 293, \dodoi{10.3847/1538-3881/adc930}

\bibitem[{E.~K. {Baines} {et~al.}(2009){Baines} {et~al.}}]{Baines2009ApJ701154}
{Baines}, E.~K., {et~al.} 2009, \bibinfo{title}{{CHARA Array angular diameter measurements},} \apj, 701, 154

\bibitem[{E.~K. {Baines} {et~al.}(2010){Baines} {et~al.}}]{Baines2010ApJ7101365}
{Baines}, E.~K., {et~al.} 2010, \bibinfo{title}{{CHARA Array angular diameter measurements},} \apj, 710, 1365

\bibitem[{P. {Berio} {et~al.}(2011){Berio} {et~al.}}]{Berio2011AA53559}
{Berio}, P., {et~al.} 2011, \bibinfo{title}{{Interferometric angular diameter measurements},} \aap, 535, A59

\bibitem[{H.~E. {Bond} {et~al.}(2015){Bond}, {Gilliland}, {Schaefer}, {Demarque}, {Girard}, {Holberg}, {Gudehus}, \& {others}}]{alfCMi_Bond2015}
{Bond}, H.~E., {Gilliland}, R.~L., {Schaefer}, G.~H., {et~al.} 2015, \bibinfo{title}{{Hubble Space Telescope Astrometry of the Procyon System},} \apj, 813, 106, \dodoi{10.1088/0004-637X/813/2/106}

\bibitem[{M. {Bottom} {et~al.}(2015){Bottom}, {Kuhn}, {Mennesson}, {Mawet}, {Shelton}, {Wallace}, \& {Serabyn}}]{delAnd_Bottom2015}
{Bottom}, M., {Kuhn}, J., {Mennesson}, B., {et~al.} 2015, \bibinfo{title}{{Resolving the Delta Andromedae Spectroscopic Binary with Direct Imaging},} \apj, 809, 11, \dodoi{10.1088/0004-637X/809/1/11}

\bibitem[{L. {Bourg\'{e}s} {et~al.}(2017){Bourg\'{e}s}, {Mella}, {Lafrasse}, {Duvert}, {Chelli}, {Le Bouquin}, {Delfosse}, \& {Chesneau}}]{Bourges2017}
{Bourg\'{e}s}, L., {Mella}, G., {Lafrasse}, S., {et~al.} 2017, \bibinfo{title}{{VizieR Online Data Catalog: JMMC Stellar Diameters Catalogue - JSDC. Version 2 (Bourges+, 2017)},} VizieR Online Data Catalog, II/346

\bibitem[{T.~S. {Boyajian} {et~al.}(2009){Boyajian} {et~al.}}]{Boyajian2009ApJ6911243}
{Boyajian}, T.~S., {et~al.} 2009, \bibinfo{title}{{CHARA Array angular diameter measurements},} \apj, 691, 1243

\bibitem[{T.~S. Boyajian {et~al.}(2012)Boyajian, von Braun, van Belle, McAlister, ten Brummelaar, Kane, Muirhead, Jones, White, Schaefer, Ciardi, Henry, López-Morales, Ridgway, Gies, Jao, Rojas-Ayala, Parks, Sturmann, Sturmann, Turner, Farrington, Goldfinger, \& Berger}]{boyajian_stellar_2012}
Boyajian, T.~S., von Braun, K., van Belle, G., {et~al.} 2012, \bibinfo{title}{{STELLAR} {DIAMETERS} {AND} {TEMPERATURES}. {II}. {MAIN}-{SEQUENCE} {K}- {AND} {M}-{STARS},} The Astrophysical Journal, 757, 112, \dodoi{10.1088/0004-637X/757/2/112}

\bibitem[{A. {Chelli} {et~al.}(2016){Chelli}, {Duvert}, {Bourg{\`e}s}, {Mella}, {Lafrasse}, {Bonneau}, \& {Chesneau}}]{Chelli2016}
{Chelli}, A., {Duvert}, G., {Bourg{\`e}s}, L., {et~al.} 2016, \bibinfo{title}{{Pseudomagnitudes and differential surface brightness: Application to the apparent diameter of stars},} \aap, 589, A112, \dodoi{10.1051/0004-6361/201527484}

\bibitem[{A. {Chowhan} {et~al.}(2026){Chowhan}, {Bedding}, {Huber}, {Joel Ong}, {Schimak}, {Li}, {Crawford}, \& {White}}]{2026MNRAS.548ag719C}
{Chowhan}, A., {Bedding}, T.~R., {Huber}, D., {et~al.} 2026, \bibinfo{title}{{CHARA interferometry and TESS asteroseismology of the core-helium burning red giant {\ensuremath{\kappa}} Cyg},} \mnras, 548, stag719, \dodoi{10.1093/mnras/stag719}

\bibitem[{A. Claret(2000)Claret}]{claret_new_2000}
Claret, A. 2000, \bibinfo{title}{A new non-linear limb-darkening law for {LTE} stellar atmosphere models. Calculations for $-5.0 \le \log [{M}/{H}] \le +1$, $2000\,\mathrm{K} \le T_{\rm eff} \le 50000\,\mathrm{K}$ at several surface gravities,} Astronomy and Astrophysics, 363, 1081, \dodoi{10.1051/0004-6361:20000073}

\bibitem[{A. Claret(2018)Claret}]{claret_mucrit_2018}
Claret, A. 2018, \bibinfo{title}{A new method to compute limb-darkening coefficients for stellar atmosphere models with spherical symmetry: the space missions TESS, Kepler, Corot, and MOST,} arXiv e-prints, arXiv:1804.10135.
\newblock \doarXiv{1804.10135}

\bibitem[{A. Claret \& S. Bloemen(2011)Claret \& Bloemen}]{claret_gravity_2011}
Claret, A., \& Bloemen, S. 2011, \bibinfo{title}{Gravity and limb-darkening coefficients for the {Kepler}, {CoRoT}, {Spitzer}, uvby, {UBVRIJHK}, and {Sloan} photometric systems,} Astronomy and Astrophysics, 529, A75, \dodoi{10.1051/0004-6361/201116451}

\bibitem[{A. {Claret} {et~al.}(2025){Claret}, {Hauschildt}, \& {Torres}}]{Claret2025A&A...699A..97C}
{Claret}, A., {Hauschildt}, P.~H., \& {Torres}, G. 2025, \bibinfo{title}{{Limb-darkening coefficients for four-term and power-2 laws for the JWST mission adopting spherical PHOENIX models at high resolution: NIRCam, NIRISS, and NIRSpec passbands},} \aap, 699, A97, \dodoi{10.1051/0004-6361/202554770}

\bibitem[{L.-P. {Coulombe} {et~al.}(2024){Coulombe}, {Roy}, \& {Benneke}}]{Coulombe2024AJ}
{Coulombe}, L.-P., {Roy}, P.-A., \& {Benneke}, B. 2024, \bibinfo{title}{{Biases in Exoplanet Transmission Spectra Introduced by Limb-darkening Parametrization},} \aj, 168, 227, \dodoi{10.3847/1538-3881/ad7aef}

\bibitem[{L. {da Silva} {et~al.}(2006){da Silva}, {Girardi}, {Pasquini}, {Setiawan}, {von der Luehe}, {de Medeiros}, {Hatzes}, {Doellinger}, \& {Weiss}}]{gamTau_daSilva2006}
{da Silva}, L., {Girardi}, L., {Pasquini}, L., {et~al.} 2006, \bibinfo{title}{{Basic physical parameters of a selected sample of evolved stars},} \aap, 458, 609, \dodoi{10.1051/0004-6361:20065105}

\bibitem[{N. {Espinoza} \& A. {Jord{\'a}n}(2015){Espinoza} \& {Jord{\'a}n}}]{Espinoza2015MNRAS.450.1879E}
{Espinoza}, N., \& {Jord{\'a}n}, A. 2015, \bibinfo{title}{{Limb darkening and exoplanets: testing stellar model atmospheres and identifying biases in transit parameters},} \mnras, 450, 1879, \dodoi{10.1093/mnras/stv744}

\bibitem[{T. {Gardner} {et~al.}(2022){Gardner}, {Monnier}, {Fekel}, {Le Bouquin}, {Scovera}, {Schaefer}, {Kraus}, {Adams}, {Anugu}, {Berger}, {Ten Brummelaar}, {Davies}, {Ennis}, {Gies}, {Johnson}, {Kervella}, {Kratter}, {Labdon}, {Lanthermann}, {Sahlmann}, \& {Setterholm}}]{Gardner2022AJ....164..184G}
{Gardner}, T., {Monnier}, J.~D., {Fekel}, F.~C., {et~al.} 2022, \bibinfo{title}{{ARMADA. II. Further Detections of Inner Companions to Intermediate-mass Binaries with Microarcsecond Astrometry at CHARA and VLTI},} \aj, 164, 184, \dodoi{10.3847/1538-3881/ac8eae}

\bibitem[{D. {Grant} \& H. {Wakeford}(2024){Grant} \& {Wakeford}}]{Grant2024JOSS}
{Grant}, D., \& {Wakeford}, H. 2024, \bibinfo{title}{{ExoTiC-LD: thirty seconds to stellar limb-darkening coefficients},} The Journal of Open Source Software, 9, 6816, \dodoi{10.21105/joss.06816}

\bibitem[{D.~B. {Guenther} {et~al.}(2005){Guenther}, {Kallinger}, {Reegen}, {Weiss}, {Matthews}, {Kuschnig}, {Marchenko}, \& {others}}]{etaBoo_Guenther2005}
{Guenther}, D.~B., {Kallinger}, T., {Reegen}, P., {et~al.} 2005, \bibinfo{title}{{Stellar Model Analysis of the Oscillation Spectrum of eta Bootis Obtained from MOST},} \apj, 635, 547, \dodoi{10.1086/497387}

\bibitem[{B. {Gustafsson} {et~al.}(2008){Gustafsson}, {Edvardsson}, {Eriksson}, {J{\o}rgensen}, {Nordlund}, \& {Plez}}]{Gustafsson2008}
{Gustafsson}, B., {Edvardsson}, B., {Eriksson}, K., {et~al.} 2008, \bibinfo{title}{{A grid of MARCS model atmospheres for late-type stars. I. Methods and general properties},} \aap, 486, 951, \dodoi{10.1051/0004-6361:200809724}

\bibitem[{R. {Hanbury Brown} {et~al.}(1974){Hanbury Brown}, {Davis}, \& {Allen}}]{HanburyBrown1974}
{Hanbury Brown}, R., {Davis}, J., \& {Allen}, L.~R. 1974, \bibinfo{title}{{The Angular Diameters of 32 Stars},} \mnras, 167, 121, \dodoi{10.1093/mnras/167.1.121}

\bibitem[{P.~H. {Hauschildt} {et~al.}(2025){Hauschildt}, {Barman}, {Baron}, {Aufdenberg}, \& {Schweitzer}}]{Hauschildt2025A&A...698A..47H}
{Hauschildt}, P.~H., {Barman}, T., {Baron}, E., {Aufdenberg}, J.~P., \& {Schweitzer}, A. 2025, \bibinfo{title}{{The NewEra model grid},} \aap, 698, A47, \dodoi{10.1051/0004-6361/202554171}

\bibitem[{D. {Hestroffer}(1997){Hestroffer}}]{Hestroffer1997}
{Hestroffer}, D. 1997, \bibinfo{title}{{Centre to limb darkening of stars. New model and application to stellar interferometry.},} \aap, 327, 199

\bibitem[{M.~M. {Hohle} {et~al.}(2010){Hohle}, {Neuhaeuser}, \& {Schutz}}]{Hohle2010}
{Hohle}, M.~M., {Neuhaeuser}, R., \& {Schutz}, B.~F. 2010, \bibinfo{title}{{Masses and luminosities of O- and B-type stars and red supergiants},} \an, 331, 349, \dodoi{10.1002/asna.200911355}

\bibitem[{I.~D. {Howarth}(2011){Howarth}}]{Howarth2011MNRAS.418.1165H}
{Howarth}, I.~D. 2011, \bibinfo{title}{{On stellar limb darkening and exoplanetary transits},} \mnras, 418, 1165, \dodoi{10.1111/j.1365-2966.2011.19568.x}

\bibitem[{P. {Kervella} {et~al.}(2022){Kervella}, {Arenou}, \& {Th{\'e}venin}}]{Kervella2022A&A...657A...7K}
{Kervella}, P., {Arenou}, F., \& {Th{\'e}venin}, F. 2022, \bibinfo{title}{{Stellar and substellar companions from Gaia EDR3. Proper-motion anomaly and resolved common proper-motion pairs},} \aap, 657, A7, \dodoi{10.1051/0004-6361/202142146}

\bibitem[{P. {Kervella} {et~al.}(2017){Kervella}, {Bigot}, {Gallenne}, \& {Th{\'e}venin}}]{Kervella2017}
{Kervella}, P., {Bigot}, L., {Gallenne}, A., \& {Th{\'e}venin}, F. 2017, \bibinfo{title}{{The radii and limb darkenings of {\ensuremath{\alpha}} Centauri A and B . Interferometric measurements with VLTI/PIONIER},} \aap, 597, A137, \dodoi{10.1051/0004-6361/201629505}

\bibitem[{E. {Knudstrup} {et~al.}(2023){Knudstrup}, {Lund}, {Fredslund Andersen}, {Rorsted}, {Perez Hernandez}, {Grundahl}, {Palle}, \& {others}}]{gamCep_Knudstrup2023}
{Knudstrup}, E., {Lund}, M.~N., {Fredslund Andersen}, M., {et~al.} 2023, \bibinfo{title}{{Solar-like oscillations in gamma Cephei A as seen through SONG and TESS. A seismic study of gamma Cephei A},} \aap, 675, A197, \dodoi{10.1051/0004-6361/202346707}

\bibitem[{Z. {Kopal}(1950){Kopal}}]{Kopal1950}
{Kopal}, Z. 1950, \bibinfo{title}{{Detailed effects of limb darkening upon light and velocity curves of close binary systems},} Harvard College Observatory Circular, 454, 1

\bibitem[{N. {Kostogryz} {et~al.}(2023){Kostogryz}, {Shapiro}, {Witzke}, {Grant}, {Wakeford}, {Stevenson}, {Solanki}, \& {Gizon}}]{Kostogryz2023}
{Kostogryz}, N., {Shapiro}, A.~I., {Witzke}, V., {et~al.} 2023, \bibinfo{title}{{MPS-ATLAS Library of Stellar Model Atmospheres and Spectra},} Research Notes of the American Astronomical Society, 7, 39, \dodoi{10.3847/2515-5172/acc180}

\bibitem[{N.~M. {Kostogryz} {et~al.}(2022){Kostogryz}, {Witzke}, {Shapiro}, {Solanki}, {Maxted}, {Kurucz}, \& {Gizon}}]{Kostogryz2022}
{Kostogryz}, N.~M., {Witzke}, V., {Shapiro}, A.~I., {et~al.} 2022, \bibinfo{title}{{Stellar limb darkening. A new MPS-ATLAS library for Kepler, TESS, CHEOPS, and PLATO passbands},} \aap, 666, A60, \dodoi{10.1051/0004-6361/202243722}

\bibitem[{N.~M. {Kostogryz} {et~al.}(2024){Kostogryz}, {Shapiro}, {Witzke}, {Cameron}, {Gizon}, {Krivova}, {Ludwig}, {Maxted}, {Seager}, {Solanki}, \& {Valenti}}]{Kostogryz2024}
{Kostogryz}, N.~M., {Shapiro}, A.~I., {Witzke}, V., {et~al.} 2024, \bibinfo{title}{{Magnetic origin of the discrepancy between stellar limb-darkening models and observations},} Nature Astronomy, 8, 929, \dodoi{10.1038/s41550-024-02252-5}

\bibitem[{R. {Kurucz}(1993){Kurucz}}]{Kurucz1993}
{Kurucz}, R. 1993, \bibinfo{title}{{ATLAS9 Stellar Atmosphere Programs and 2 km/s grid.},} Robert Kurucz CD-ROM, 13

\bibitem[{S. {Lacour} {et~al.}(2008){Lacour}, {Meimon}, {Thi{\'e}baut}, {Perrin}, {Verhoelst}, {Pedretti}, {Schuller}, {Mugnier}, {Monnier}, {Berger}, {Haubois}, {Poncelet}, {Le Besnerais}, {Eriksson}, {Millan-Gabet}, {Ragland}, {Lacasse}, \& {Traub}}]{Lacour2008A&A...485..561L}
{Lacour}, S., {Meimon}, S., {Thi{\'e}baut}, E., {et~al.} 2008, \bibinfo{title}{{The limb-darkened Arcturus: imaging with the IOTA/IONIC interferometer},} \aap, 485, 561, \dodoi{10.1051/0004-6361:200809611}

\bibitem[{J.-B. Le~Bouquin {et~al.}(2024)Le~Bouquin, Anugu, Davies, Gardner, Ibrahim, \& Monnier}]{lebouquin_2024}
Le~Bouquin, J.-B., Anugu, N., Davies, C.~L., {et~al.} 2024, CHARA MIRC-X and MYSTIC Data Reduction Pipeline, Zenodo, \dodoi{10.5281/zenodo.12735292}

\bibitem[{B.-C. {Lee} {et~al.}(2014){Lee}, {Han}, {Park}, {Mkrtichian}, \& {Kim}}]{Lee2014}
{Lee}, B.-C., {Han}, I., {Park}, M.-G., {Mkrtichian}, D.~E., \& {Kim}, K.-M. 2014, \bibinfo{title}{{Low-amplitude and long-period radial velocity variations in giants HD 3574, 63 Cygni, and HD 216946},} \aap, 566, A124, \dodoi{10.1051/0004-6361/201321863}

\bibitem[{Z. {Magic} {et~al.}(2015){Magic}, {Chiavassa}, {Collet}, \& {Asplund}}]{Magic2015}
{Magic}, Z., {Chiavassa}, A., {Collet}, R., \& {Asplund}, M. 2015, \bibinfo{title}{{The Stagger-grid: A grid of 3D stellar atmosphere models. IV. Limb darkening coefficients},} \aap, 573, A90, \dodoi{10.1051/0004-6361/201423804}

\bibitem[{Z. {Magic} {et~al.}(2013){Magic}, {Collet}, \& {Asplund}}]{Magic2013}
{Magic}, Z., {Collet}, R., \& {Asplund}, M. 2013, \bibinfo{title}{{The Stagger-grid: A Grid of 3D Stellar Atmosphere Models},} in EAS Publications Series, Vol.~63, EAS Publications Series, ed. G.~{Alecian}, Y.~{Lebreton}, O.~{Richard}, \& G.~{Vauclair}, 367--371, \dodoi{10.1051/eas/1363041}

\bibitem[{A. {Massarotti} {et~al.}(2008){Massarotti}, {Latham}, {Stefanik}, \& {Fogel}}]{Massarotti2008}
{Massarotti}, A., {Latham}, D.~W., {Stefanik}, R.~P., \& {Fogel}, J. 2008, \bibinfo{title}{{Rotational and Radial Velocities for a Sample of 761 HIPPARCOS Giants and the Role of Binarity},} \aj, 135, 209, \dodoi{10.1088/0004-6256/135/1/209}

\bibitem[{P.~F.~L. {Maxted}(2018){Maxted}}]{Maxted2018A&A...616A..39M}
{Maxted}, P.~F.~L. 2018, \bibinfo{title}{{Comparison of the power-2 limb-darkening law from the STAGGER-grid to Kepler light curves of transiting exoplanets},} \aap, 616, A39, \dodoi{10.1051/0004-6361/201832944}

\bibitem[{P.~F.~L. {Maxted}(2023){Maxted}}]{Maxted2023}
{Maxted}, P. F.~L. 2023, \bibinfo{title}{{Limb darkening measurements from TESS and Kepler light curves of transiting exoplanets},} \mnras, 519, 3723, \dodoi{10.1093/mnras/stac3741}

\bibitem[{A. {M{\'e}rand}(2022){M{\'e}rand}}]{Merand2022SPIE12183E..1NM}
{M{\'e}rand}, A. 2022, \bibinfo{title}{{Flexible spectro-interferometric modelling of OIFITS data with PMOIRED},} in Society of Photo-Optical Instrumentation Engineers (SPIE) Conference Series, Vol. 12183, Optical and Infrared Interferometry and Imaging VIII, ed. A.~{M{\'e}rand}, S.~{Sallum}, \& J.~{Sanchez-Bermudez}, 121831N, \dodoi{10.1117/12.2626700}

\bibitem[{A. {M{\'e}rand} {et~al.}(2010){M{\'e}rand}, {Kervella}, {Barban}, {Josselin}, {ten Brummelaar}, {McAlister}, {Coud{\'e} du Foresto}, {Ridgway}, {Turner}, {Sturmann}, {Sturmann}, {Goldfinger}, \& {Farrington}}]{Merand2010A&A...517A..64M}
{M{\'e}rand}, A., {Kervella}, P., {Barban}, C., {et~al.} 2010, \bibinfo{title}{{Interferometric radius and limb darkening of the asteroseismic red giant {\ensuremath{\eta}} Serpentis with the CHARA Array},} \aap, 517, A64, \dodoi{10.1051/0004-6361/200912103}

\bibitem[{J.~D. {Monnier} {et~al.}(2007){Monnier}, {Zhao}, {Pedretti}, {Thureau}, {Ireland}, {Muirhead}, {Berger}, {Millan-Gabet}, {Van Belle}, {ten Brummelaar}, {McAlister}, {Ridgway}, {Turner}, {Sturmann}, {Sturmann}, \& {Berger}}]{Monnier2007}
{Monnier}, J.~D., {Zhao}, M., {Pedretti}, E., {et~al.} 2007, \bibinfo{title}{{Imaging the Surface of Altair},} Science, 317, 342, \dodoi{10.1126/science.1143205}

\bibitem[{D. {Mourard} {et~al.}(2022){Mourard}, {Berio}, {Pannetier}, {Nardetto}, {Allouche}, {Bailet}, {Dejonghe}, {Geneslay}, {Jacqmart}, {Lagarde}, {Lecron}, {Morand}, {Rousseau}, {Salabert}, {Spang}, {Albrecht}, {Anugu}, {Bourg{\`e}s}, {ten Brummelaar}, {Creevey}, {Deheuvels}, {Domiciano de Souza}, {Gies}, {Ligi}, {Mella}, {Perraut}, {Schaefer}, \& {Wittkowski}}]{Mourard2022}
{Mourard}, D., {Berio}, P., {Pannetier}, C., {et~al.} 2022, \bibinfo{title}{{CHARA/SPICA: a six-telescope visible instrument for the CHARA Array},} in Society of Photo-Optical Instrumentation Engineers (SPIE) Conference Series, Vol. 12183, Optical and Infrared Interferometry and Imaging VIII, 1218308, \dodoi{10.1117/12.2628881}

\bibitem[{H.~R. {Neilson} \& J.~B. {Lester}(2011){Neilson} \& {Lester}}]{Neilson2011}
{Neilson}, H.~R., \& {Lester}, J.~B. 2011, \bibinfo{title}{{Limb darkening in spherical stellar atmospheres},} \aap, 530, A65, \dodoi{10.1051/0004-6361/201116623}

\bibitem[{H.~R. {Neilson} \& J.~B. {Lester}(2013){Neilson} \& {Lester}}]{Neilson2013}
{Neilson}, H.~R., \& {Lester}, J.~B. 2013, \bibinfo{title}{{Spherically-symmetric model stellar atmospheres and limb darkening. I. Limb-darkening laws, gravity-darkening coefficients and angular diameter corrections for red giant stars},} \aap, 554, A98, \dodoi{10.1051/0004-6361/201321502}

\bibitem[{J.~A. {Patel} \& N. {Espinoza}(2022){Patel} \& {Espinoza}}]{Patel2022AJ}
{Patel}, J.~A., \& {Espinoza}, N. 2022, \bibinfo{title}{{Empirical Limb-darkening Coefficients and Transit Parameters of Known Exoplanets from TESS},} \aj, 163, 228, \dodoi{10.3847/1538-3881/ac5f55}

\bibitem[{R.~M. Roettenbacher {et~al.}(2016)Roettenbacher, Monnier, Korhonen, Aarnio, Baron, Che, Harmon, Kővári, Kraus, Schaefer, Torres, Zhao, ten Brummelaar, Sturmann, \& Sturmann}]{roettenbacher_no_2016}
Roettenbacher, R.~M., Monnier, J.~D., Korhonen, H., {et~al.} 2016, \bibinfo{title}{No {Sun}-like dynamo on the active star ζ {Andromedae} from starspot asymmetry,} Nature, 533, 217, \dodoi{10.1038/nature17444}

\bibitem[{J. {Rudrasingam} {et~al.}(2026){Rudrasingam}, {Bedding}, {Pope}, {Pedersen}, {Lund}, {White}, \& {Hey}}]{alfCas_Rudrasingam2026}
{Rudrasingam}, J., {Bedding}, T.~R., {Pope}, B. J.~S., {et~al.} 2026, \bibinfo{title}{{Halo Photometry and Asteroseismology for 98 of the Brightest Stars Observed by TESS},} \mnras, 547, \dodoi{10.1093/mnras/stag413}

\bibitem[{Z. {Rustamkulov} {et~al.}(2023){Rustamkulov}, {Sing}, {Mukherjee}, {May}, {Kirk}, {Schlawin}, {Line}, {Piaulet}, {Carter}, {Batalha}, {Goyal}, {L{\'o}pez-Morales}, {Lothringer}, {MacDonald}, {Moran}, {Stevenson}, {Wakeford}, {Espinoza}, {Bean}, {Batalha}, {Benneke}, {Berta-Thompson}, {Crossfield}, {Gao}, {Kreidberg}, {Powell}, {Cubillos}, {Gibson}, {Leconte}, {Molaverdikhani}, {Nikolov}, {Parmentier}, {Roy}, {Taylor}, {Turner}, {Wheatley}, {Aggarwal}, {Ahrer}, {Alam}, {Alderson}, {Allen}, {Banerjee}, {Barat}, {Barrado}, {Barstow}, {Bell}, {Blecic}, {Brande}, {Casewell}, {Changeat}, {Chubb}, {Crouzet}, {Daylan}, {Decin}, {D{\'e}sert}, {Mikal-Evans}, {Feinstein}, {Flagg}, {Fortney}, {Harrington}, {Heng}, {Hong}, {Hu}, {Iro}, {Kataria}, {Kempton}, {Krick}, {Lendl}, {Lillo-Box}, {Louca}, {Lustig-Yaeger}, {Mancini}, {Mansfield}, {Mayne}, {Miguel}, {Morello}, {Ohno}, {Palle}, {Petit dit de la Roche}, {Rackham}, {Radica}, {Ramos-Rosado}, {Redfield}, {Rogers}, {Shkolnik}, {Southworth}, {Teske}, {Tremblin},
  {Tucker}, {Venot}, {Waalkes}, {Welbanks}, {Zhang}, \& {Zieba}}]{Rustamkulov2023}
{Rustamkulov}, Z., {Sing}, D.~K., {Mukherjee}, S., {et~al.} 2023, \bibinfo{title}{{Early Release Science of the exoplanet WASP-39b with JWST NIRSpec PRISM},} \nat, 614, 659, \dodoi{10.1038/s41586-022-05677-y}

\bibitem[{K. Schwarzschild(1906)Schwarzschild}]{Schwarzschild1906}
Schwarzschild, K. 1906, \bibinfo{title}{Über das Gleichgewicht der Sonnenatmosphäre,} Nachrichten von der Gesellschaft der Wissenschaften zu Göttingen, Mathematisch-Physikalische Klasse, 41

\bibitem[{B.~R. {Setterholm} {et~al.}(2023){Setterholm}, {Monnier}, {Le Bouquin}, {Anugu}, {Ennis}, {Jocou}, {Ibrahim}, {Kraus}, {Anderson}, {Chhabra}, {Codron}, {Farrington}, {Flores}, {Gardner}, {Gutierrez}, {Lanthermann}, {Majoinen}, {Mortimer}, {Schaefer}, {Scott}, {ten Brummelaar}, \& {Vargas}}]{Setterholm2023}
{Setterholm}, B.~R., {Monnier}, J.~D., {Le Bouquin}, J.-B., {et~al.} 2023, \bibinfo{title}{{MYSTIC: a high angular resolution K-band imager at CHARA},} Journal of Astronomical Telescopes, Instruments, and Systems, 9, 025006, \dodoi{10.1117/1.JATIS.9.2.025006}

\bibitem[{C. {Soubiran} {et~al.}(2016){Soubiran}, {Le Campion}, {Brouillet}, \& {Chemin}}]{Soubiran2016}
{Soubiran}, C., {Le Campion}, J.-F., {Brouillet}, N., \& {Chemin}, L. 2016, \bibinfo{title}{{The PASTEL catalogue: 2016 version},} \aap, 591, A118, \dodoi{10.1051/0004-6361/201628497}

\bibitem[{J. {Southworth}(2008){Southworth}}]{Southworth2008}
{Southworth}, J. 2008, \bibinfo{title}{{Homogeneous studies of transiting extrasolar planets - I. Light-curve analyses},} \mnras, 386, 1644, \dodoi{10.1111/j.1365-2966.2008.13145.x}

\bibitem[{I. {Tallon-Bosc} {et~al.}(2024){Tallon-Bosc}, {Berger}, {Matter}, {Bourg{\`e}s}, {Mella}, {Benisty}, {Chelli}, {Domiciano de Souza}, {Duch{\^e}ne}, {Le Bouquin}, {Leftley}, {Millour}, {Montarg{\`e}s}, {Perraut}, {Soulez}, {Tallon}, {Thi{\'e}baut}, {Berdeu}, {Cruzal{\`e}bes}, {Haubois}, {Meilland}, {Mourard}, {Nardetto}, {Salabert}, \& {Verbiese}}]{Tallon-Bosc2024}
{Tallon-Bosc}, I., {Berger}, J.-P., {Matter}, A., {et~al.} 2024, \bibinfo{title}{{JMMC: a Service for current \& future optical interferometers},} in SF2A-2024: Proceedings of the Annual meeting of the French Society of Astronomy and Astrophysics, ed. M.~{B{\'e}thermin}, K.~{Bailli{\'e}}, N.~{Lagarde}, J.~{Malzac}, R.~M. {Ouazzani}, J.~{Richard}, O.~{Venot}, \& A.~{Siebert}, 215--216

\bibitem[{T.~A. ten Brummelaar {et~al.}(2005)ten Brummelaar, McAlister, Ridgway, Bagnuolo, Turner, Sturmann, Sturmann, Berger, Ogden, Cadman, Hartkopf, Hopper, \& Shure}]{ten_brummelaar_first_2005}
ten Brummelaar, T.~A., McAlister, H.~A., Ridgway, S.~T., {et~al.} 2005, \bibinfo{title}{First {Results} from the {CHARA} {Array}. {II}. {A} {Description} of the {Instrument},} The Astrophysical Journal, 628, 453, \dodoi{10.1086/430729}

\bibitem[{G.~T. van Belle \& K. von Braun(2009)van Belle \& von Braun}]{van_belle_directly_2009}
van Belle, G.~T., \& von Braun, K. 2009, \bibinfo{title}{Directly {Determined} {Linear} {Radii} and {Effective} {Temperatures} of {Exoplanet} {Host} {Stars},} The Astrophysical Journal, 694, 1085, \dodoi{10.1088/0004-637X/694/2/1085}

\bibitem[{K. {Verma} {et~al.}(2024){Verma}, {Maxted}, {Singh}, {Ludwig}, \& {Sable}}]{Verma2024}
{Verma}, K., {Maxted}, P. F.~L., {Singh}, A., {Ludwig}, H.-G., \& {Sable}, Y. 2024, \bibinfo{title}{{A regularization technique to precisely infer limb darkening using transit measurements: can we estimate stellar surface magnetic fields?},} \mnras, 534, 3893, \dodoi{10.1093/mnras/stae2344}

\bibitem[{G. {Wade} {et~al.}(2025){Wade}, {Oksala}, {Neiner}, {Boucher}, \& {Barron}}]{alfPer_Wade2025}
{Wade}, G., {Oksala}, M., {Neiner}, C., {Boucher}, E., \& {Barron}, J. 2025, \bibinfo{title}{{Magnetic field monitoring of four massive A-F supergiants},} arXiv e-prints.
\newblock \doarXiv{2507.00282}

\bibitem[{M. {Wittkowski} {et~al.}(2001){Wittkowski}, {Hummel}, {Johnston}, {Mozurkewich}, {Hajian}, \& {White}}]{Wittkowski2001}
{Wittkowski}, M., {Hummel}, C.~A., {Johnston}, K.~J., {et~al.} 2001, \bibinfo{title}{{Direct multi-wavelength limb-darkening measurements of three late-type giants with the Navy Prototype Optical Interferometer},} \aap, 377, 981, \dodoi{10.1051/0004-6361:20011124}

\end{thebibliography}
\end{document}